\documentclass[aps,prd,nofootinbib,a4paper,showpacs,floatfix,flushbottom,showkeys]{revtex4-1}

\usepackage[utf8]{inputenc}
\usepackage[T1]{fontenc}
\usepackage{slashed}
\usepackage{times}

\usepackage{amssymb,amsfonts,amsmath,amsthm}
\usepackage{dsfont,bbm}
\usepackage{dcolumn}

\usepackage{graphicx}
\usepackage[caption=false]{subfig}
\usepackage{color}
\usepackage{hhline}

\usepackage{hyperref}
\usepackage{dcolumn}
\usepackage{tikz}

\usepackage[vcentermath]{youngtab}
\usepackage{slashed}
\usepackage{multirow}

\newcommand{\bite}{\begin{itemize}}
\newcommand{\eite}{\end{itemize}}

\newcommand{\identidad}{{\bf 1}}
\newcommand{\bea}{\begin{eqnarray}}
\newcommand{\eea}{\end{eqnarray}}
\newcommand{\res}[1]{Section~\ref{#1}}
\newcommand{\ret}[1]{Table~\ref{#1}}
\newcommand{\refig}[1]{Fig.~\ref{#1}}
\newcommand{\cO}{{\cal O}}
\newcommand{\be}{\begin{equation}}
\newcommand{\ee}{\end{equation}}

\begin{document}

\def\Regensburg{Institute for Theoretical Physics, University of Regensburg, 93040 Regensburg, Germany}

\author{Paula~P\'erez-Rubio}
\email{paula.perez-rubio@ur.de}
\affiliation{\Regensburg}

\author{Sara Collins}
\affiliation{\Regensburg}

\author{Gunnar S.~Bali}
\altaffiliation[Adjunct Faculty: ]{Tata Institute of Fundamental Research, Homi Bhabha Road, Mumbai 400005, India}
\affiliation{\Regensburg}

\date{\today}
\title{Charmed baryon spectroscopy and light flavour symmetry from lattice QCD}

\begin{abstract}
We determine the ground state and first excited state masses of singly
and doubly charmed spin 1/2 and 3/2 baryons with positive and negative
parity.  Configurations with $N_f=2+1$ non-perturbatively improved
Wilson-clover fermions were employed, with the same quark action also
being used for the valence quarks, including the charm. The spectrum
is calculated for pion masses in the range $M_\pi \sim 259-460$~MeV at
a lattice spacing $a\sim 0.075$ fm.  Finite volume effects are studied
comparing lattices with two different linear spatial extents
($1.8\,{\rm fm}$ and $2.4\,{\rm fm}$). The physical point is
approached from the SU(3) limit keeping the flavour averaged light
quark mass fixed. The baryon masses are extrapolated using expansions in
the strange-light quark mass difference. Most particles fall into the
expected SU(3) multiplets with well constrained extrapolations, the
exceptions having a possibly more complex internal structure. Overall
agreement is found with experiment for the masses and splittings of
the singly charmed baryons. As part of the calculation an analysis of
the lower lying charmonium, $D$ and $D_s$ spectra was performed in
order to assess discretisation errors. The gross spectra are
reproduced, including the $D^*_{s0}$, $D_{s1}$ and $D_1$ mesons, while at
this single lattice spacing hyperfine splittings come out $10-20$~MeV
too low.
\end{abstract}

\maketitle

\section{\label{1}Introduction}

Heavy baryons combine relativistic light quarks and non-relativistic
heavy quarks and may have similarities with heavy light mesons and, in
the case of doubly charmed baryons, also quarkonia. These particles
can be understood using a number of theoretical tools, in particular,
Heavy Quark Effective
Theory~(HQET)~\cite{Isgur:1989vq,Eichten:1989zv}, non-relativistic
QCD~(NRQCD)~\cite{Caswell:1985ui,Bodwin:1994jh} and potential
NRQCD~(pNRQCD)~\cite{Pineda:1997bj,Brambilla:1999xf}.  Currently, 19
charmed baryons and 8 bottomed baryons are present in the PDG summary
tables\footnote{With 3 star status or
  higher.}\cite{Agashe:2014kda}. Shortly after the discovery of the
$J/\psi$, the first charmed baryon was detected in 1975, the
$\Lambda_c^+$, at the BNL \cite{Cazzoli:1975et}, 
followed by the discovery of
$\Sigma_c^{++}$ in 1976 at FNAL \cite{Knapp:1976qw} and the first bottomed baryon, the
$\Lambda_b$ in 1981 at CERN \cite{Basile:1981wr}.  In the last decade, bottomed baryons
were studied at the Tevatron and more recently at the LHC, whereas
charmed baryons were mainly discovered at the B-factories. While
masses, lifetimes, widths and form factors have been measured, direct
spin and parity identification is often still missing (they are
assigned from quark model considerations).  This situation will
probably improve with the study of angular distributions of particle
decays in LHC data thanks to the large statistics. In the future, the
Belle-II experiment at SuperKEKB and the PANDA experiment at the FAIR facility will
study singly charmed baryons.  With respect to doubly charmed baryons,
the prospects are not so promising.  Only SELEX published evidence for
the $\Xi^+_{cc}(3520)$ \cite{Mattson:2002vu,Ocherashvili:2004hi} and
its isospin partner $\Xi^{++}_{cc}(3460)$ \cite{Russ:2002bw}, with no
further confirmation by any other experiment.  Besides the
experimental findings, various theoretical approaches have been
employed.  Among these are studies based on quark models
\cite{Copley:1979wj,Capstick:1986bm,Roncaglia:1995az,SilvestreBrac:1996bg,
  Ebert:2002ig,Ebert:2005xj,Roberts:2007ni,Garcilazo:2007eh,Valcarce:2008dr},
HQET \cite{Jenkins:1993ta}, QCD sum rules
\cite{Bagan:1991sc,Bagan:1992tp,Wang:2002ts,Wang:2007sqa,Zhang:2008rt,Zhang:2008pm} and lattice
QCD, using the quenched approximation
\cite{Bowler:1996ws,Flynn:2003vz,Mathur:2002ce,Lewis:2001iz,Chiu:2005zc}
and, more recently, including sea quarks
\cite{Na:2007pv,Liu:2009jc,Briceno:2012wt,
  Alexandrou:2012xk,Basak:2012py,Durr:2012dw,Namekawa:2013vu,Alexandrou:2014sha,Brown:2014ena,Padmanath:2015jea}.

While most previous studies have focused on post- and prediction of
the charmed baryon spectrum, in this work we investigate the light
flavor dependence of the singly and doubly charmed states. Observed
spectra of mesons and baryons have long been understood in terms of
flavor symmetry, with, for the example of three light quark
flavors~($N_f=3$ corresponding to up, down and strange), the mesons
falling into singlets and octets  and the baryons into octets
and decuplets. While SU(2) isospin symmetry is reasonably well
respected in nature, SU(3) flavor symmetry is not. The pattern of
symmetry breaking for the latter can be derived by expanding about the
flavor symmetric limit in the strange-light quark mass
difference~($\delta m_{\ell}=m_s-m_{\ell}$). This leads to the Gell-Mann--Okubo
relations~\cite{GellMann:1962xb,Okubo:1961jc}, which are found to hold
within a few per cent. Enlarging the symmetry group to SU(4) to
include the charm quark provides indications of what charmed mesons
and baryons should exist, however, the mass spectra may be best
explained by treating charm quarks as spectators and considering the
SU(3) symmetry breaking pattern as for the light hadrons.

In the past, studying the flavor structure of hadrons on the lattice
was  mostly  restricted to approaching the physical point, keeping
the strange quark mass approximately constant~(and consistent with
experiment) while reducing the up/down quark mass. Extrapolations to
the physical point, if necessary, are guided by chiral perturbation
theory, which is of uncertain validity in the range where lattice
results are generated depending on the
observable~\cite{Durr:2014oba}. Recently, the QCDSF collaboration
adopted an alternative
strategy~\cite{Bietenholz:2010jr,Bietenholz:2011qq} where, starting
from the $N_f=3$ theory, one approaches the physical point keeping the
average quark mass fixed. This enables one to derive the quark mass
dependence of physical quantities \`a la Gell-Mann--Okubo. The
extrapolation to the physical point is limited by the order of the
expansion required relative to the number and precision of the data
points available to fix the corresponding coefficients. The
incorporation of an additional valence quark, for example, the charm,
into this framework is described
in~\cite{Bietenholz:2011qq}. In this work, we study the
flavor dependence of singly and doubly charmed baryons at a single
lattice spacing for both positive and negative parity states treating
the charm quark as a spectator, with the view to performing a larger
scale analysis on CLS ensembles~\cite{Bruno:2014jqa} including a
continuum extrapolation. For the associated Gell-Mann--Okubo relations
to be applicable, the charmed baryons must fall into the expected
SU(3) multiplets. With only a few exceptions we find this to be the
case. The SU(4) representations naturally suggest interpolators for
creating~(and destroying) the baryon states. We also compare these
interpolators with another basis derived from HQET. 
Similar results for both bases are obtained for
positive parity states, while differences are found for negative
parity states. For the latter, 
some channels are obscured by the presence of two-particle
scattering states of the same quantum numbers. We carefully study 
this as well.

The article is organized as follows. A description of our
computational setup is given in Section~\ref{simdetails}. This
includes a discussion of the leading ${\rm O}(a^2)$ discretization
effects, which could be significant as the simulations were
performed on ensembles at a single lattice spacing $a\sim 0.075$~fm
for which the charm quark mass in lattice units is around $0.4$. We
attempt to quantify these effects by comparing results for the lower
lying charmonium and $D/D_s$ states to experiment. In \res{3} we
motivate our basis of interpolators for the correlation
functions, followed by a discussion of the methodology used to extract the mass
spectrum and the efficacy of our interpolator basis in
\res{varmethod}. The charm quark is partially quenched in this
study~(it does not appear in the sea) and as such must be tuned to
reproduce experiment on each ensemble. This procedure is described in
\res{kappacharm}. We remark that for transparency and in order to
maximize the predictive power of our simulations, we try to make use
of as little experimental input as possible.  For instance, we predict
the absolute meson and baryon masses using one and the same value of
the charm quark mass, rather than quoting splittings relative to a
reference mass, such as that of the  $\eta_c$ meson. Comparing 
two volumes, we  also investigate finite volume effects
in \res{finitesize}. The main
result of the paper, the extrapolation to the physical point of the
charmed baryon spectrum is given in Sections~\ref{extrapmass} and
\ref{extrapfinal} and for the mass splittings in
Section~\ref{extrapdiff}, before comparing with other recent lattice
determinations in \res{compothers}.  We finish with some concluding
remarks in \res{6}.  Additional details are provided in  Appendices
\ref{Appmesons}~(meson effective masses), \ref{AppD}~(finite volume
effects), \ref{AppB} and \ref{derive_gmo}~(derivation of the
Gell-Mann--Okubo expressions) and \ref{AppC}~(fit ranges and
extrapolations).

\section{\label{simdetails}Simulation  details}

We have employed SLiNC ({\bf S}tout {\bf Li}nk {\bf N}on-perturbative
{\bf C}lover)~\cite{Cundy:2009yy} $N_f=2+1$ gauge configurations,
generated by the QCDSF
collaboration~\cite{Bietenholz:2010jr,Bietenholz:2011qq}.  The gluonic
action is tree level Symanzik improved and the fermionic action has a
single level of stout smearing in the hopping terms and unsmeared
links in the clover term. The clover coefficient was determined
non-perturbatively.  The quark masses were chosen by finding the SU(3)
symmetric point where the average octet pion mass, $X_\pi=\sqrt{
(M_\pi^2+2M_K^2)/3}$, coincides with experiment. $M_\pi$
and $M_K$ correspond to the pion and kaon masses, respectively. The
strange and light sea quark masses are then varied so as to approach
the physical point keeping the singlet quark mass, $\overline{ m} =
\frac 13 (2m_{\ell} + m_s) $ fixed up to ${\rm O}(a)$ corrections. The
ensembles used in this work include several pion masses and two
volumes for a single lattice spacing, see Table~\ref{configs}. Note
that the physical SU(3) symmetric value $M_\pi=M_K\approx 411$~MeV was
somewhat missed. Below, we discuss our strategy for correcting for the
``wrong'' trajectory in the mass plane and also quantify the size of
discretization errors in spectral quantities. In order to reduce
auto-correlations, a single measurement was performed per
configuration where the position 
of the source was randomly chosen 
and consecutive configurations are separated by two
trajectories.

\begin{table}[ht!]
\begin{center}
\begin{tabular}{cccccccccccc}
\hline
\hline
$\kappa_\ell$ & $\kappa_s$ & $\kappa_{c_1}$ & $\kappa_{c_2}$ &  
$L/a\times T/a$ &  $M_{\pi}$ (MeV)&$M_{K}$ (MeV)& $LM_\pi$ & $N_{\rm meas}$ & 
$a_{w_0} $ (fm)& $a_{r_0}$ (fm) & $a_{1{\rm S} - 1{\rm P}}$ (fm) \\ 
\hline
$0.12090$  & $0.12090$  & $0.11065$ & $0.1116$ &$24\times 48$  & $471$ & $471$ & $4.3$ & $2747$ &$0.0756(10)$&$0.076(2)$ &\\  
$0.12100$  & $0.12070$  & $0.11065$ & $0.1116$ &$24\times 48$  & $394$ & 493 & $3.6$ & $1018$   &              &$0.076(2)$ &\\  
$0.12104$  & $0.12062$  & $0.11065$ & $0.1116$ &$24\times 48$  & $364$ & 507 & $3.3$ & $934$    &              &$0.076(2)$ &\\  
$0.12090$  & $0.12090$  & $0.1110 $ & $0.1116$ &$32\times 64$  & $461$ & 461 & $5.6$ & $875$    &$0.0748(9)$ &             &$0.072(4)$\\ % k_c=0.111998 
$0.12104$  & $0.12062$  & $0.1110 $ & $0.1116$ &$32\times 64$  & $355$ & 499 & $4.3$ & $989$    &$0.0742(8)$ &             &$0.068(4)$\\ % k_c=0.113047 
$0.121145$ & $0.120413$ & $0.1110 $ & $0.1116$ &$32\times 64$ & $259$  & 530 & $3.2$ & $885$    &$0.0746(9)$ &             &$0.075(4)$\\  % k_c=0.111493
\hline
\hline
\end{tabular}
\caption{\label{configs} Details of the ensembles used. $\kappa_\ell$,
  $\kappa_s$ and $\kappa_{c1/c2}$ correspond to the light~($u/d$),
  strange and charm quark mass parameters, respectively, and the gauge
  coupling is the same in all cases, $\beta = 10/g^2=5.5$. $N_{\rm meas}$
  indicates the number of measurements used for the analysis. 
  The last three 
  columns indicate the values of the lattice spacing for 
  the different ensembles, determined through $w_0$, $r_0$ and
  the $1$S - $1$P charmonium splitting. Note that the lattice spacing 
  determinations were performed using different numbers of configurations
  than those used in the spectroscopy analysis.}
\end{center}
\end{table}
\vspace{-1cm}
\begin{figure}[ht!]
\includegraphics[width=0.5\textwidth]{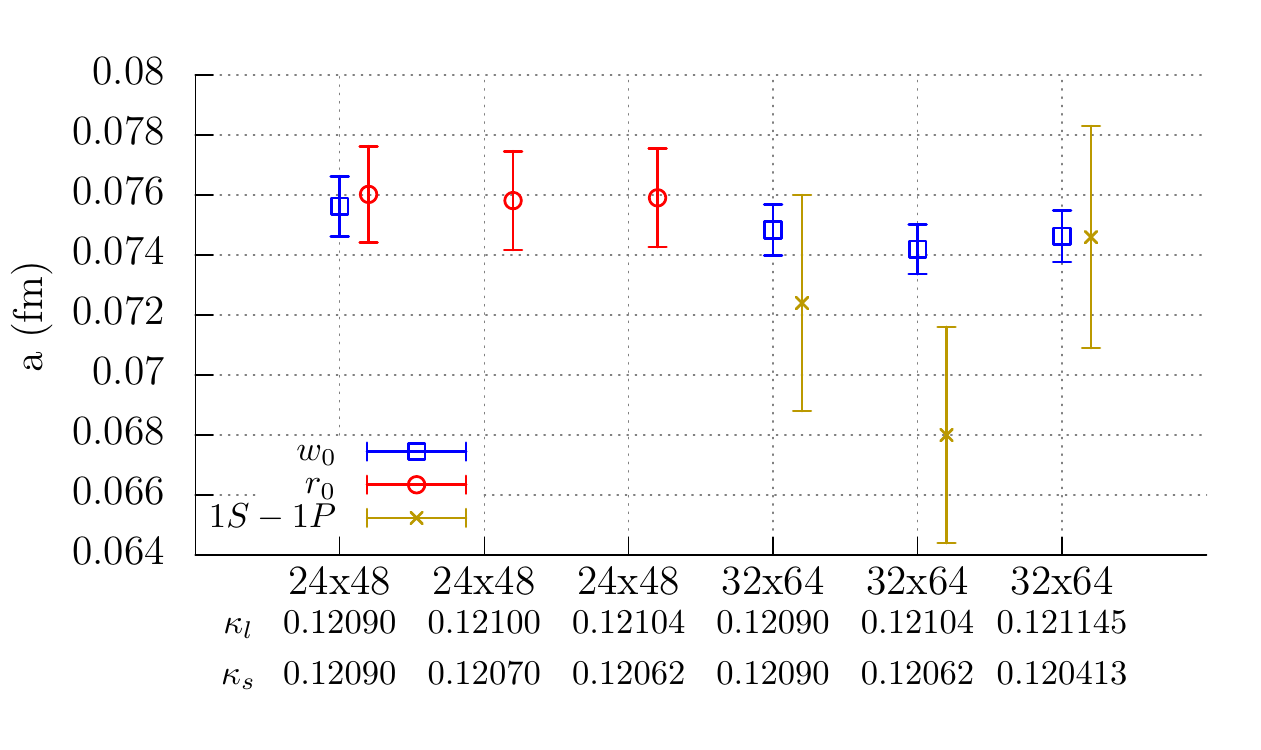}
\vspace{-0.4cm}
\caption{\label{scale} The lattice spacing determined via $w_0$, $r_0$ and
the 1P-1S charmonium splitting for each ensemble.}
\end{figure}

The valence charm quark is treated relativistically using the SLiNC
action. We employed two different values of the mass parameter for
each set of configurations, given in Table~\ref{configs}. These were
chosen so that we could interpolate to the physical charm quark mass
determined by comparing the spin-averaged $1$S charmonium mass to
experiment~(see Section~\ref{kappacharm}). In order to convert
dimensionless lattice results to physical units the lattice spacing
must be determined. We considered three quantities to set the scale:
the spin-averaged 1P-1S charmonium splitting, the Sommer scale,
$r_0$~\cite{Sommer:1993ce} (where we used $r_0=0.50(1)$ fm) and the
Wilson flow observable $w_0$~\cite{Borsanyi:2012zs}, which is related
to $t_0$~\cite{Luscher:2010iy}. \refig{scale} shows the results for
each quantity, where available. The values obtained are reasonably
consistent across the ensembles and also from the different
observables.  We take $a= 0.075$~fm without quoting an error 
\footnote{Note that we take the continuum limit value of $w_0$ at physical quark masses. 
Therefore the precise lattice spacing is ambiguous up to a few percent.}, 
since the
systematics due to the lack of a continuum limit extrapolation will
induce a larger uncertainty.

\begin{figure}[ht!]
\includegraphics[width=0.5\textwidth]{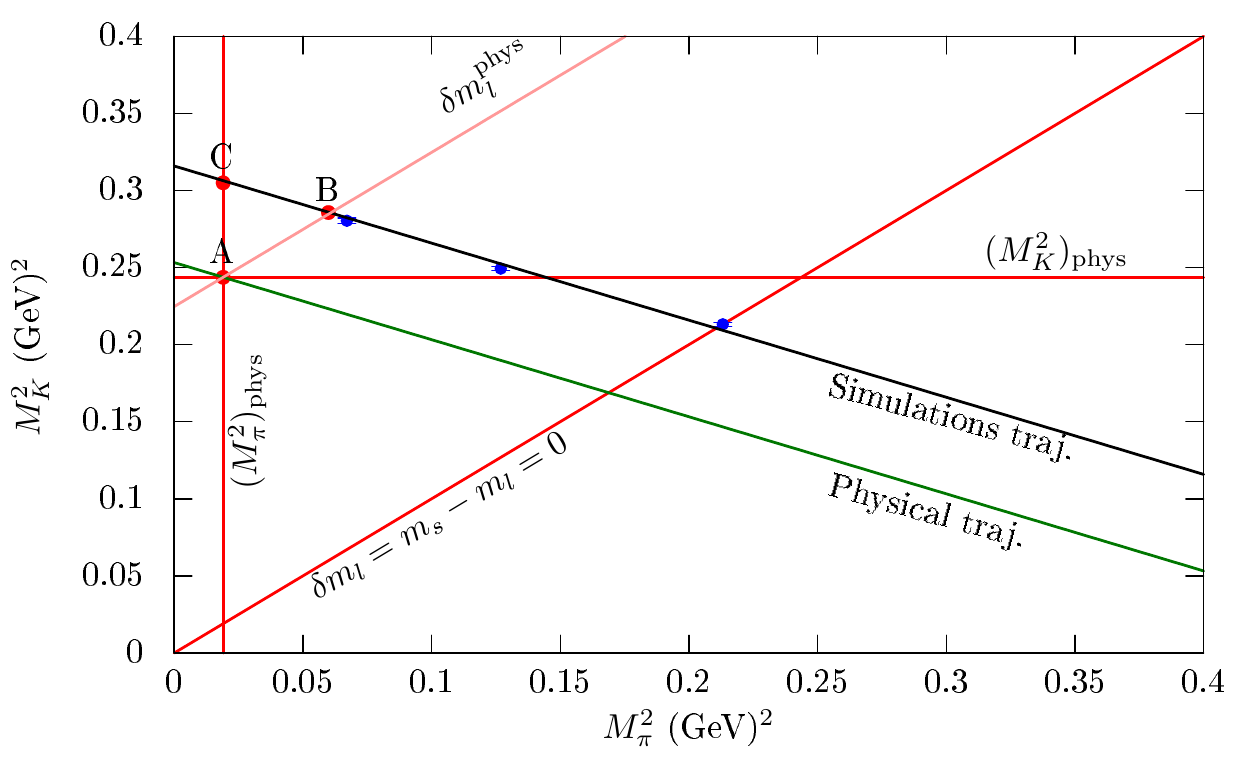}
\caption{\label{simtraj} The pion and kaon masses squared in physical
  units, using $a=0.075$~fm, for our ensembles~(blue circles) compared
  to the physical trajectory. The point A indicates the physical pion
  and kaon masses, while B indicates the point on the simulation
  trajectory corresponding to the physical value for $\delta
  m_{\ell}=m_s-m_{\ell}$ and C shows the (unphysically heavy) kaon mass
  corresponding to physical pion mass~(but unphysical $\delta m_{\ell}$).
%, i.e. .
}
\end{figure}

Our value of $a$ is about $10\%$ smaller than the value given in
Refs.~\cite{Bietenholz:2010jr,Bietenholz:2011qq} ($a\sim 0.083$ fm)
which was set from the flavor singlet baryon mass QCDSF
obtained. However, it is consistent with a later determination by
QCDSF~\cite{Horsley:2013wqa} ($a \sim 0.073(2)$ fm).  Note that the
smaller lattice spacing means our values for the pion masses given in
Table~\ref{configs} differ from those of
Refs.~\cite{Bietenholz:2010jr,Bietenholz:2011qq}. In addition, the
average octet pion mass, $X_\pi$, is larger than the experimental
value by approximately 50 MeV~($V=32^3\times 64$). Recall that
$\overline{m}$ is kept fixed in our simulations and
$X_\pi=X_\pi(\overline{m})$ is approximately constant as $\delta
m_{\ell}=m_s-m_{\ell}$ is varied from the flavor symmetric
point~($\delta m_{\ell}=0$) to the physical value $\delta
m_{\ell}^{\rm phys}$, where $(M_K^{\rm sim})^2- (M^{\rm sim}_\pi)^2 =
(M_K^{\rm phys})^2- (M^{\rm phys}_\pi)^2 \approx 0.225 $
GeV$^{2}$. This means that an extrapolation to the physical pion mass
will result in an unphysically heavy kaon, as illustrated in
Fig.~\ref{simtraj}~(point C).  SU(3) mass multiplets are extrapolated
via Taylor expansions in $\delta m_{\ell}$ starting from the flavor
symmetric point such that an individual hadron mass has the dependence
$M=M_0(\overline{m})+c\delta m_{\ell}+{\rm O}(\delta
m_{\ell}^2)$. Note that the linear coefficient of this expansion~($c$)
does not depend on $\overline{m}$ and, up to quadratic corrections,
$M_0(\overline{m})\approx X_{\rm multiplet}$. The latter quantity denotes the
flavor average for the given multiplet.  In order to make contact
with the physical theory we extrapolate masses within each SU(3)
multiplet to $\delta m^{\rm phys}_\ell$~(point B in
Fig.~\ref{simtraj}) and then shift these masses by an estimate of
$X_{\rm multiplet}^{\rm phys}-X_{\rm multiplet}^{\rm
  sim}$~(i.e. moving from B to A in Fig.~\ref{simtraj}). This
procedure is described in more detail in Section~\ref{extrapmass}.

\vspace{-0.4cm}
\begin{table}[ht!]
   \begin{center}
      \begin{tabular}{c|c}   
         \hline
         \hline
         Particles & Operators \\
         \hline
         $\eta_c,D, D_s$ &$ \bar q_1 \gamma_5 q_2$ \\
         $J/\psi, D^*, D_s^*$&$ \bar q_1 \gamma_i q_2$ \\
         $\chi_{c0}(1P), D_0, D_{s0}$ & $\bar q_1 q_2$\\
         $D_1,D_{s1}$ & $\bar q_1 \gamma_i\gamma_5q_2$, $\bar q_1 \epsilon_{ijk} \gamma_j\gamma_k q_2$  \\
         $\chi_{c1}(1P)$ & $\bar q_1 \gamma_i\gamma_5q_2$\\
         $h_c(1P)$ & $\bar q_1 \epsilon_{ijk} \gamma_j\gamma_k q_2$\\
         \hline \hline
      \end{tabular}
      \caption{\label{meson_op} The meson interpolators employed
to calculate the charmonium, $D$ and $D_s$ spectra.}   
   \end{center}   
\end{table}
\begin{figure}[ht!]
\includegraphics{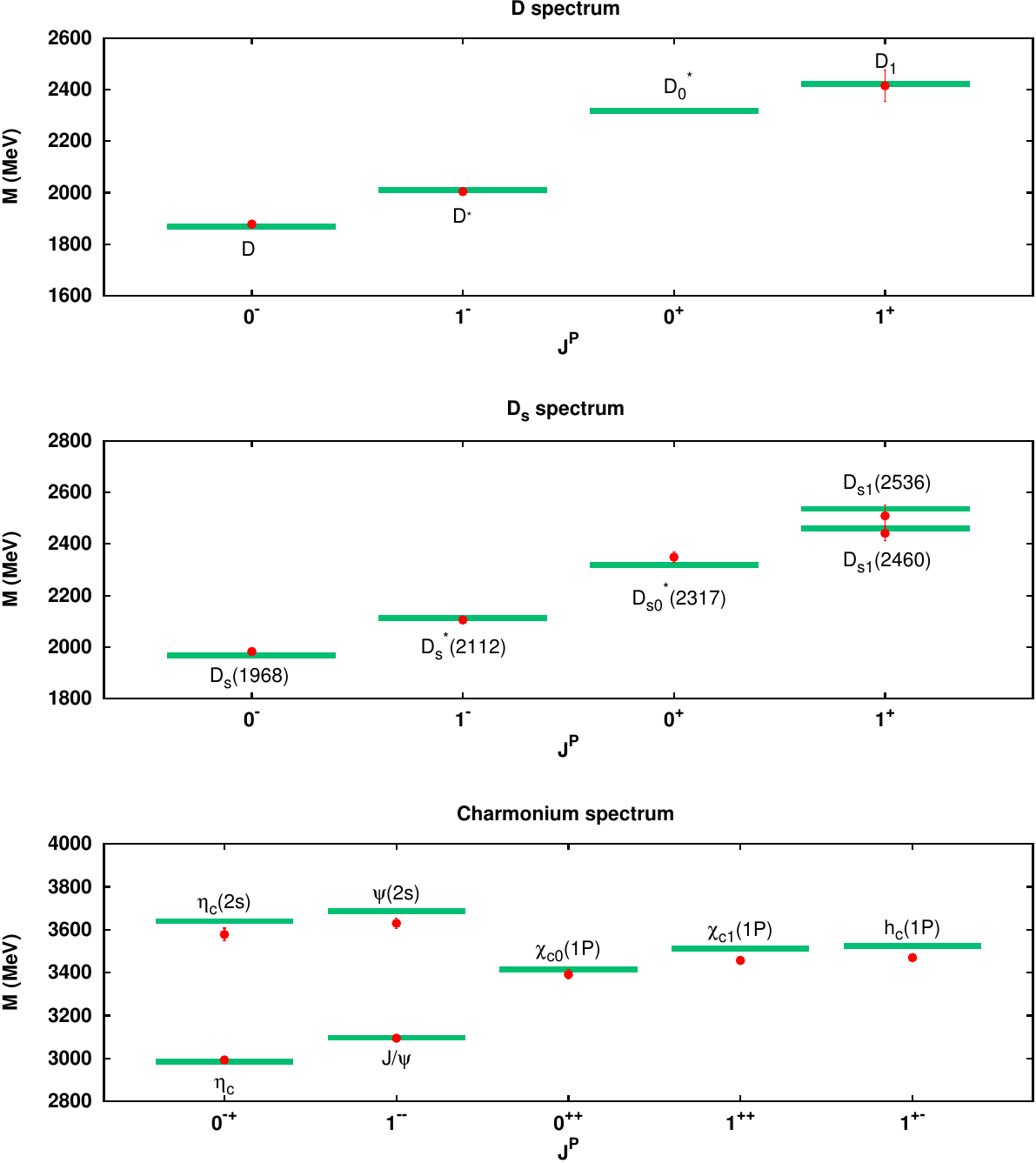}
\caption{\label{meson_spec} The low lying $D$ (top), $D_s$ (middle)
  and charmonium (bottom) spectra at the physical point for the
  $V=32^3\times 64$ ensembles. See Sections~\ref{extrapmass} and \ref{extrapfinal} for details
 on how the results at the physical point are obtained.
}
\end{figure}

For the SLiNC action the leading discretization effects are of ${\rm
  O}\left(a^2\right)$. The combination ${\rm O}\left((m_ca)^2\right)$
may be significant for some quantities given that $am_c\sim 0.4$ in
our simulation. In order to gauge the size of the corresponding
systematic uncertainty we have computed the low-lying charmonium,
$D_s$ and $D$ spectra for the $V=32^3\times 64$ ensembles using the
interpolators given in \ret{meson_op}. Correlators
constructed from these interpolators with quarks smeared over different
spatial extents were combined with the variational method~(see
Section~\ref{varmethod}) to extract the ground state. The
extrapolation to the physical point is discussed in
Sections~\ref{extrapmass} and \ref{extrapfinal}, the results are presented in
Fig.~\ref{meson_spec} and Table~\ref{meson_spec_table}.

\begin{table}[ht!] 
\begin{center} 
\begin{tabular}{ccc||ccc} 
\hline 
\hline 
Channel & $J^{PC}$ & M (GeV) & Channel & $J^{PC}$ & M (GeV)\\
\hline
 $\eta_c$     &$0^{-+}$&  $2.9929(12 )$& $D_s(1968)$     &$0^-$& $1.9824 (85) $    \\
 $\eta_c(2s)$ &$0^{-+}$&  $3.5778(296)$& $D^*_s(2112)$   &$1^-$& $2.1054 (95) $    \\ 
 $J/\psi$     &$1^{--}$&  $3.0944(19 )$& $D^*_{s0}(2317)$&$0^+$& $2.3490 (191)$    \\
 $\psi(2s)$   &$1^{--}$&  $3.6294(246)$& $D_{s1}(2460)$  &$1^+$& $2.4415 (271)$    \\
 $\chi_{c0}$  &$0^{++}$&  $3.3914(228)$& $D_{s1}(2536)$  &$1^+$& $2.5092 (431)$    \\
 $\chi_{c1}$  &$1^{++}$&  $3.4570(192)$& $D $ &$0^-$& $1.8778(106)$                \\
 $h_c$        &$1^{+-}$&  $3.4697(150)$& $D^*$&$1^-$& $2.0041(142)$                \\
 &&                                    & $D_1$&$1^+$& $2.4147(610)$                \\
\hline 
\hline 
\end{tabular} 
\caption{\label{meson_spec_table} The low lying open and hidden
   charm meson spectra at the physical point for the $V=32^3\times 64$ ensembles.}
\vspace{-0.8cm}
\end{center} 
\end{table}

Overall the radial and orbital excitations are reasonably well
reproduced. This is to be expected since the typical energy scale for
the 1P--1S and 2S--1S splittings in heavy-light systems is
$\overline{\Lambda}\sim 0.5$~GeV, which is much smaller than the
inverse lattice spacing. Similarly, for the charmonium ground state
the corresponding energy scale is $m_c v_c^2\sim 0.5$~GeV for
$v_c^2\sim 0.4$~\cite{Bali:1998pi}. We find for
    charmonium that the $\eta_c$(2S), $\psi$(2S), $\chi_{c1}$ and
    $h_c$ states lie $50-60$~MeV below experiment corresponding to
    $2-3.7$ standard deviations. However, as discussed
    below, the 2S and 1P fine structure splittings are reproduced,
    albeit with large errors. It is likely that other systematics, in
    particular, finite volume effects are important for these radial and
    orbital excitations. These systematics are not investigated for
    mesons in this paper, however, finite volume effects for the
    charmed baryons are discussed in Section~\ref{finitesize}.

For the $0^+$ and $1^+$
heavy-light mesons that are above or close to strong decay thresholds
one needs to consider the relevant scattering states.  We are able to
resolve two closely lying states for the $D_{s1}$ by using two
interpolators~(see Table~\ref{meson_op}) in addition to multiple
smearings in the variational method. However, a proper finite volume
analysis would be required to identify the true nature of the higher
lying state as the $D^*K$ threshold lies between the $D_{s1}(2460)$
and the $D_{s1}(2546)$.  For the $D_{1}$ the same analysis showed the
lowest eigenvalue of the variational method to be clearly consistent
with $D^*\pi$, and the next level to be compatible with 
experiment~(the latter shown in Fig.~\ref{meson_spec}), higher eigenvalues
were much larger in mass. As for the $D_{s1}$, from heavy quark
symmetry one expects two states close
together~\cite{Nowak:1992um,Bardeen:1993ae,Ebert:1994tv} and in
experiment there is the $D^0_1(2420)$~(width
27.4(2.5)~MeV~\cite{Agashe:2014kda}) and the $D^0_1(2430)$~(width
$384^{+130}_{-110}$~MeV~\cite{Agashe:2014kda}). However, a larger
basis, including interpolators with derivatives, would be required to
resolve the additional level. For the $0^+$, only one state is
expected and our results from a single scalar operator are compatible
with experiment for the $D^*_{s0}$ within $2\sigma$. Exchanging the
strange quark for an up/down quark, the lowest eigenvalue of the
$D^*_{0}$ is consistent with $D\pi$. Unfortunately, we were not able
to reliably extract the second eigenvalue for this channel and so we
do not include a value in Fig.~\ref{meson_spec}.  A more extensive
analysis of open and hidden charm mesons with a larger set of
interpolators, including also $J=2$ states, will be presented in a
forthcoming publication. We note that studies of open charmed 
meson channels near thresholds including four quark interpolators 
have already been performed, see Refs.~\cite{Mohler:2012na,Lang:2014yfa}.

Fine structure splittings are dominated by higher energy scales~(${\rm
  O}(m_c v_c)$ and ${\rm O}(m_c)$ in charmonium and the $D/D_s$
systems, respectively) and are therefore more sensitive to
discretization effects.  We find $M_{J/\psi} - M_{\eta_c}=100.1(1.5)$~MeV,
$M_{D_s^*} - M_{D_s} = 122.0(2.6)$~MeV and $M_{D^*}-M_{D}=126.8(10.1)$~MeV. These 
splittings are approximately $13$ MeV, $22$~MeV and $15$~MeV below 
experiment, respectively. Assuming lattice spacing effects are the 
main cause for the discrepancies, we estimate $10-20$~MeV as the 
likely size of this systematic in the charmed baryon spectra~(in
particular for spin splittings)\footnote{We note that there is an additional 
uncertainty arising from the scale setting of a few percent in the
splittings.}. For most observables the total error
of our final results after extrapolation to the physical point is of a 
comparable or larger size.  For the radial and orbital excitations in
charmonium, the wave function will be broader, suggesting
smaller lattice spacing effects.
We find the 2S hyperfine splitting, $\psi-\eta_c=48(20)$~MeV
and the 1P splitting, $\chi_{c1}-h_c=11(19)$~MeV, compared to
$46.7$~MeV and $14.8$~MeV, respectively, in experiment.

\section{\label{3}Charmed baryons and interpolators}

\begin{figure}[ht!]
   \includegraphics{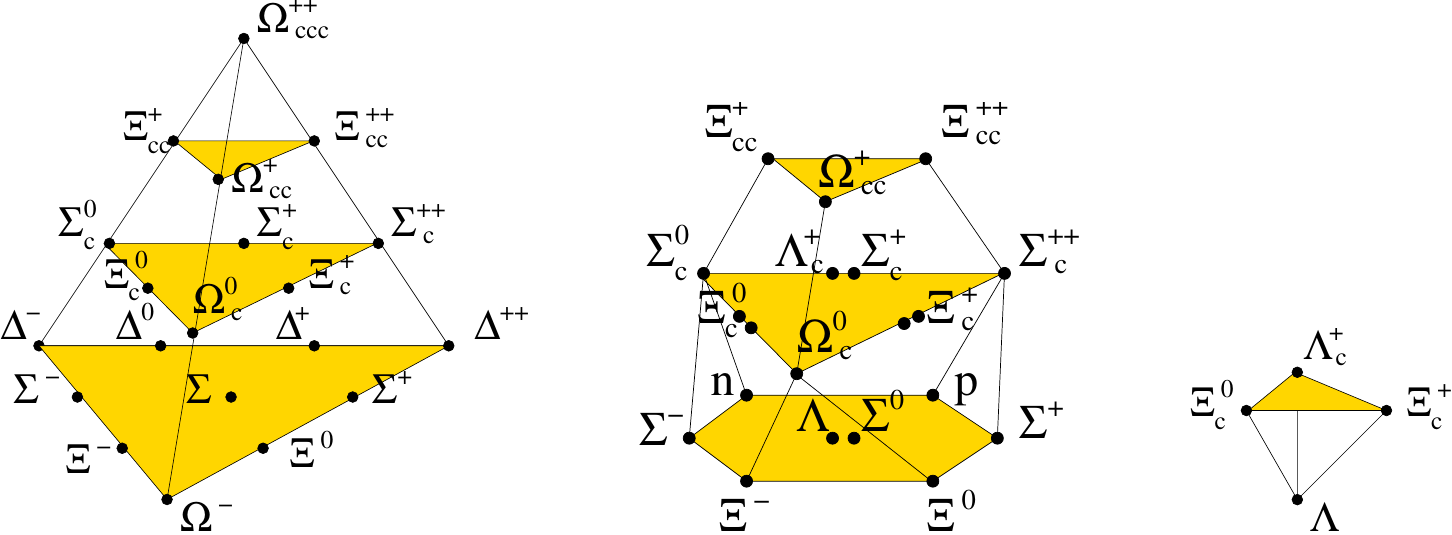}
\caption{\label{multiplets} The SU(4) irreducible representations
  for baryons: (left) the totally flavor symmetric $20{_{\rm
      S}}$-plet, (middle) the mixed symmetry $20_{{\rm M}}$-plet and
  (right) the totally anti-symmetric anti-quadruplet.  }
\end{figure}

The simplest way to see which charmed baryons are likely to exist is to
consider the irreducible representations of the tensor product of
three SU(4) fundamental representations:
\begin{equation}\label{irred_rep}
{\bf \hspace{1.1mm}4\hspace{1.1mm}\otimes \hspace{1.1mm}4\hspace{1.1mm} \otimes
\hspace{1.1mm}4 \hspace{1.1mm}= \hspace{4mm}20_{ S}\hspace{3.5mm} \oplus
\hspace{0.4mm} 20_{ M} \hspace{1.2mm} \oplus \hspace{0.8mm}20_{ M}
\hspace{1.2mm}\oplus\hspace{1.6mm} \overline 4_A.\hspace{0.3mm}}
\end{equation}
This flavor symmetry is not respected in nature, however, the
number of baryons and their flavor quantum numbers may be
reproduced.
\refig{multiplets} displays the totally symmetric $20_S$-plet, the
mixed symmetry $20_M$-plet and the total anti-symmetric
anti-quadruplet, $\overline{4}_A$. For each state within a multiplet
the total spin and parity is the same and resonances may also be
expected to fall into this pattern.  The ground
state $20_S$ and $20_M$-plets have positive parity~(consistent with the
light baryon members of the multiplets) and $J=\frac 32$ and $J=\frac
12$, respectively, while the $\overline{4}_A$ ground states, in the non-relativistic
limit, require non-zero orbital angular momentum and have negative
parity $(J^P=\frac12^-)$.  Orbital angular
momentum is also needed for the negative parity counterparts to $20_S$
and $20_M$.  The lowest mass members of the SU(4)  multiplets are the familiar
decuplet, octet and singlet of SU(3) symmetry. The singly charmed
baryons can form SU(3) sextets and anti-triplets~(${\bf 3\otimes
  3=6_S+\overline{3}_A} $), where the restriction of a totally
anti-symmetric wavefunction means that there is no anti-triplet 
within the $20_S$-plet.  The doubly charmed baryons are all in SU(3) triplets. 

In nature, where SU(3) flavor symmetry is broken, the physical
states may not fall exactly into the expected multiplets, for example,
the $\Xi_c$~($\Xi_c^\prime$) may contain a small contribution from the
sextet~(anti-triplet). Furthermore, the above picture does not include
possible non-quark-model states. However, the observed charmed baryon
spectra, with either measured $J^P$ or assignment from potential model
predictions~(for a review see, for example, Ref.~\cite{Crede:2013kia}) seem
to reproduce the expected pattern: replacing a strange quark by a light quark,
splittings are similar within a given SU(3) multiplet,
consistent with a constituent quark model picture.  We remark that
since we are simulating QCD in the isospin limit~($m_u=m_d$ and
omitting QED effects) then all states within an isospin multiplet 
are degenerate.

The internal structure of baryons containing light~($q\in\{\ell,s\}$)
and heavy quarks~($Q$) can be addressed in terms of HQET and, for the
$QQq$ case, pNRQCD, starting from the static limit.  At finite heavy
quark masses also NRQCD is valid for doubly heavy systems. In the case
of $Qqq$ baryons, the heavy quark $Q$ provides a color source for the
two light quarks. In the $m_{Q} \rightarrow \infty$ limit, the light
quarks have a definite total angular momentum and a total spin $s_d=0$
or $1$, corresponding to a flavor anti-symmetric or symmetric
structure, respectively. In this limit, the spin splittings between
baryons with $s_d=1$ and $s_d=0$ vanish since they are of ${\rm
  O}(\overline{\Lambda}^2/m_Q)$, where $\overline{\Lambda}$ is the
energy scale of the light degrees of freedom.

\begin{figure}[ht!]
 \includegraphics{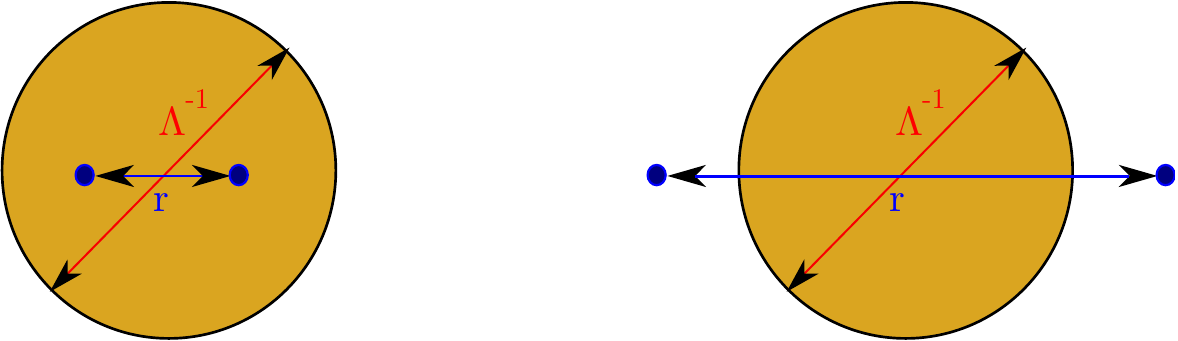}
\caption{\label{doubly_charmed} { Internal structure 
of a doubly heavy baryon: HQET picture~(left hand side), 
Quarkonium-like~(right hand side). The blue circles represent the heavy quarks, $Q$, and
$r$ is the average $Q-Q$ separation. In the HQET picture $r\ll \overline{\Lambda}^{-1}$\hspace{-1mm}, 
while in the charmonium picture $r > \overline{\Lambda}^{-1}$\hspace{-2.5mm}. }}
\end{figure}

For doubly
heavy baryons, in what we call the HQET picture, the two heavy quarks form
a diquark of small spatial extension that interacts with the light quark in analogy
to heavy-light mesons~(if the $QQ$ diquark is in a color anti-triplet),
shown in~\refig{doubly_charmed}. HQET corresponds to pNRQCD in the
limit of the distance between the two heavy quarks, $r\rightarrow 0$. Assuming such
a $QQ$ diquark, to leading non-trivial order in $1/m_Q$,
one can show that $QQq$ baryon spin-splittings,
$M_{QQq}(J=\frac 32)-M_{QQq}(J=\frac 12)$, are  $3/4$ times the
corresponding $\overline{Q}q$ fine structure splitting,
$M_{\overline{Q}q}(J=1)-M_{\overline{Q}q}(J=0)$~\cite{Brambilla:2005yk}.

Alternatively, the doubly heavy baryons could be comprised of a heavy
and a light quark in a~(color-anti-triplet) diquark and, together with
the remaining heavy quark, one has a charmonium-like system, also shown
in \refig{doubly_charmed}. In this case the level splittings can be
understood in terms of pNRQCD and NRQCD, for example, the $QQq$
spin splittings can be related to the charmonium fine structure
splitting.  However, this is not so straightforward as for the HQET
picture since the light quark within the $Qq$ diquark cannot be
considered as spatially localized.  It is possible that the HQET
picture works best for the lower lying states~(where $r \ll
\overline{\Lambda}^{-1}$), while a charmonium-like picture is
applicable for higher excitations with $r > \overline{\Lambda}^{-1}$.
pNRQCD includes both possibilities.

The expected internal structure of a particle informs the choice of
lattice interpolator employed, since we want to have a good overlap with
the physical state.  Any interpretation that is valid in the heavy
quark limit may not work particularly well for charm quarks and will
be more applicable to baryons involving bottom quarks. With this in
mind, we implemented two sets of interpolators, those based on
HQET for $J=\frac 12$ and $\frac 32$~(\ret{op_hqet}, where the 
diquark is formed from two light quarks or two heavy quarks in the case of
singly and doubly charmed baryons, respectively) and also those
arising from SU(4) symmetry~(\ret{op_su4}) for $J=\frac 12$. 
A comparison of the efficacy of these interpolators is made in
Section~\ref{varmethod}.

\vspace{-0.7cm}
\begin{table}[ht!]
\begin{center}
\setlength{\tabcolsep}{0.7mm}
\renewcommand{\arraystretch}{1.2}
\begin{tabular}{rcc|cc|cc}
\hline
\hline
\multicolumn{7}{c}{{ HQET singly charmed baryon interpolators}} \\
\hline
  $ S$& $ I$& $ s_d$ & $ (qq)Q$ & $\mathcal O$  &\textbf{$J= \frac12$}&\textbf{$J= \frac32$}\\
\hline
$0$ & $0$                  & $0$& $(\ell\ell')c$ & 
${\mathcal O}_{5\gamma}=\epsilon_{abc}(\ell^{aT}C\gamma_5\ell^{'b})c_\gamma^c  $&$\Lambda_c$ &  \\
$0$ & $1$                  & $1$& $(\ell\ell)c$ &
${\mathcal O}_{\mu\gamma}=\epsilon_{abc}(\ell^{aT}C\gamma_\mu \ell^b)c_\gamma^c $&$\Sigma_c$  &$\Sigma^*_c$   \\
$-1$& $\frac12$ & $0$& $(\ell s)c$ &
${\mathcal O}_{5\gamma}=\epsilon_{abc}(\ell^{aT}C\gamma_5 s^b)c_\gamma^c $&$\Xi_c$     &      \\
$-1$& $\frac12$ & $1$& $(\ell s)c$ &
${\mathcal O}'_{\mu\gamma}=\epsilon_{abc}(\ell^{aT}C\gamma_\mu s^b)c_\gamma^c $&$\Xi'_c$    &$\Xi^*_c$     \\
$-2$& $0$                  & $1$& $(ss)c$ &
${\mathcal O}_{\mu\gamma}=\epsilon_{abc}(s^{aT}C\gamma_\mu s^b)c_\gamma^c $&$\Omega_c$  &$\Omega^*_c$   \\
\hline
\multicolumn{7}{c}{{ HQET doubly charmed baryon interpolators}} \\
\hline
$0$ & $\frac 12$                  & $1$& $(cc)\ell$ &    
${\mathcal O}_{\mu\gamma}=\epsilon_{abc}(c^{aT}C\gamma_\mu c^{b})\ell_\gamma^c  $&$\Xi_{cc}$ & $\Xi^*_{cc}$  \\
$-1$& 0           & $1$& $(cc)s$ &
${\mathcal O}_{\mu\gamma}=\epsilon_{abc}(c^{aT}C\gamma_\mu c^b)s_\gamma^c $&$\Omega_{cc}$  &$\Omega^*_{cc}$   \\
\hline
\hline
\end{tabular}
\caption{\label{op_hqet} Quantum numbers of singly and doubly charmed
  baryons and interpolators, ${\cal O}_{A\gamma}$ in the HQET
  picture, where $\gamma$ is the spin index.  $\ell$ and $\ell'$ stand
  for up and down quarks, $c$ for charm and $s$ for strange. $S,I$
  and $s_d $ are strangeness, isospin, and diquark total spin quantum
  numbers, respectively.  }
\vspace{-1.7em}
\end{center}
\end{table}

\begin{table}[ht!]
\begin{center}
\begin{tabular}{ccc|l|c}
\hline
\hline
\multicolumn{5}{c}{$SU(4)$ singly charmed baryon interpolators} \\
\hline
$SU(4)$-plet & { S} & { I}& ${\cal O}$ & $J=\frac12$ \\\hline
{\sc 20$_{M}$}   & 0 & 0 & $ {\mathcal O}_{5\gamma}= \frac{1}{\sqrt{6}}\epsilon^{abc} 
  \left\{ 2 ( \ell^{aT}C\gamma_5 \ell^{'b}_2) 
  c^c_{\gamma}  +
  ( c^{aT}C\gamma_5\ell^{'b})
  \ell^c_{\gamma} - ( c^{aT}C\gamma_5 \ell^b)
   \ell^{'c}_{\gamma}  
  \right\}$  & $\Lambda_c$ \\
& -1 & $\frac 12$ & $ {\mathcal O}_{5\gamma}= \frac{1}{\sqrt{6}}\epsilon^{abc} 
  \left\{ 2 ( s^{aT}C\gamma_5 \ell^{b}_2) 
  c^c_{\gamma}  +
  ( c^{aT}C\gamma_5\ell^{b})
  s^c_{\gamma} - ( c^{aT}C\gamma_5 s^b)
   \ell^{c}_{\gamma}  
  \right\}$  & $\Xi_c$ \\
&0 & 1  & ${\mathcal O}_{5\gamma}=\epsilon^{abc} 
  ( c^{aT}C\gamma_5 \ell^b
  )\ell^c_{\gamma}$ &  $\Sigma_c$ \\
 & -1 & $\frac 12$ & ${\mathcal O}_{5\gamma} = \frac{1}{\sqrt{2}}\epsilon^{abc} 
  \left\{  ( s^{aT}C\gamma_5 c^b) 
  \ell^{c}_{\gamma}  +
  ( \ell^{aT}C\gamma_5 c^b)
  s^c_{\gamma} \right\}$ & $\Xi^{\prime}_c$ \\
&-2 & 0  & ${\mathcal O}_{5\gamma}=\epsilon^{abc} 
  ( c^{aT}C\gamma_5 s^b
  )s^c_{\gamma}$ &  $\Omega_c$ \\\hline
  \multicolumn{5}{c}{{ $SU(4)$ doubly charmed baryon interpolators}} \\\hline
{\sc 20$_{M}$}  & 0  & $\frac 12$ &
  ${\mathcal O}_{5\gamma}=\epsilon^{abc} 
    ( \ell^{aT}C\gamma_5 c^b
    )c^c_{\gamma}$ & $\Xi_{cc}$ \\
 & -1 & 0 &
  ${\mathcal O}_{5\gamma}=\epsilon^{abc} 
    ( s^{aT}C\gamma_5 c^b
    )c^c_{\gamma}$  &  $\Omega_{cc}$ \\\hline
\hline
\end{tabular}
\caption{\label{op_su4} 
As in \ret{op_hqet} for the interpolators from SU(4) symmetry.
In this case we did not consider $J=\frac 32$ states. 
}
\vspace{-1.5cm}
\end{center}
\end{table}

The interpolators given in the tables do not have definite parity and
one needs to use the projection operator $P^\pm=\frac12 (1\pm
\gamma_4)$ to obtain positive or negative parity states. Since, in the
non-relativistic limit, negative parity requires non-zero orbital
angular momentum, interpolators including derivatives may improve the 
overlap with the physical states.  Further exploration of the best
basis will be performed in a future study. An additional projection is
required for the HQET interpolators of type ${\cal O}_{\mu}$ and
${\cal O}^\prime_{\mu}$~(see Table~\ref{op_hqet}) as these contain
both $J=\frac 12$ and $J=\frac 32$ components. The two contributions
are disentangled using projection operators~(at zero momentum): 

\bea
(P^{3/2})_{ij} &=& \delta_{ij} - \frac 13 \gamma_i\gamma_j, \quad
i,j\in\{1,2,3\}, \nonumber \\ (P^{1/2})_{ij} &=& \frac 13
\gamma_i\gamma_j.  \eea We note that the interpolators for the $\Xi_c$
and $\Xi_c^\prime$ states will, in principle, have an overlap with
both physical states since they are not distinguished by a conserved
quantum number. In the HQET case, the two interpolators differ in terms of
the total spin of the diquark~($s_d$) which is not conserved at a finite
heavy quark mass and for the SU(4) case the interpolators are in different
SU(3) flavor multiplets, however, the flavor symmetry is, of course,
broken. We have not taken possible mixing between the states 
created by the $\Xi_c$ and $\Xi_c^\prime$ interpolators into account. 
This would require calculating
the corresponding cross-correlation functions. However, such a study
was performed in Ref.~\cite{Brown:2014ena} with high statistics and no
significant mixing was found.  This suggests that the approximate SU(3) symmetry holds
sufficiently well to suppress the mixing.

\section{\label{varmethod}Variational method and fitting procedure}

We compute the mass spectrum in the conventional way by calculating 
two-point correlation functions created from sets of baryonic
interpolators $\cO_{i,\gamma}$, with spin index $\gamma$. These correlation
functions contain contributions from all states with the same quantum
numbers given by the interpolators
\bea [C(t)]_{ij} &=& P^\pm_{\bar \gamma \gamma}\left\langle \cO_{i,\gamma}(t)
\overline{\cO}_{j,\bar\gamma}(0)\right\rangle = \sum_n \langle
0|\cO_{i,\gamma} (0)| n\rangle \langle n| \overline{\cO}_{j,\bar
  \gamma}(0)| 0\rangle e^{-M_nt}, \eea 
where $P^\pm_{\bar \gamma \gamma}$ is the parity projection operator.
We employ the variational
method~\cite{Michael:1985ne,Luscher:1990ck} in order to reliably
extract the ground state and first excited state.  For
each interpolator given in Section~\ref{3} we generate a $3\times 3$ matrix of
correlators by varying the smearing applied to the quark fields. 
Extracting the eigenvalues, $\lambda^\alpha(t,t_0)$, of the
generalized eigenvalue problem at large times, the lowest two states
are cleanly separated:
\be
C^{1/2}(t_0) C(t) C^{-1/2}(t_0) v^{\alpha} (t, t_0) = 
\lambda^\alpha (t,t_0) v^\alpha(t,t_0), \quad \lambda^\alpha(t,t_0) 
\propto e^{-M_\alpha(t-t_0)} \left[ 1 + {\rm O}
(e^{-\Delta M_{\alpha'\alpha} t}) \right], \label{genprob}
\ee
where $\Delta M_{\alpha'\alpha} = M_{\alpha'} -M_\alpha$, and $\alpha'
> \alpha$. The time-slice $t_0$ can be varied to reduce the excited
state contamination\footnote{For $t>t_0 \ge t/2$, only $\alpha' >N$ will
  contribute where $N$ is the rank of the
  basis~\cite{Blossier:2009kd}, a limit that we do not consider
  here.} to $\lambda^\alpha(t,t_0)$ and the eigenvectors
$v^\alpha(t,t_0)$.  We use Wuppertal smearing
\cite{Gusken:1989ad,Gusken:1989qx} with APE smoothed
links~\cite{Falcioni:1984ei,Albanese:1987ds} where the number of
Wuppertal iterations is varied to obtain one smearing combination
which leads to a spatially extended interpolator with a good overlap
with the ground state and two other combinations which have
significant, but different, overlap with excited states. For the
singly charmed baryons, the overlap of the interpolator with the
physical states was most sensitive to the light quark smearing. The
number of Wuppertal iterations for the heavy quark was fixed to
$n=150$ and for the light quark $n\in\{5,25,150\}$ was realized. For
the doubly charmed baryons the situation is reversed and $n= 150$ was
used for the light quark and $n\in \{5,25,150\}$ for the heavy quarks.

\begin{figure}[ht!]
\includegraphics{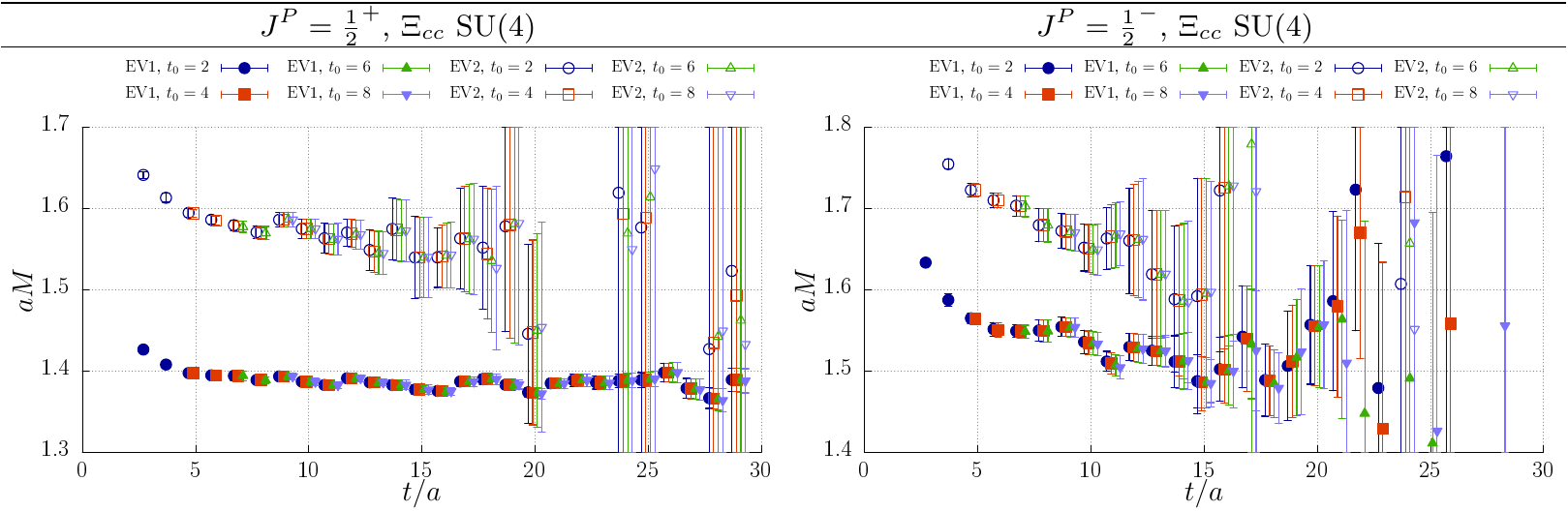}
\caption{\label{t0_dep} Effective masses for the first~(EV1) and
   second~(EV2) eigenvalues of $\Xi_{cc}$ for positive 
   and negative parity obtained using different values for $t_0$
   when solving the generalized eigenvalue problem~Eq.~(\ref{genprob})
   for SU(4) interpolators on the symmetric ensemble with $V=32^3\times
  64$.}  
\end{figure}

The eigenvalues $\lambda^{\alpha=1,2}(t,t_0)$ were fitted~(separately)
with single exponentials\footnote{We assume the baryon masses to be heavy enough for the 
backward propagating particle to have a negligible influence.} for fixed $t_0\ge 2a$ in the range $t_i$ to
$t_f$, taking the correlations between time-slices into account. The
third eigenvalue was discarded as $M_3$ cannot be cleanly separated
from higher excited states. The final fit ranges chosen, compiled in
\ret{tablefitrange}~(Appendix~\ref{AppC}), have reduced correlated 
$\chi^2$ values   $\chi^2/{\rm
  dof}<2 $ and in most cases $\chi^2/{\rm dof}\sim 1 $. For these fit
ranges the masses extracted were stable within errors as $t_i$ was
further increased. 
No significant dependence on $t_0$ was found 
(see, for example, Fig.~\ref{t0_dep}) and we
take $t_0=2a$.  The statistical errors were evaluated using the
jackknife method combined with binning.  Measurements were performed
on every other trajectory for each ensemble and the errors were stable
for $n_{\rm bin} \ge 2-4$. We made the conservative choice of
$n_{\rm bin}=4$, which is consistent with $n_{\rm bin}>4\tau_{\rm int}$, where
$\tau_{\rm int}$ is the integrated auto-correlation time.  The latter 
was estimated via the
$\Gamma$-method~\cite{Wolff:2003sm,Schaefer:2010hu} to be between
$0.5$ and $0.7$, depending on the state.

\begin{figure}[ht!]
\includegraphics{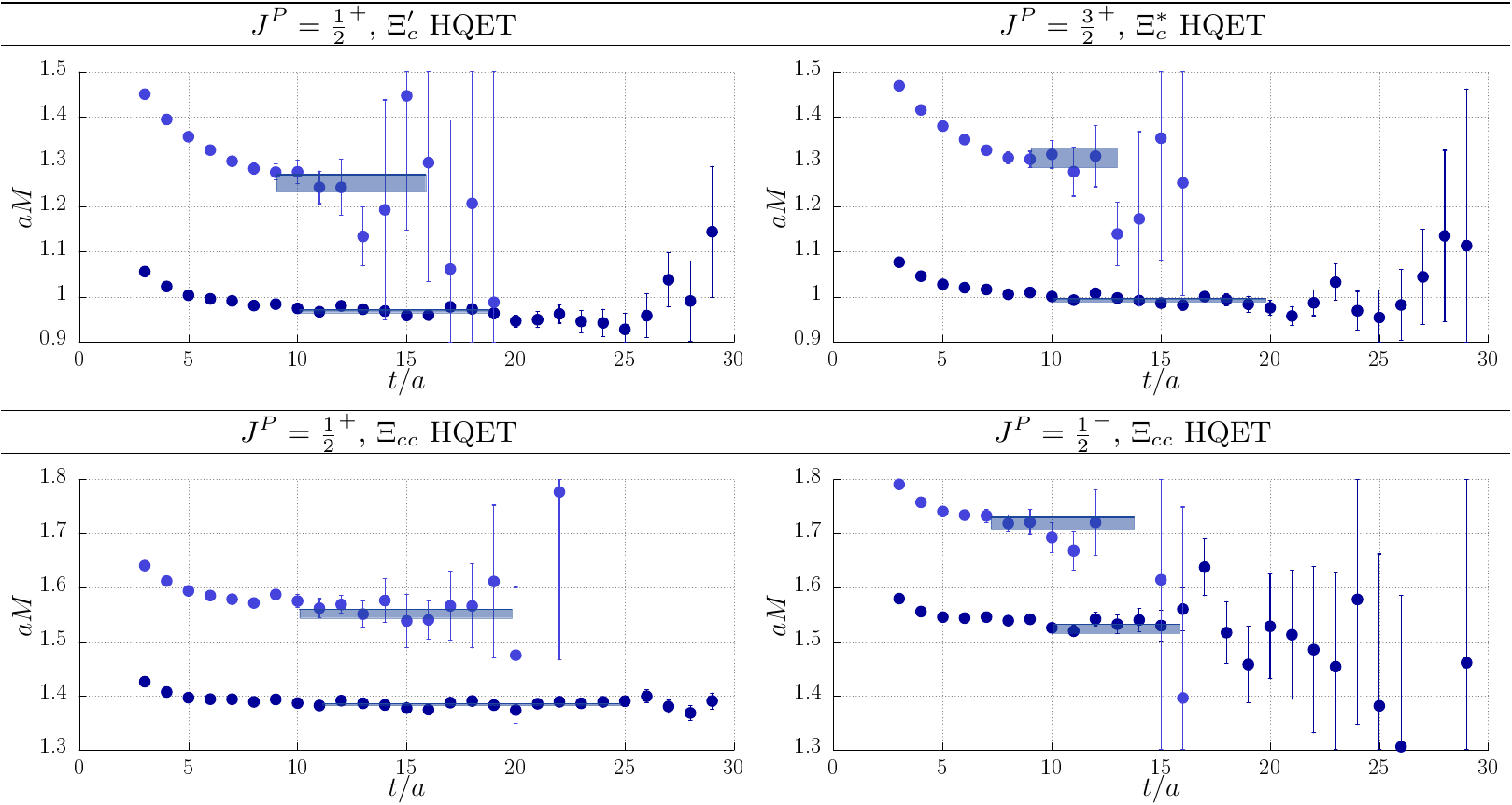}
\caption{\label{sample_fits} Effective masses for the first and second
  eigenvalues for  HQET interpolators on the
  symmetric ensemble with $V=32^3\times 64$. The filled
  regions indicate the fit ranges chosen and the fit results,
  including the statistical errors.}
\end{figure}

Effective masses,
\be
M_\alpha(t) = \frac{1}{2a} \log \left( \frac{\lambda^\alpha(t-a,t_0)}
{\lambda^\alpha(t+a,t_0)}\right),
\ee
and fits for a representative sample of states are
shown in \refig{sample_fits} for the HQET interpolators on the symmetric
ensembles with $V=32^2\times 64$ corresponding to $M_\pi = M_K= 461$~MeV.
For the positive parity states we are able to extract a
reasonable signal for the first two eigenvalues for both
$J=\frac 12$ and $\frac 32$.  For the negative parity states, which
are statistically noisier, clear ground state signals were obtained
for $J=\frac 12$ and $\frac 32$, however, the first excited state was
only reliably extracted for the doubly charmed baryons.

\begin{figure}[ht!]
\includegraphics{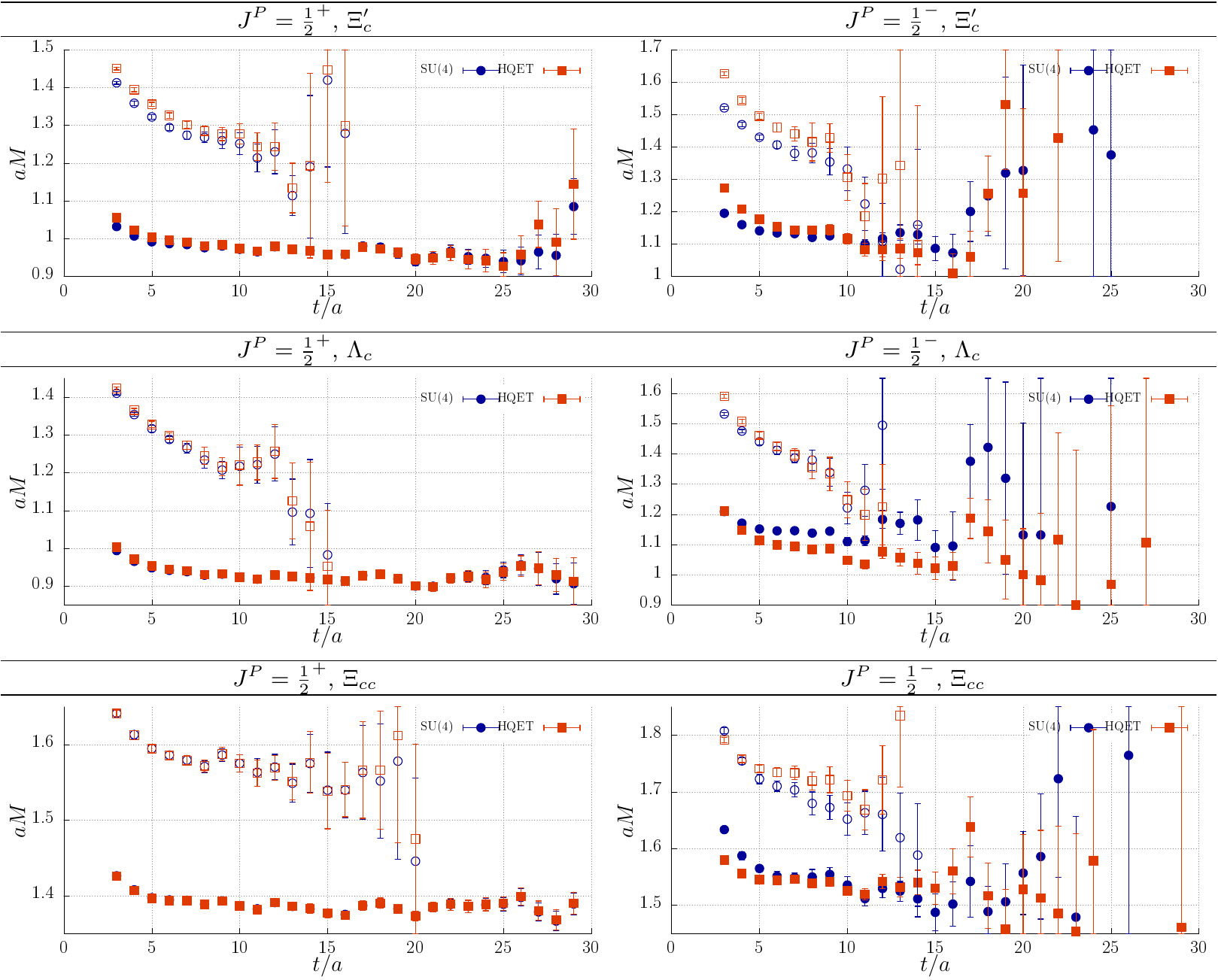}
\caption{\label{sample_effm} Comparison of the effective masses for SU(4) and HQET interpolators
  employed on the ensemble at the symmetric point with $V=32^3\times 64$. The 
  ground~(first excited) state is indicated by filled~(open) symbols. }
  \vspace{-2mm}
\end{figure}

For the spin-$1/2$ particles we can compare the HQET and SU(4)
interpolators. Effective masses for a state from each multiplet are shown
in~\refig{sample_effm}. The positive parity particles display 
a consistent picture: the SU(4) interpolators have a slightly better
overlap with the desired states for the sextet baryons 
$(\Sigma_c, \Xi'_c,\Omega_c)$, are only marginally
better for the anti-triplet baryons $(\Lambda_c, \Xi_c)$ and no difference is observed for
the doubly charmed triplet $(\Xi_{cc},\Omega_{cc})$. The latter is due to the corresponding
correlation functions only differing by terms which are suppressed in
the non-relativistic limit, when expressed through upper and lower components
of the quark spinors. This is not the case for singly charmed baryons
nor for negative parity.
Indeed, a more striking pattern emerges for negative parity. For the
sextets, the SU(4) interpolators have a much better overlap with the ground
state while for the triplet and (even more markedly) the anti-triplet
the HQET interpolators are clearly better.  For the first excitation the SU(4)
interpolators provide a better signal, although there is no clear
single state dominance for the singly charmed baryons.

\begin{figure}[ht!]
\includegraphics{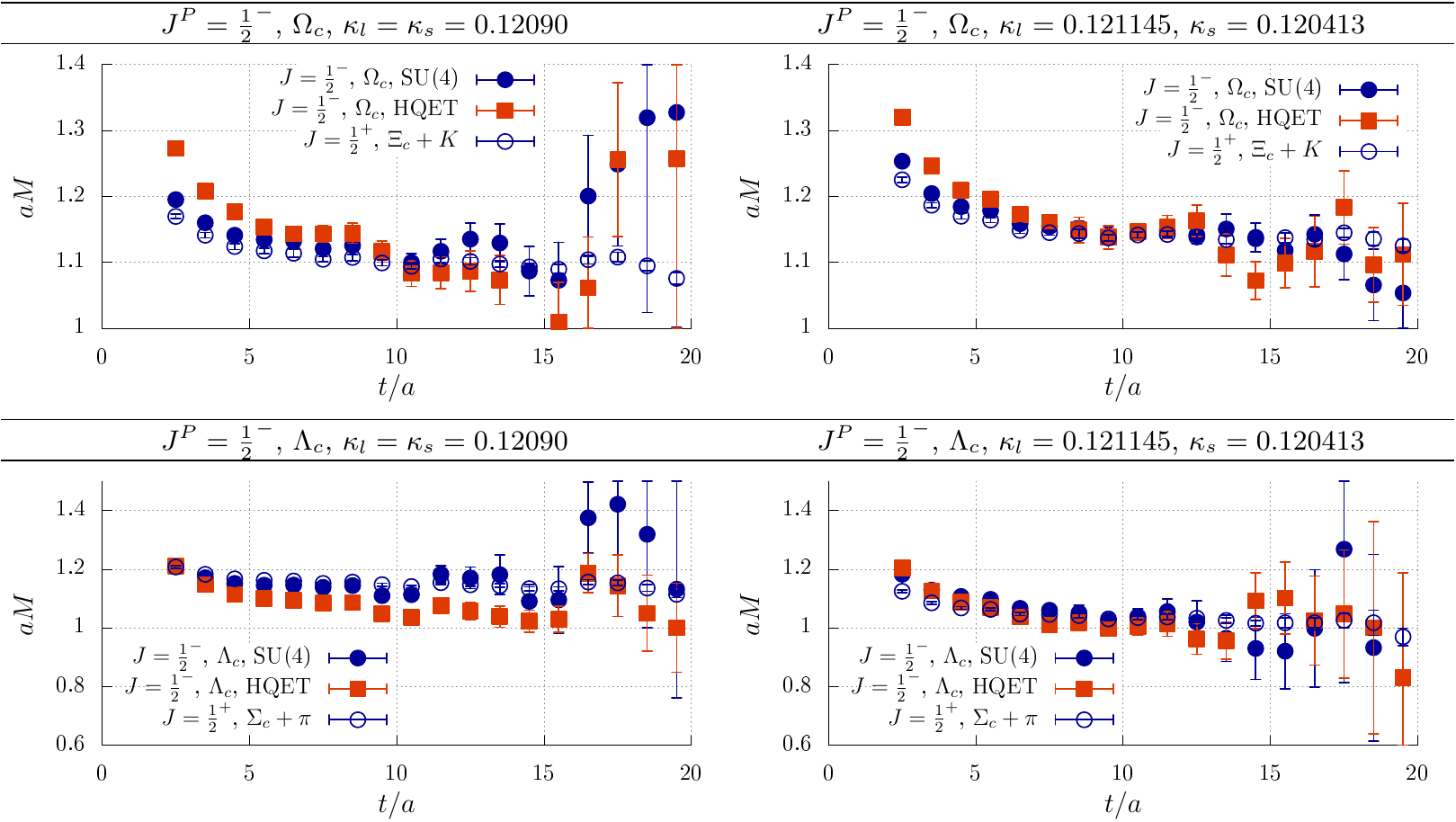}
\caption{\label{sample_scattering} Comparison of negative 
parity effective masses with the relevant strong 
decay thresholds for (top) $\Omega_c$ and (bottom) $\Lambda_c$ 
on (left) the symmetric ensemble and (right) 
the most asymmetric ensemble for $V=32^3\times 64$.  }
\vspace{-0.6cm}
\end{figure}

These comparisons are made with the same smearings applied to both sets of
interpolators. Changing or optimizing the smearing in the individual cases
may change these conclusions.  The SU(4) and HQET interpolators
in each channel both belong to the same (lattice) hypercubic
representation: either the dimension two $G_1$ representation for the
interpolators denoted as spin-1/2 in Tables~\ref{op_hqet} and~\ref{op_su4},
which in the continuum gives $J=\frac 12, \frac 72,\ldots$ or the
dimension four $H$ representation for the spin-3/2 interpolators
corresponding to continuum $J=\frac 32, \frac 52,\ldots$.  In the
limit of very large statistics the SU(4) and HQET interpolators should
give the same energy for each state. For positive parity consistency
is seen, while for negative parity we use the interpolator with the best
overlap for our setup. Similar behavior is seen throughout for the
smaller volume, $V = 24^3\times 48$.

Note that care has to be taken in order to avoid the misidentification
of multi-particle states, for example, a positive parity baryon plus a
pion as a negative parity baryon. We have studied this systematically
for the ground state negative parity channels. All but one of our
$J^P=\frac32^-$ ground states are smaller in mass than two-particle
states of the same quantum numbers, consisting either of a $\frac32^+$
baryon or a $\frac12^+$ baryon plus a pseudoscalar meson in a S-wave
or a P-wave, respectively.  None of the $\frac12^-$ states contain
$D$ or $D_s$ mesons -- all combinations of these mesons with
$\frac12^+$ baryons of the same combined isospin, strange and
charmness are heavier in mass.  There are, however, quite a few
S-wave decay channels with thresholds close to our mass estimates
that require careful study, in particular
\begin{align}
   &\Lambda_c\rightarrow \Sigma_c+\pi\,, \qquad\hspace{0.7mm}  \Sigma_c\rightarrow\Lambda+\pi\,,\qquad
   \hspace{2.2mm}\Sigma_c\rightarrow\Sigma_c+\pi\,,\qquad  \Xi_c^*\rightarrow \Xi^*_c+\pi\,,\nonumber\\
&\Xi_c\rightarrow\Lambda_c+K\,,\qquad\Xi_c\rightarrow\Sigma_c+K\,,\qquad\Xi_c\rightarrow\Xi_c+\pi\,,\qquad
\Xi_c\rightarrow\Xi_c'+\pi\,,\nonumber\\
&\Xi'_c\rightarrow\Lambda_c+K\,,\qquad \Xi'_c\rightarrow\Sigma_c+K\,,\qquad\Xi'_c\rightarrow\Xi_c+\pi\,,\qquad
\Xi'_c\rightarrow\Xi_c'+\pi\,,\nonumber\\
&\Omega_c\rightarrow\Xi_c+K\,,\qquad\Omega_c\rightarrow\Xi'_c+K\,,\qquad
\Xi_{cc}\rightarrow\Xi_{cc}+\pi\,,\quad\hspace{1mm}\Omega_{cc}\rightarrow\Xi_{cc}+K\,,\nonumber
\end{align}
where the baryons on the left hand sides are implied to have negative
parity and those on the right hand sides positive parity. Neither have we 
made the isospin nor charges explicit.
Comparing all these channels at our three different sea quark mass
combinations leads to the following conclusions. The ground state
negative parity $\Omega_c$ state, obtained both from the SU(4) and
HQET inspired interpolators, is always degenerate with the sum of the
$\Xi_c$ and a kaon. We show the corresponding effective masses 
at $M_\pi = X_\pi=461$~MeV and at $M_\pi = 259$~MeV in
Fig.~\ref{sample_scattering}. Likewise, we identify our negative
parity $\Xi_c'$ signal as a $\Lambda_c$ plus kaon scattering
state. 

Particularly interesting is the behavior of the negative parity
$\Lambda_c$, also depicted in Fig.~\ref{sample_scattering} 
for two ensembles: while
effective masses obtained from the SU(4) interpolator are degenerate
with the combined mass of a $\Sigma_c$ and a pion, the effective
masses from the HQET interpolator are systematically lower, at least
at our two heavier pion mass points. This suggests the HQET
interpolator to significantly overlap with a physical state lower in
mass than the scattering state while the SU(4) interpolator in this
case strongly couples to the close-by two-particle state. All the
other negative parity singly and doubly charmed baryons seem to be
relatively stable under strong decays or, in the few cases where their
masses are higher than those of the potential decay products, at least
the interpolators we employ are insensitive to the presence of these
scattering states. A similar analysis for the first excitation~(for
both parities) is challenging due to the number of relevant
two- and even three-particle states.

The lower lying meson spectrum in the charmonium, $D$ and $D_s$
sectors was also determined in order to estimate the size of
discretization errors~(as discussed in Section~\ref{simdetails}) and
to enable the tuning of the charm quark mass parameter~(see the next
Section). Except for the $1^+$ $D/D_s$ channel, we employed $3\times
3$ correlation matrices and applied the variational method, as
described above, extracting the ground state mass and in some cases
the excited state for each meson. Spatially extended interpolators
where constructed using $n\in\{5,25,150\}$ iterations of Wuppertal
smearing for the charm quark, the light quarks were not
smeared\footnote{Note that for meson interpolators without derivatives
  the smearing can be ``transformed'' from one quark to the
  other.}. Effective masses and fit results for a sub-set of states
are displayed in Fig.~\ref{sample_mesons} in Appendix~\ref{Appmesons}
for the symmetric larger volume ensemble. Fitting ranges for all
channels are given in Table~\ref{tablefitmesons} of
Appendix~\ref{AppC}. For the $1^+$ channel, two interpolators were
implemented~(see Table~\ref{meson_op}) and a $6\times 6$ correlation
matrix was computed. This enabled us to resolve pairs of closely lying
states, as illustrated in
Fig.~\ref{mesons_a1_b1}~(Appendix~\ref{Appmesons}).  In terms of
multi-particle states, the $0^+$ and $1^+$ $D/D_s$ mesons are
above/close to the $D\pi/DK$ and $D^*\pi/D^*K$ experimental thresholds,
respectively. We discuss our results for these channels, 
also shown in  Fig.~\ref{meson_spec}, in Section~\ref{extrapmass}.

\section{\label{kappacharm} Tuning of  $\kappa_{\rm charm}$}

\begin{table}[ht!]
\begin{center}
\begin{tabular}{cccc}
\hline
\hline
$\kappa_l$ & $\kappa_s$ & $L/a\times T/a$ &  $\kappa_{\rm charm}$    \\
\hline
$0.12090$  & $0.12090$   &$32\times 64$  & $0.1114801(67)$  \\ 
$0.12104$  & $0.12062$   &$32\times 64$  & $0.1114869(43)$  \\ 
$0.121145$ & $0.120413$  &$32\times 64$  & $0.1114908(48)$  \\ 
\hline
\hline
\end{tabular}
\caption{\label{final_kappa_value} Final values for $\kappa_{\rm charm}$
  for each ensemble determined using the $1S$ charmonium mass with full
  statistics.}
\end{center}
\end{table}

\begin{figure}[ht!]
\includegraphics[width=0.7\textwidth]{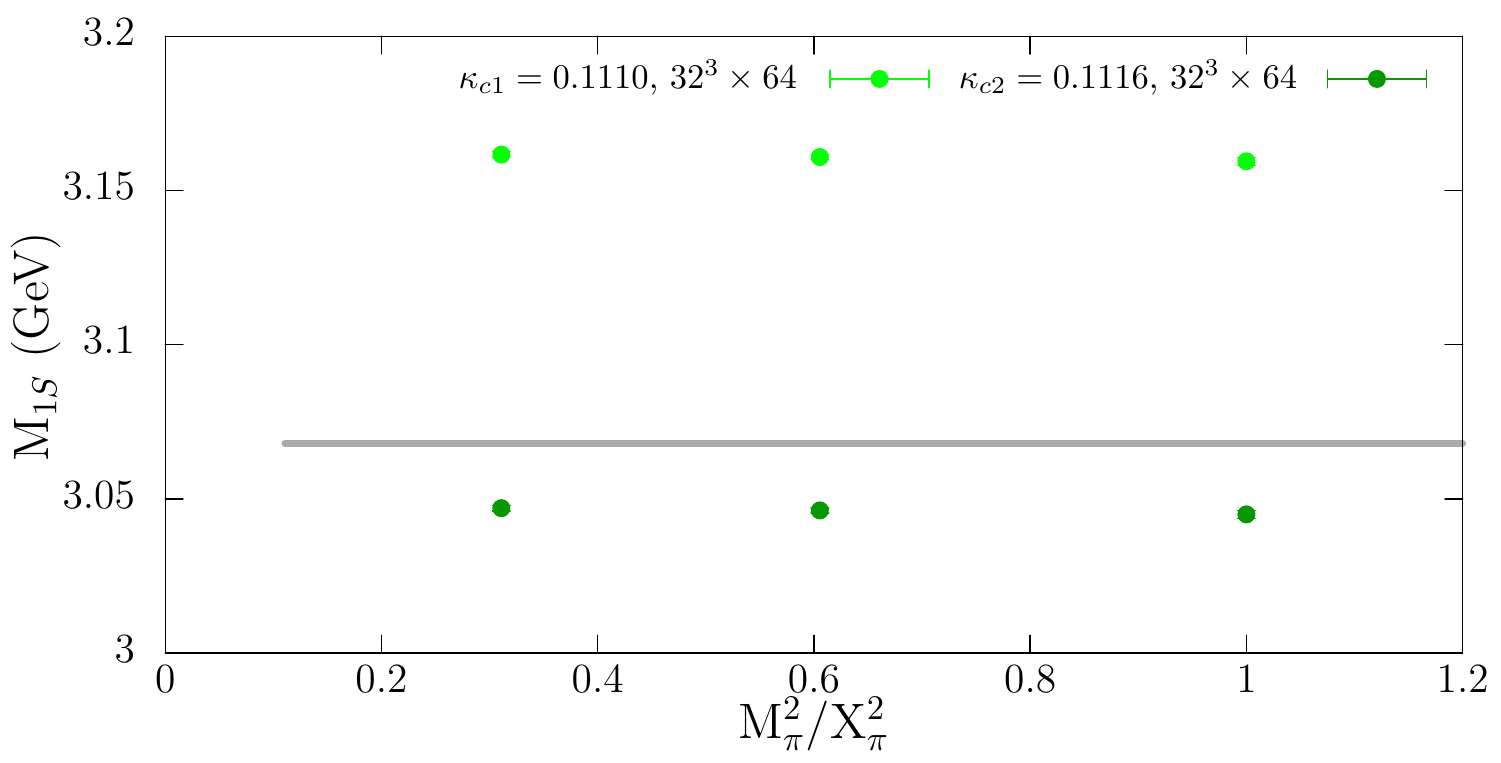}
\caption{\label{fig_1S_D1sfav} The charmonium $M_{\rm 1S}$ mass
  determined with $\kappa_{c1/c2}$. The horizontal
  line indicates the experimental value.
}
\end{figure}

The charm quark mass parameter, $\kappa_{\rm charm}$, was tuned by
requiring that the $1{\rm S}$ spin-averaged charmonium mass, $M_{\rm
  1S} = \frac14 M_{\eta_c} + \frac 34 M_{J/\psi} $ is equal to the
experimental value. For this quantity the dependence on the light sea
quarks is sub-leading and furthermore the associated discretization
errors are likely to be reduced since contributions from fine
structure interactions are removed.  The tuning was performed in two
stages on each $32^3\times 64$ ensemble. Firstly, a coarse tuning
involving 200 configurations separated by 2 trajectories determined a
preliminary value for $\kappa_{\rm charm}$. Two values of the mass
parameter, $\kappa_{c1}$ and $\kappa_{c2}$~(see \ret{configs}), were
chosen to bracket $\kappa_{\rm charm}$, spaced closely enough that a
linear interpolation in $1/\kappa$ to the physical point would be
sufficient.  In the second stage the meson and baryon spectrum was
determined with full statistics for $\kappa_{c1/c2}$ and the final
value of $\kappa_{\rm charm}$ determined from $M_{\rm 1S}$. This  mass
 is displayed in \refig{fig_1S_D1sfav} for 
 $\kappa_{c1/c2}$ and the results for $\kappa_{\rm charm}$ are
listed in \ret{final_kappa_value}.  The meson and baryon masses were
then interpolated to the physical charm quark mass.
\refig{fig_1S_D1sfav} shows there is little dependence of $M_{\rm 1S}$ 
at fixed $\kappa_{c1/c2}$ on
the sea quark masses. This translates into very similar values for
$\kappa_{\rm charm}$ across the different ensembles. The same
 values were employed for the smaller, $24^3\times
48$, ensemble where the light/strange quark masses are matched. For
the intermediate small volume, an interpolation was performed.

\section{Results}

\subsection{\label{finitesize} Finite size effects}

With two volumes available with spatial extents, $L=24a$ and $L=32a$,
corresponding to $1.8\,{\rm fm}$ and $2.4\,{\rm fm}$, respectively, a
study of finite volume effects of the charmed baryon spectrum can be
attempted. The same quark mass parameters have been used for two sets of
ensembles, at the symmetric point~($\kappa_{\rm sym}=0.12090$) and the
asymmetric ensemble with the light-strange combination~($\kappa_{\rm
  asym}= 0.12104, 0.12062 $), see Table~\ref{configs}. Differences
between the masses extracted from the two volumes, $\Delta M =
M_{24}-M_{32}$, can be computed directly; the results are given in
Tables~\ref{finitevols_pos} and \ref{finitevols_neg} in
Appendix~\ref{AppD}. The mass splittings are
only significantly non-zero~(where we take $2.5\sigma$ to indicate
significance) at the symmetric point since we have much higher
statistics in this case for the smaller volume. However, similar or
larger magnitudes for the difference on the asymmetric
ensemble~(although possibly less significant) indicate a trend. In
addition, we look for consistency between the two types of interpolators,
HQET and SU(4), where available and that any difference decreases if a
light quark is replaced by a strange quark within a multiplet~(for the
asymmetric ensemble).

\begin{figure}[ht!]
\includegraphics{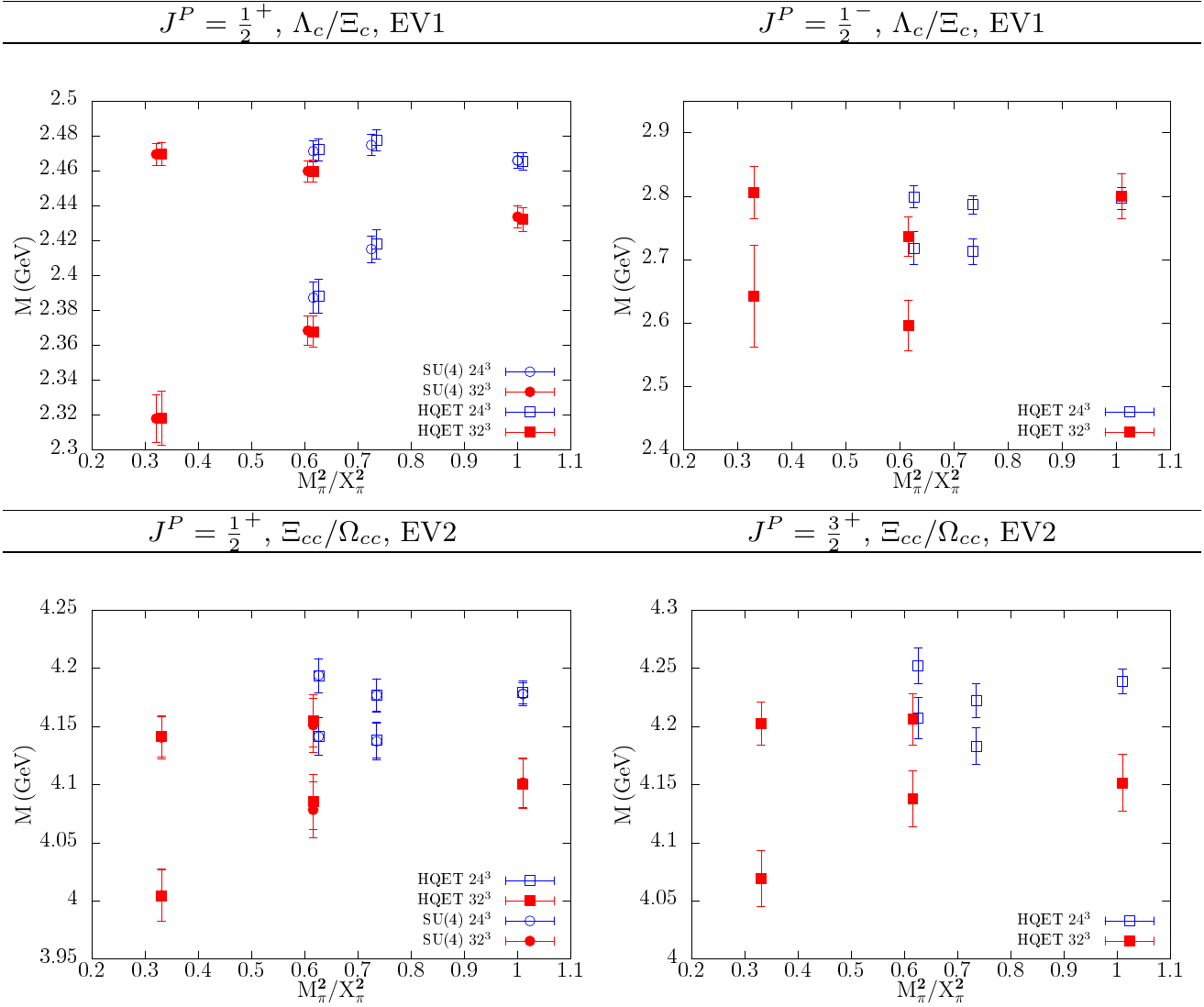} 
\caption{\label{vol_effects} Comparison of the $V=24^3\times 48$ and
  $32^3\times 64$ results as a function of $M_\pi^2/X^2_\pi$ for a
  selection of multiplets exhibiting finite volume effects, where
  $1-M_\pi^2/X^2_\pi\propto\delta m_{\ell}+{\rm O}(\delta
  m_{\ell}^2)$. Top left and top right, the first eigenvalue~(EV1) of
  the positive and negative parity singly charmed anti-triplet,
  respectively. Bottom left and bottom right, the second
  eigenvalue~(EV2) of the positive parity doubly charmed triplet for
  $J=\frac 12$ and $\frac 32$, respectively. }
\end{figure}

In general, one expects larger finite volume effects for singly
charmed baryons compared to doubly charmed ones, similarly, for
excited states compared to ground states. Whether the effects can be
observed depends on the size of the statistical errors. Our results
fit into the expected pattern. For the positive parity multiplets, we
find masses are reduced by about $10$--$40$~MeV when changing
the spatial extent from $1.8$~fm to $2.4$~fm for the ground state 
sextets and the anti-triplet, while the ground state
doubly charmed triplets are unchanged.  The statistical errors for the
excited states of the singly charmed multiplets are large and no
significant finite size effects are visible apart from a small 
increase for the anti-triplet. 
However, in the doubly charmed case, with much smaller errors, clear differences of around
$40$--$87$~MeV emerge, again decreasing the mass with increasing volume.
The negative parity states are less affected with only the ground
state anti-triplet showing a notable decrease between $L=24a$ and
$32a$. The $J=\frac 12$ sextet and $J=\frac 32$ triplet have small
positive shifts.  Note that the $J=\frac 32$ negative parity sextet is
not included in Table~\ref{finitevols_neg} as we were not able to
extract a reliable signal in these channels for the smaller volume.
The finite volume effects for a subset of states are shown in
\refig{vol_effects}.

An estimate of the size of the remaining systematic in the $L=32a$
results can be made by assuming the finite volume dependence of the
masses is of the form $M_L=M_{\infty}+cF(L)$, where $M_{\infty}$ is
the infinite volume mass and $F(L)=e^{-M_\pi L}$ is the leading order
effect in the asymptotic regime and $F(L)=1/L^3$ otherwise. This leads
to
\begin{equation}
\Delta M_\infty =
M_{32}-M_\infty= (M_{24}-M_{32})(F(24a)/F(32a)-1)^{-1} = \Delta M\Delta F,
\end{equation}
where $\Delta F=0.4$ and $0.6$ for the symmetric~($\kappa_{\rm sym}$) and asymmetric
ensemble~($\kappa_{\rm asym}$) in the asymptotic case, respectively, and
$\Delta F=0.7$ in the $1/L^3$ regime. Considering the values of
$\Delta M$ in Table~\ref{finitevols_pos}, this leads to the infinite
volume masses at $M_\pi=355$~MeV being, for example, approximately
$9$--$22$~MeV smaller for the $\Sigma_c$, $11$--$14$~MeV smaller for the
$\Lambda_c$ and $34$--$48$~MeV smaller for the excited state $\Xi_{cc}$
and $\Xi_{cc}^*$. At this pion mass with $L=32a$, $LM_\pi=4.3$, one
would expect the lower bounds to apply. These estimates
should be compared to the final values for the masses given in
Table~\ref{tablechiex}, discussed in  Section~\ref{extrapfinal} .

Since the lightest pion mass for the $L=24a$ ensembles is quite large, 
$M_\pi=364$~MeV, we do not attempt to extrapolate these results
to the physical point and they are not included in any further analysis.

\subsection{\label{extrapmass} Extrapolation to the physical masses }

An important advantage of approaching the physical point using the
QCDSF strategy, starting from the flavor symmetric point and keeping
the average light quark mass fixed, is that the light quark mass
dependence of physical quantities is well constrained as long as
flavor symmetry violations are reasonably small.  The dependence of
meson and baryon masses on the symmetry breaking parameter, $\delta
m_{\ell}=m_s-m_{\ell}$, was derived in Ref.~\cite{Bietenholz:2011qq} and extending
this method to include partially quenched quarks was
discussed in Refs.~\cite{Bietenholz:2011qq,Horsley:2013wqa}. Here, we follow an analogous
approach treating the charm quark as a spectator.  We classify the
charmed baryons according to the SU(3) flavor symmetry multiplets for
the one or two light quarks present. In the
isospin limit, we have the following channels \bite
\item{sextet: $(\Sigma_c, \Xi'_c, \Omega_c)$ for $J=\frac 12$ and 
$(\Sigma^*_c, \Xi^*_c, \Omega^*_c)$ for $J=\frac 32$, }
\item{triplet: $(\Xi_{cc},\Omega_{cc})$ for $J=\frac 12$ 
   and $(\Xi^*_{cc},\Omega^*_{cc})$ for $J=\frac 32$,}
\item{anti-triplet: $(\Lambda_c,\Xi_c)$ for $J=\frac 12$.}  \eite The
  light quark mass dependence of each multiplet can be derived by
  considering the matrix element $\langle B_R|M|B_R\rangle$.  The mass
  matrix $M = \overline{m}\identidad  - \sqrt3 \delta m_{\ell} T_8 = \overline{ m} H_0 +
  \delta m_{\ell}H_8$ where $T_8$ is the SU(3) generator and $B_R$
  is the baryon multiplet in representation
  $R$. $H_8$ transforms like the eighth component of the octet
  representation and since $\overline{ m}$ is kept constant, this is the only
  part of the mass matrix which is changing as we approach the
  physical limit. Performing a perturbative expansion in $\delta m_{\ell}$
  we have
\begin{equation}
\langle B_R|M |B_R\rangle  = M_0^R +\delta m_{\ell}\langle B_R|H_8|B_R\rangle 
+ \delta m_{\ell}^2\sum_{R^\prime \neq R} 
\frac{|\langle B_R|H_8|B_{R^\prime}\rangle|^2}{M_0^R-M_0^{R^\prime}}+\ldots,\label{gmo_pert}
\end{equation}
where $M_0^R= \overline{ m} \langle B_R|H_0|B_R\rangle$ is the mass of the multiplet
$R$ in the flavor symmetric limit. In the light baryon sector the
expansion to first order leads to the Gell-Mann--Okubo~(GMO) relation
\cite{GellMann:1962xb,Okubo:1961jc}. The expressions for the linear
terms for the charmed baryon multiplets are derived in Appendix
\ref{derive_gmo}. Here, we comment that the matrix elements are the
flavor singlet terms arising from the decomposition of the direct
products of ${\bf \overline{R}\otimes T_H \otimes R}$, where ${\bf T_H}$ is
the representation of $H$. This means that the number of terms
appearing at each order is given by the number of times the trivial
representation appears in the direct product. At the lowest
order~(${\bf \overline{R}\otimes \identidad \otimes R}$) only one term
arises, and for the sextet, triplet and anti-triplet that we are
considering there is also only one term at first order~(${\bf
  \overline{R}\otimes 8 \otimes R}$). Hence, within each multiplet there is
  only one coefficient at ${\rm O}(\delta m_{\ell})$ and one can extrapolate all
states within a multiplet~(to this order) using just two
parameters. However, at second order the trivial representation
appears three~(two) times for the sextet~(triplet/anti-triplet) such
that there are the same number of new coefficients as particles within a
multiplet.

Below we display the expressions for the light quark mass dependence of the
charmed baryons masses derived in Appendix~\ref{derive_gmo}~(Eqs.
~(\ref{sextet}), (\ref{triplet}) and (\ref{antitriplet})). We use the GMO relations
for the light mesons~\cite{Bietenholz:2010jr,Bietenholz:2011qq}
\begin{eqnarray}
M_\pi^2 &=&(M^{\bf 8}_0)^2 - \frac 23 D\delta m_{\ell}\ + {\rm O}(\delta m_{\ell}^2), \\
M_K^2 &=&  (M^{\bf 8}_0)^2 + \frac 13 D\delta m_{\ell}\ + {\rm O}(\delta m_{\ell}^2), 
\end{eqnarray} 
and $X^2_\pi = \frac13(M_\pi^2+2M_K^2) = (M^{\bf 8}_0)^2 + {\rm O}(\delta m_{\ell}^2)$ to replace $\delta m_{\ell}$ with
$({X_\pi^2 - M_\pi^2})/{X_\pi^2}$ ignoring ${\rm O }(\delta m_{\ell}^2) $
terms. Such higher order terms can be absorbed into the coefficients of
the quadratic terms below.
\begin{description}
\item[Sextet, $R={\bf 6}$]{
\bea \label{ext_sextet}
M_{\Sigma_c} &=& M_0^{{\bf 6}}  - \tfrac 23 A_1\frac{X_\pi^2 -M_\pi^2}{X_\pi^2}
+ A_2 \left(\frac{X_\pi^2 -M_\pi^2}{X_\pi^2} \right)^2, \\
M_{\Xi^\prime_c} &=& M_0^{{\bf 6}}  + \tfrac 13 A_1\frac{X_\pi^2 -M_\pi^2}{X_\pi^2} + 
A_3\left(\frac{X_\pi^2 -M_\pi^2}{X_\pi^2} \right)^2, \\
M_{\Omega_c} &=& M_0^{{\bf 6}} + \tfrac 43 A_1\frac{X_\pi^2 -M_\pi^2}{X_\pi^2} +
A_4\left(\frac{X_\pi^2 -M_\pi^2}{X_\pi^2}\right)^2.
\eea
}
\item[Triplet, $R={\bf 3}$]{
  \bea \label{ext_triplet}
  M_{\Xi_{cc}} &=& M_{0}^{{\bf 3}} - \tfrac 13 B_1 \frac{X_\pi^2 -M_\pi^2}{X_\pi^2}+ 
  B_2\left(\frac{X_\pi^2 -M_\pi^2}{X_\pi^2}\right)^2,\label{tripleta}
  \\
M_{\Omega_{cc}} &=& M_{0}^{{\bf 3}} + \tfrac 23 B_1\frac{X_\pi^2 -M_\pi^2}{X_\pi^2}  + 
B_3\left(\frac{X_\pi^2 -M_\pi^2}{X_\pi^2}\right)^2.\label{tripletb}
\eea

   }
\item[Anti-Triplet, $R={\bf \bar{3}}$]{
  \bea 
M_{\Lambda_{c}} &=& M_{0}^{{\bf \bar{3}}} - \tfrac 23 C_1\frac{X_\pi^2 -M_\pi^2}{X_\pi^2} + C_2\left(\frac{X_\pi^2 -M_\pi^2}{X_\pi^2}\right)^2,\label{ext_antitriplet}
 \\
M_{\Xi_{c}} &=& M_{0}^{{\bf \bar{3}}} + \tfrac 13 C_1\frac{X_\pi^2 -M_\pi^2}{X_\pi^2} + C_3\left(\frac{X_\pi^2 -M_\pi^2}{X_\pi^2}\right)^2.\label{ext_antitripletb}
\eea
   }   
\end{description}
Note that for the sextet and triplet there are equivalent expressions
for the $J=\frac 32$ states with different coefficients. In
Refs.~\cite{Bietenholz:2011qq,Horsley:2013wqa} analogous
parameterizations are given~(but to higher order in their
framework)\footnote{Note that their expansion is given in terms of
  $m_q - \overline m$, which, unlike $\delta
  m_\ell=m_s-m_\ell$, changes when shifting $\overline{m}$.}.  The
above forms can be applied to extrapolate our results to the physical
point if the charmed baryons are regular three quark states where the
charm is a spectator and the particles fall into the SU(3) flavor
multiplets. As for any expansion the coefficients of the flavor
symmetry breaking terms should be small and the extrapolation will be
well constrained whenever only the first few terms are significant.

\begin{figure}[ht!]
\vspace{0.7em}
\includegraphics{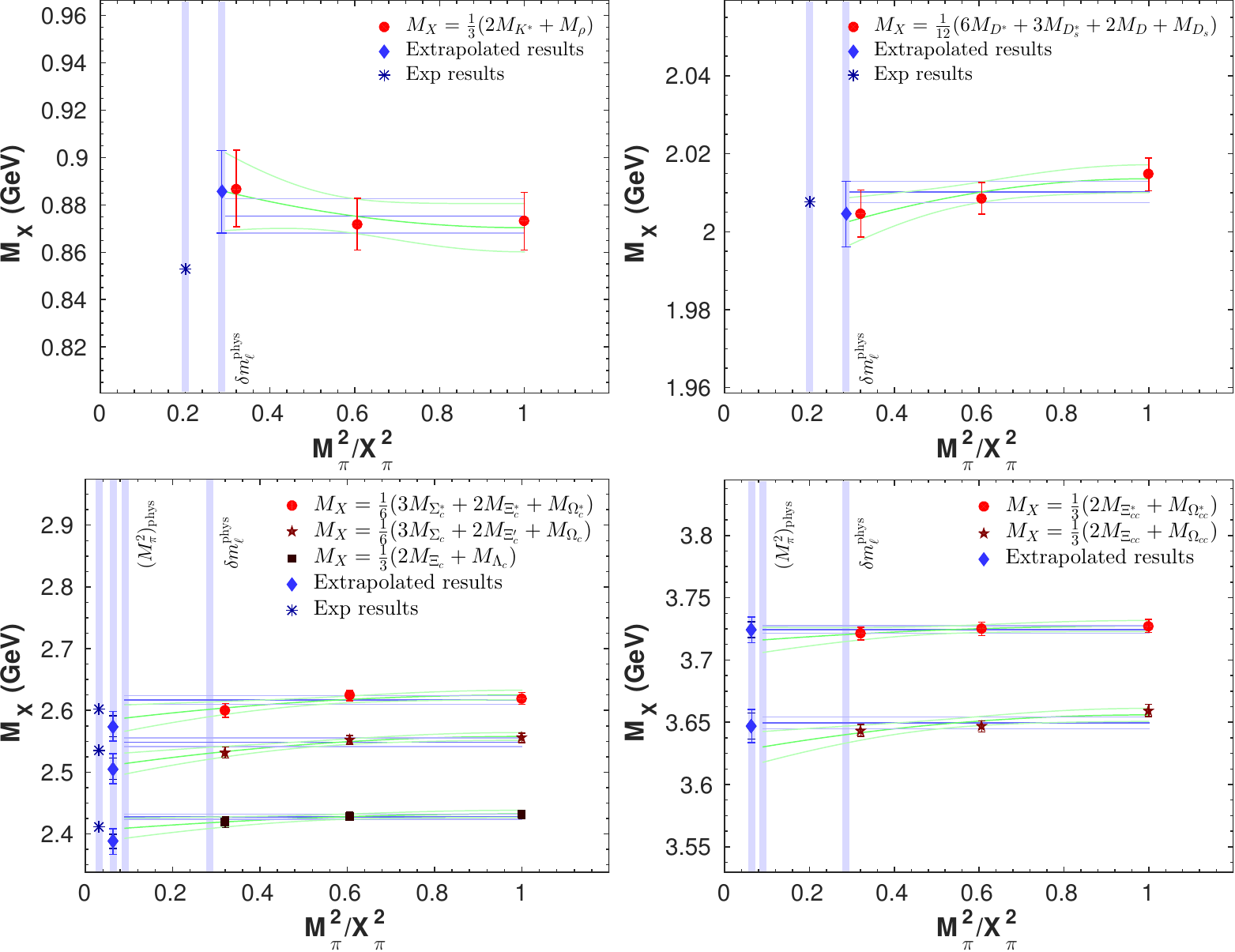}  
\caption{\label{fig_flsinglet}{Flavor invariant quantities as a
    function of $M_\pi^2/X^2_\pi$ on the $32^3\times 64$ ensembles.
    Note that $1-M_\pi^2/X^2_\pi\propto \delta m_{\ell}+{\rm O}(\delta
    m_{\ell}^2)$.  (Top left) the light vector meson and~(top right)
    the spin averaged 1S heavy-light meson. Fits to a constant and a
    constant plus quadratic term in $\delta m_{\ell}$ are indicated
    along with the extrapolated values~(blue diamonds) obtained at the
    point corresponding to physical $\delta m_{\ell}=\delta
    m^{\rm phys}_\ell$. The error on the extrapolated results incorporates
    both fits using Eqs.~(\ref{mave}) and~(\ref{deltamave}). The
    experimental results are also shown. Similarly for (bottom left)
    singly charmed baryons and (bottom right) doubly charmed baryons
    for HQET inspired interpolators.  For the charmed baryons, the final
    values are obtained by extrapolation to $\delta m^{\rm phys}_{\ell}$ and
    then shifting by $\Delta X_{\rm multiplet}$, see the text. Two errors are
    displayed, showing the uncertainty with and without the error on
    $\Delta X_{\rm multiplet}$ included in quadrature. }}
\end{figure}

In this study we have the additional complication of needing to
correct for the fact that the simulation trajectory misses the
physical point in the $m_s-m_{\ell}$ mass plane~(see
Fig.~\ref{simtraj}). The value of
$\overline{m}=\overline{m}^{\rm phys}+\Delta \overline{m}$ is off by a
small amount $\Delta \overline{m}$ and this means, in the expressions
above, the masses $M_0$ also differ from the physical ones by
$M_0=M_0^{\rm phys}+\Delta M_0$. The linear $\delta m_{\ell}$ expansion
coefficients should be unaffected. Hence, to make contact with the
physical point we extrapolate to the physical $\delta m_{\ell}$
defined as $M_\pi^2=(M_K^{\rm sim})^2-(M_K^{\rm phys})^2+(M_\pi^{\rm
  phys})^2$ and then shift the masses within each multiplet by $\Delta
X_{\rm multiplet} = X_{\rm multiplet}^{\rm phys}-X_{\rm
  multiplet}^{\rm sim}\sim \Delta M_0^{\rm multiplet}$, where $X_{\rm
  multiplet}\sim M_0^{\rm multiplet}$ is the flavor average for each
  multiplet.  This procedure should be correct up to 
  ${\rm O} (\Delta \overline m \delta m_{\ell}^2)$ corrections, see 
  Eq.~(\ref{gmo_pert}).
For the singly charmed baryons we could choose to determine
$\Delta X_{\rm multiplet}$ using the flavor averaged $J=\frac 12^+$
states and the corresponding experimental results. 
However, this would
reduce predictability and the shift would still need to be estimated
for the doubly charmed baryons. Instead we look for other multiplets
that are likely to have similar $\Delta X_{\rm multiplet}$. Assuming
that all the singly charmed multiplets are shifted by the same amount,
we choose the flavor averaged light vector channel, $\frac
13(2M_{K^*}+M_\rho)$, to estimate $\Delta X_{{\bf
6}}\approx \Delta X_{{\bf\bar{3}}}$. For the doubly charmed triplet we use the spin
and flavor averaged heavy-light combination,
$\frac{1}{12}(6M_{D^*}+3M_{D_s^*}+2M_D+M_{D_s})$.  Treating the charm
as a spectator, the $D$ and $D_s$ channels should form triplets and obey
modified versions of Eqs.~(\ref{tripleta}) and~(\ref{tripletb}).

\begin{figure}[ht!]
\vspace{0.7em}
\includegraphics[width=0.65\textwidth]{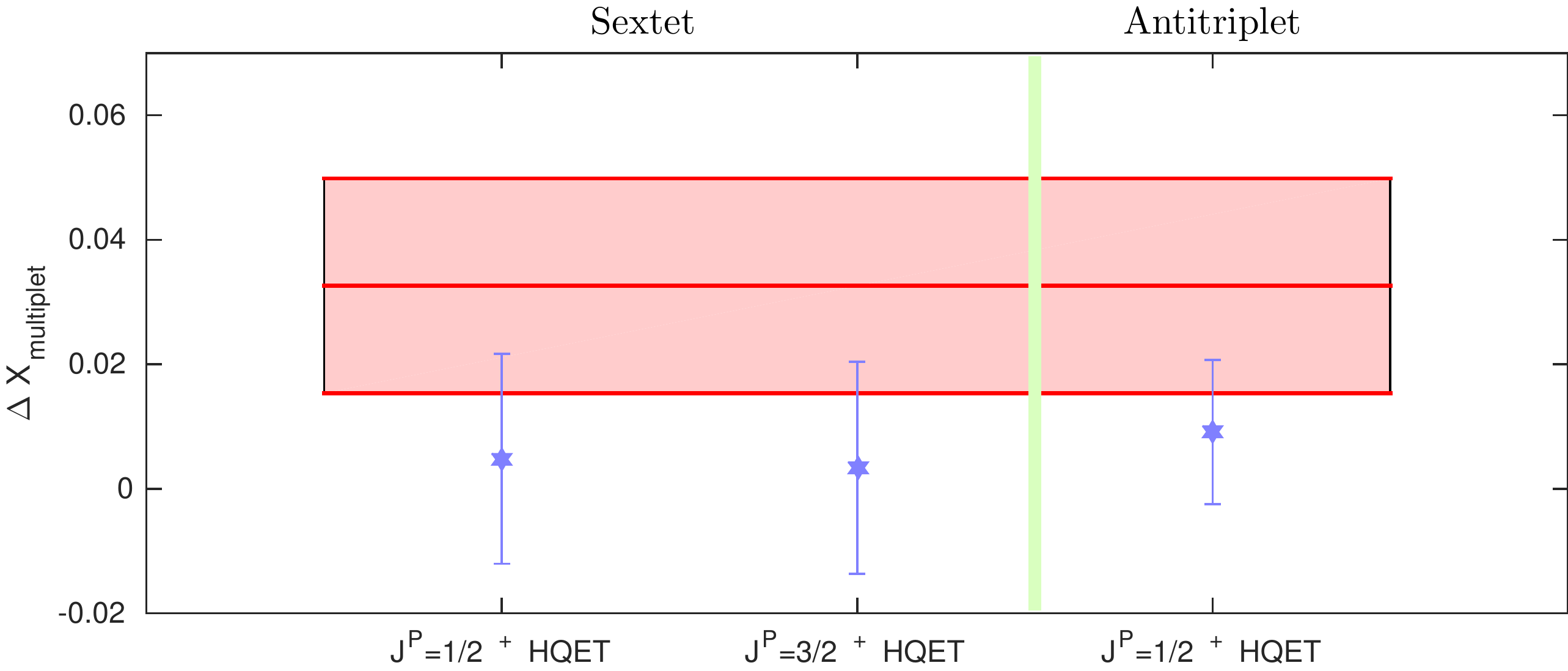}  
\caption{\label{cmp_shift}{ The shift $\Delta X_{\rm multiplet}$
    calculated by comparing different flavor averaged mass
    combinations with experiment both in the singly charmed
    sector~(blue stars) and the light vector meson sector~(shaded
    band).}} 
\end{figure}

Figure~\ref{fig_flsinglet} shows the light vector and $D/D_s$ flavor
singlet combinations as  functions of $M_\pi^2/X_\pi^2$. As expected
for quantities which depend on $\delta m_{\ell}$ to second order, there is
very little change with the light quark mass. Note also that the
extrapolation to $\delta m_{\ell}^{\rm phys}$ is relatively short. Fitting to a
constant~(fit 1) and a constant plus quadratic term~(fit 2) we obtain
extrapolated values, $M_{\rm ave}\pm \Delta M_{\rm ave}$, using the
following formulae
\begin{eqnarray}
   M_{\rm ave} &=&\frac12\left({\max_{i \in \{1,\dots n \}}(M_i+\Delta M_i) +
\min_{i \in \{1, \cdots n\}}(M_i-\Delta M_i) }\right),\label{mave}\\ \Delta
M_{\rm ave} & =&  \frac12\left({\max_{i \in \{1,\dots n \}}(M_i+\Delta M_i) -
\min_{i \in \{1, \cdots n\}}(M_i-\Delta M_i) }\right),\label{deltamave}
\end{eqnarray}
where  in this case $n=2$, $M_i\pm\Delta M_i$ is the result from fit $i=1,2$. The difference
of $M_{\rm ave}$ to experiment gives the estimate of $\Delta X_{\rm
  multiplet}$. $\Delta M_{\rm ave}$ is taken as the error in the
shift. For the $D/D_s$ case the shift is consistent with
zero, $-3(8)$~MeV, while for the light vector it is $33(17)$~MeV.

Figure~\ref{fig_flsinglet} also shows the positive parity charmed
baryon flavor singlet combinations. Here, the data are extrapolated to
$\delta m^{\rm phys}_\ell$ and the fit error is calculated as above.
The appropriate shift, $\Delta X_{\rm multiplet}$ is applied, where
the associated error is added in quadrature. Encouragingly, the final
values are in agreement with experiment for the sextet and
anti-triplet, consistent with $\Delta X_{\rm multiplet}$ being small
and of a similar magnitude for the different
multiplets. This is emphasized in
    Fig.~\ref{cmp_shift} which compares $\Delta X_{\rm multiplet}$
    determined from $\frac 13(2M_{K^*}+M_\rho)$ with the shift
    obtained using the flavor averaged masses for the sextet and
    anti-triplet and the corresponding experimental values for the
    singly charmed baryons. 

As for the
mesons, there is very little dependence o $\delta m_{\ell}$ for all
flavor average masses in Fig.~\ref{fig_flsinglet}. Ensembles on the
correct trajectory, close to the physical point will be needed to
quantify the size of the leading, i.e.  ${\rm O}(\delta m_\ell^2)$,
flavor symmetry violations. For our data the coefficients of the
quadratic terms are zero within $2.5\,\sigma$.

\subsection{\label{extrapfinal} Baryon and meson masses at the physical point}

\begin{figure}[t!]
\includegraphics{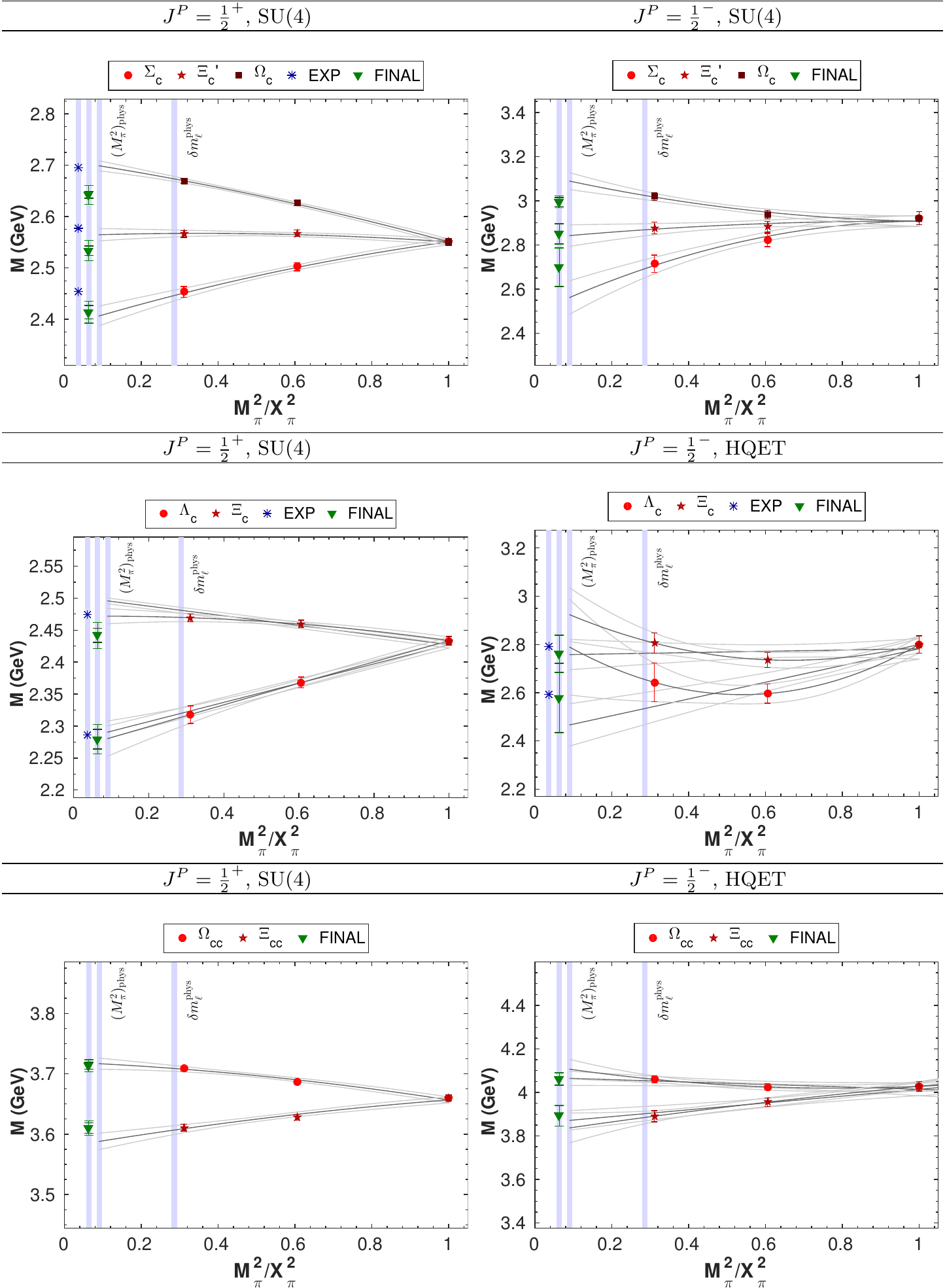}
\caption{\label{fan_plots} Extrapolation to $\delta m_{\ell}^{\rm phys}$
  for (left) positive parity and (right) negative parity charmed
  baryons on $V=32^3\times 64$ ensembles. The final~(extrapolated)
  results are displayed as in Fig.~\ref{fig_flsinglet}, with the shift
  $\Delta X_{\rm multiplet}$ included. The type of fit displayed~(GMO
  or LC) corresponds to that used to determine the final values, see
  Table~\ref{tablefitrange}.  For improved visibility, where GMO2, LC1 and LC2
  fits were used, we only show the GMO2 fit.  The type of interpolators
  chosen for each multiplet~(HQET/SU(4)) is given above each
  plot. Where available the experimental results are displayed on the very left.}
  \vspace{-0.5cm}
\end{figure}

Turning to the extrapolation of the individual baryon masses our
strategy is as follows.  For a given multiplet, we employ the
Gell-Mann--Okubo relations with and without the quadratic term,
denoted GMO1 and GMO2, respectively. If the $\chi^2/{\rm dof} \lesssim
2$ for both fits, the final value for the mass at the physical point
is computed using Eqs.~(\ref{mave}) and~(\ref{deltamave}) and
performing the shift, $\Delta X_{\rm multiplet}$, as discussed
above\footnote{Whenever $\chi^2/{\rm dof}>1 $, the error $\Delta
  M_{1,2}$ is inflated by the factor $\sqrt{\chi^2/{\rm dof}}$.}.  The
mass values within a multiplet on a given ensemble are highly
correlated. These correlations have been taken into account in the
fits.  Overall the GMO expressions describe the data well, however, in
a small number of cases either the linear~(GMO1) fit produced a
$\chi^2/{\rm dof}>2$ or both the linear and quadratic~(GMO2) fits.  If
the masses do not fall into the GMO pattern, we can extrapolate using
the functional forms Eqs.~(\ref{ext_sextet})--(\ref{ext_antitripletb})
but allowing the coefficients of the linear terms to differ for each
member of a multiplet~(i.e.  only $M_0$ is constrained to be the
same).  Such phenomenological fits are always possible
    if $\overline{m}$ is kept fixed in the simulation. Again we
considered these fits including and excluding quadratic terms, denoted
as less-constrained 1 and 2~(LC1 and LC2), respectively. $M_{\rm ave}$
and $\Delta M_{\rm ave}$ are calculated as above, where for the cases
where we combined GMO2, LC1 and LC2 extrapolated results we take the
minimum and maximum in Eqs.~(\ref{mave}) and~(\ref{deltamave}) of all
three results\footnote{Note that for the LC2 fit there are as many
  data points as there are unknown coefficients, and thus there are no
  remaining degrees of freedom.}. \refig{fan_plots} shows examples of
the extrapolations. The corresponding values of
  the coefficients extracted from the fits are given in
  Table~\ref{tablecoefficients}~(Appendix \ref{AppC}). A summary of
those fits included in the determination of the results at the
physical point is provided in
Table~\ref{tablefitrange}~(Appendix~\ref{AppC}) and the final values
are given in Table~\ref{tablechiex}.

\begin{figure}[t!]
\includegraphics{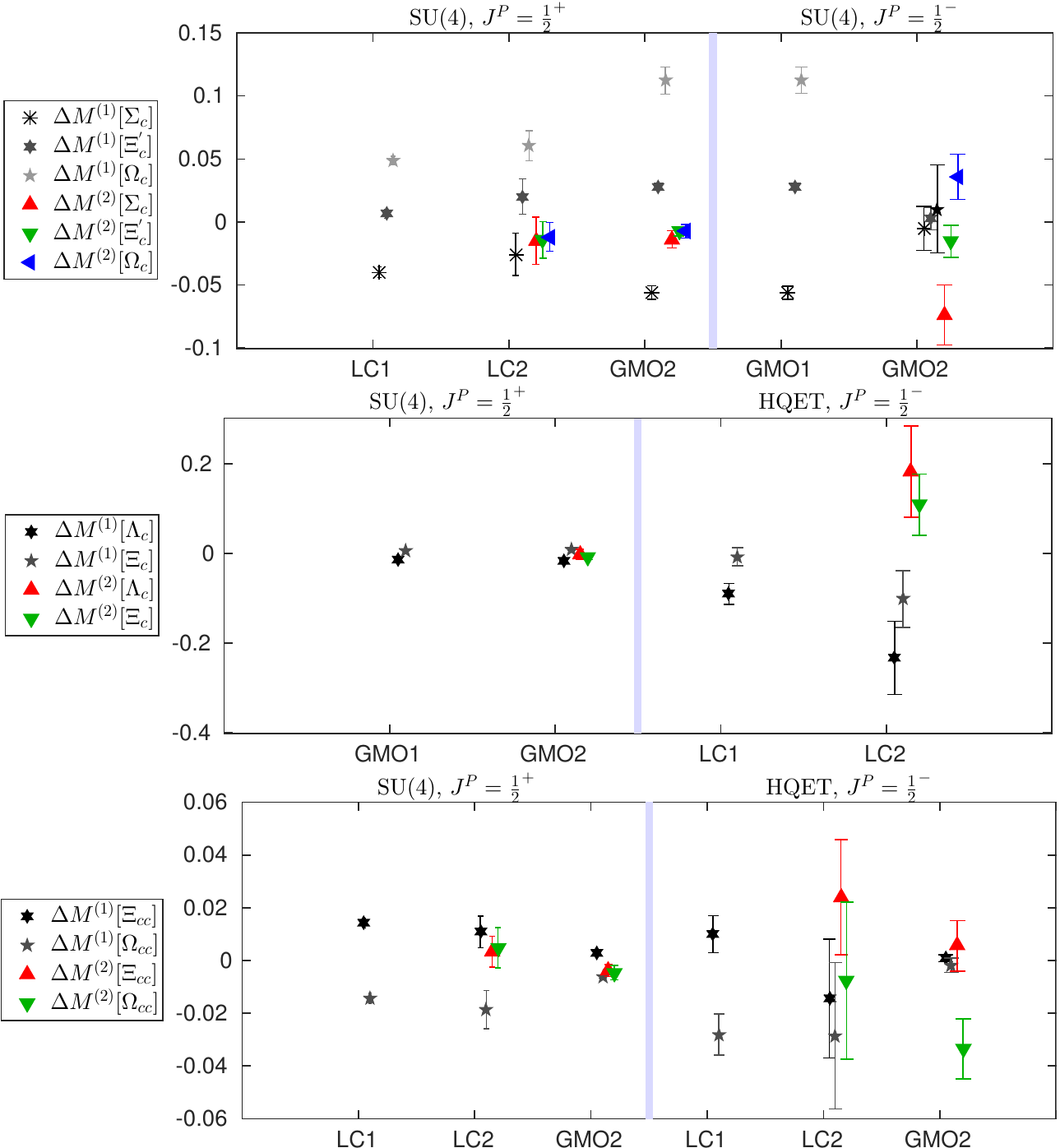}
\caption{\label{coefficients} Contributions of the first and second
  order terms in the expansions in $\delta m_\ell$ evaluated at
  $\delta m_\ell^{\rm phys}$ for each multiplet displayed in
  Fig.~\ref{fan_plots}.  The contributions are given as a percentage of
  the baryon mass of each multiplet in the SU(3) flavour limit~($M_0$), i.e. $M_B =
  M_0( 1 + \Delta M^{(1)} + \Delta M^{(2)})$ for each baryon mass
  $M_B$.} 
  \vspace{-0.5cm}
\end{figure}

As mentioned above, the GMO expressions fit most data well, in
particular the positive parity states. In the cases where the
quadratic terms are statistically significant, the coefficients of
these terms are generally small.  The size of the
    linear and quadratic contributions at $\delta m_\ell^{\rm phys}$
    are quantified in Fig.~\ref{coefficients} for the ground state
    $J=\frac{1}{2}$ multiplets. The first order terms, for positive
    parity, are small while the second order terms are smaller
    still. For negative parity, both terms are reasonably small
    but of  similar sizes. A larger set of ensembles
    with pion masses closer to the physical point are needed in order
    to reliably extract the coefficients and quantify the size of the
    individual terms. Interestingly, the $J^P=\frac 12^-$
    anti-triplet~(shown in \refig{fan_plots}) and the $J=\frac 32^-$
    triplet could not be fitted to the GMO formulae. Indeed,
    Fig.~\ref{coefficients} shows that for a quadratic fit the linear
    and quadratic contributions to the $J^P=\frac 12^-$ anti-triplet are both
    around $10-20\%$, albeit with large errors. This may indicate they
have a more complicated internal structure. We also remark that, as
discussed in Section~\ref{varmethod}, our ground state spin-1/2
$\Omega_c$ and $\Xi_c^\prime$ negative parity results are likely to be
scattering states. In the sextet, only the results for $\Sigma_c$ away
from the symmetric limit represent single particle states. From
\refig{fan_plots} we can see that a linear extrapolation to $\delta
m_\ell^{\rm phys}$ of these two points would give consistent results
with the GMO fit. The errors on the final values in
Table~\ref{tablechiex} are of similar sizes or larger than our
estimate of the size of the discretization errors, in the range of
$10-20$~MeV. Apart from the positive parity singly charmed
baryons, the shift $\Delta X_{\rm multiplet}$ does not have a
significant impact on the results.

\begin{table}[ht!]
\begin{center}
\begin{tabular}{cc|c|c|c|c}\hline
\hline
         &        &    \multicolumn{2}{c}{$M_1$ (GeV)} &       \multicolumn{2}{|c}{$M_2$ (GeV)}\\
Particle & Parity &   SU(4) & HQET & SU(4) & HQET \\ 
\hline
\hline
     $\Sigma_{c}$ & $+$ &   $2.414(13)(22)$&   $2.434(20)(26)$&     $3.090(94)(95)$& $3.144(115)(116)$\\ 
 $\Xi_{c}^\prime$ & $+$ &   $2.534(09)(20)$&   $2.543(12)(21)$&     $3.177(67)(69)$& $3.226(72)(74)$\\ 
     $\Omega_{c}$ & $+$ &   $2.642(07)(18)$&   $2.648(09)(19)$&     $3.268(47)(50)$& $3.294(50)(53)$\\ \hline 
    $\Lambda_{c}$ & $+$ &   $2.280(15)(23)$&   $2.280(17)(24)$&     $3.183(75)(77)$& $3.181(76)(78)$ \\
        $\Xi_{c}$ & $+$ &   $2.442(11)(21)$&   $2.442(11)(20)$&     $3.234(37)(41)$& $3.245(36)(40)$ \\ \hline 
       $\Xi_{cc}$ & $+$ &   $3.610(09)(12)$&   $3.610(09)(12)$&     $4.017(39)(40)$& $4.018(40)(41)$\\ 
    $\Omega_{cc}$ & $+$ &   $3.713(06)(10)$&   $3.713(06)(10)$&     $4.157(31)(32)$& $4.158(31)(32)$\\  \hline
   $\Sigma^*_{c}$ & $+$ &              &   $2.506(18)(25)$& &      $3.234(121)(122)$\\
    $\Xi^{*}_{c}$ & $+$ &              &   $2.608(13)(22)$& &      $3.295(87)(89)$\\
   $\Omega^*_{c}$ & $+$ &              &   $2.709(11)(21)$& &      $3.355(64)(66)$\\\hline
     $\Xi^*_{cc}$ & $+$ &              &   $3.694(07)(11)$& &      $4.078(36)(37)$\\
  $\Omega^*_{cc}$ & $+$ &              &   $3.785(06)(10)$& &      $4.215(26)(28)$\\\hline\hline
     $\Sigma_{c}$ & $-$ &       $2.698(87)(89)$&   &  &\\ 
 $\Xi_{c}^\prime$ & $-$ &       $2.850(44)(47)^{**}$&    &  &\\ 
     $\Omega_{c}$ & $-$ &       $2.995(20)(26)^{**}$&    &  &\\ \hline
    $\Lambda_{c}$ & $-$ &       &       $2.578(144)(145)$&  &\\ 
        $\Xi_{c}$ & $-$ &       &       $2.761(77)(79)$&  &\\ \hline
       $\Xi_{cc}$ & $-$ &       &       $3.892(47)(48)$& $4.306(45)(44)$  & \\ 
    $\Omega_{cc}$ & $-$ &       &       $4.061(28)(29)$& $4.394(42)(41)$  & \\ \hline
   $\Sigma_{c}^*$ & $-$ &        &      $2.740(48)(51)$&  &\\ 
      $\Xi_{c}^*$ & $-$ &        &      $2.891(32)(36)$&  &\\ 
   $\Omega_{c}^*$ & $-$ &        &      $3.016(32)(37)$&  &\\ \hline
     $\Xi_{cc}^*$ & $-$ &        &      $3.989(58)(58)$&  & $4.447(57)(57)$ \\ 
  $\Omega_{cc}^*$ & $-$ &        &      $4.132(42)(43)$&  & $4.567(55)(56)$\\ \hline

\hline
\end{tabular}
\caption{\label{tablechiex} Mass estimates at the physical point for
  the ground and first excited states of the singly and doubly charmed
  positive and negative parity baryons. The mean is computed using
  Eq.~(\ref{mave}) for the fits indicated in Table~\ref{tablefitrange}
  and includes the shifts $\Delta X_{\rm multiplet}$. The first error is
  due to varying the fit parameterization and is calculated using
  Eq.~(\ref{deltamave}), while the second error includes the
  uncertainty associated with $\Delta X_{\rm multiplet}$. Note that for
  the negative parity channels we only give results for the operators
  for which the mass could reliably be extracted. The superscript
  $^{**}$ indentifies likely scattering states. The results for the
  first excitations may also contain such states.}
\end{center}
\end{table}

\begin{figure}[ht!]
\includegraphics{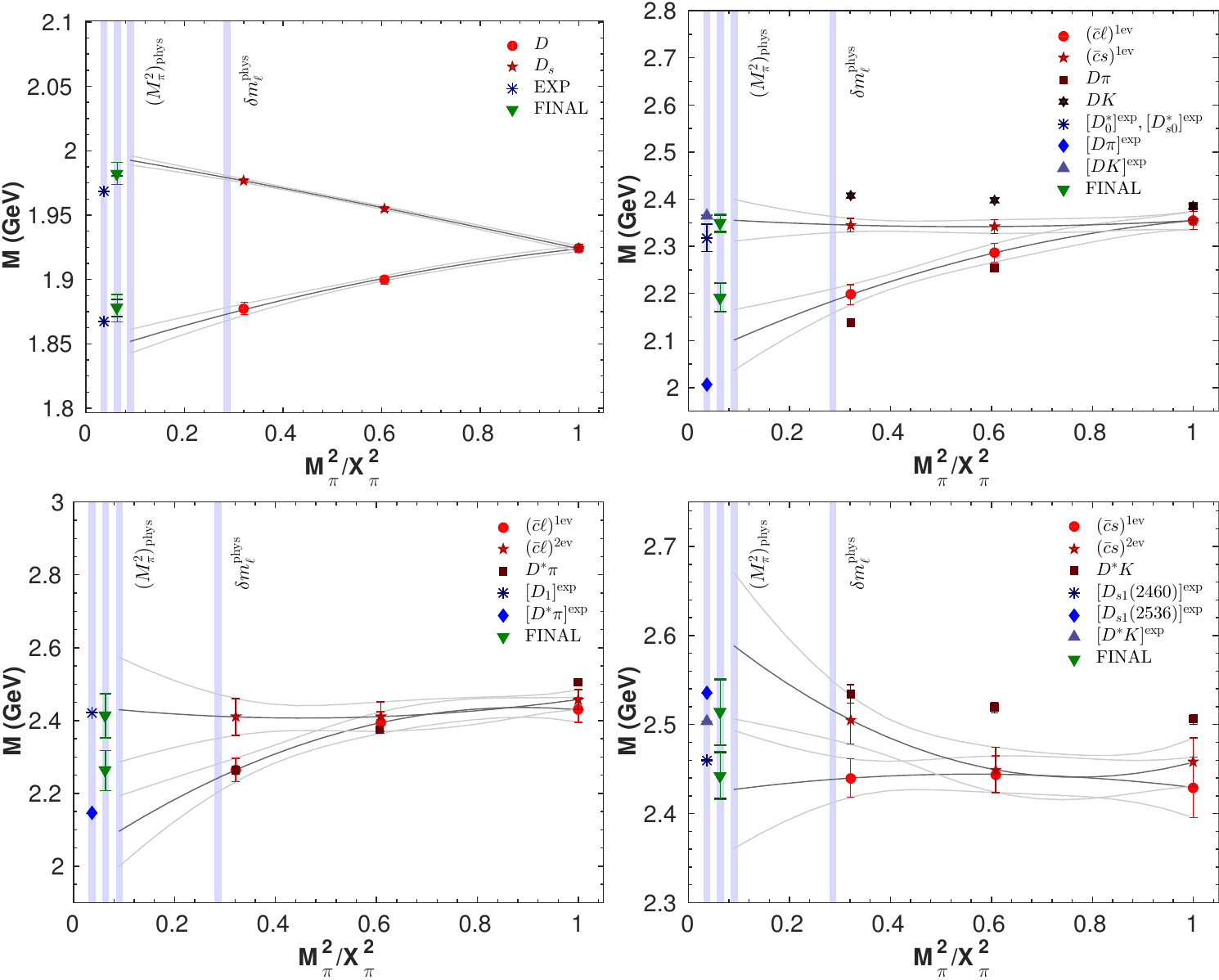}
\caption{\label{fan_mesons} Extrapolations to the physical point for
  (top left) the $D$ and $D_s$ mesons, (top right) the $D_0^*$ and
  $D_{s0}^*$ mesons, (bottom left) the lowest two eigenvalues for
  $D_{1}$ and (bottom right) the lowest two eigenvalues for $D_{s1}$.
  For simplicity one fit is displayed in each case. The full set of
  fits used to calculate the final values are given in
  Table~\ref{tablefitmesons}~(Appendix~\ref{AppC}).  These values include the
  shift $X_{\rm multiplet}$~(and corresponding error) determined from the
  spin and flavor averaged $D/D_s$ mass. For the $D^*_0$, $D_1$ and
  $D_{s1}$ the results for the relevant non-interacting two particle
  states are also shown.}
\end{figure}

Finally, we present the extrapolations for the mesons. The charmonium
channels are fitted individually as for the other flavor singlets
discussed above (to a constant and a constant plus quadratic term). 
Figure~\ref{fan_mesons} displays the extrapolations
for the heavy-light mesons,
Table~\ref{tablefitmesons}~(Appendix~\ref{AppC}) details the fit
functions used.  Since the heavy-light mesons form triplets, the
expressions Eqs.~(\ref{tripleta})  and (\ref{tripletb}) apply.
While the lowest lying $0^-$ and $1^-$ states exhibit
GMO dependence on $\delta m_{\ell}$, this does not hold for the higher
states. In particular, in the $0^+$ channel, the lowest eigenvalue for
the $\bar{c}\ell$ interpolator decreases rapidly towards $\delta
m_{\ell}^{ \rm phys}$ and is very close in value to the corresponding
non-interacting $D\pi$ state on each ensemble. In contrast, replacing
the light quark with a strange quark, the $D^*_{s0}$ is below $DK$ and
is consistent with experiment at the physical point. Unfortunately, we
were not able to reliably extract higher eigenvalues in these
channels, nor did expanding the interpolator basis to include
$\bar{c}\gamma_4 q$ prove successful. Further work, including interpolators
with derivatives will be performed in the future. A similar
picture emerges for the $1^+$, however, here we are able to resolve
closely lying levels by using two interpolators~(see
Table~\ref{meson_op}). For the $D$ meson case, the lowest eigenvalue
coincides with $D^*\pi$, while the next level is much less sensitive
to the light quark mass and is compatible with experiment at the
physical point, leading us to associate this level with the
$D_1$. Switching to the $D_s$ sector, the lowest  level is below the
$D^*K$ threshold and is consistent with the $D_{s1}(2460)$.
The next level is both consistent with $D^* K$ and with $D_{s1}(2536)$. 
A careful finite volume study is needed to
clarify this issue.

\subsection{\label{extrapdiff} Mass differences}

\begin{figure}[ht!]
\includegraphics{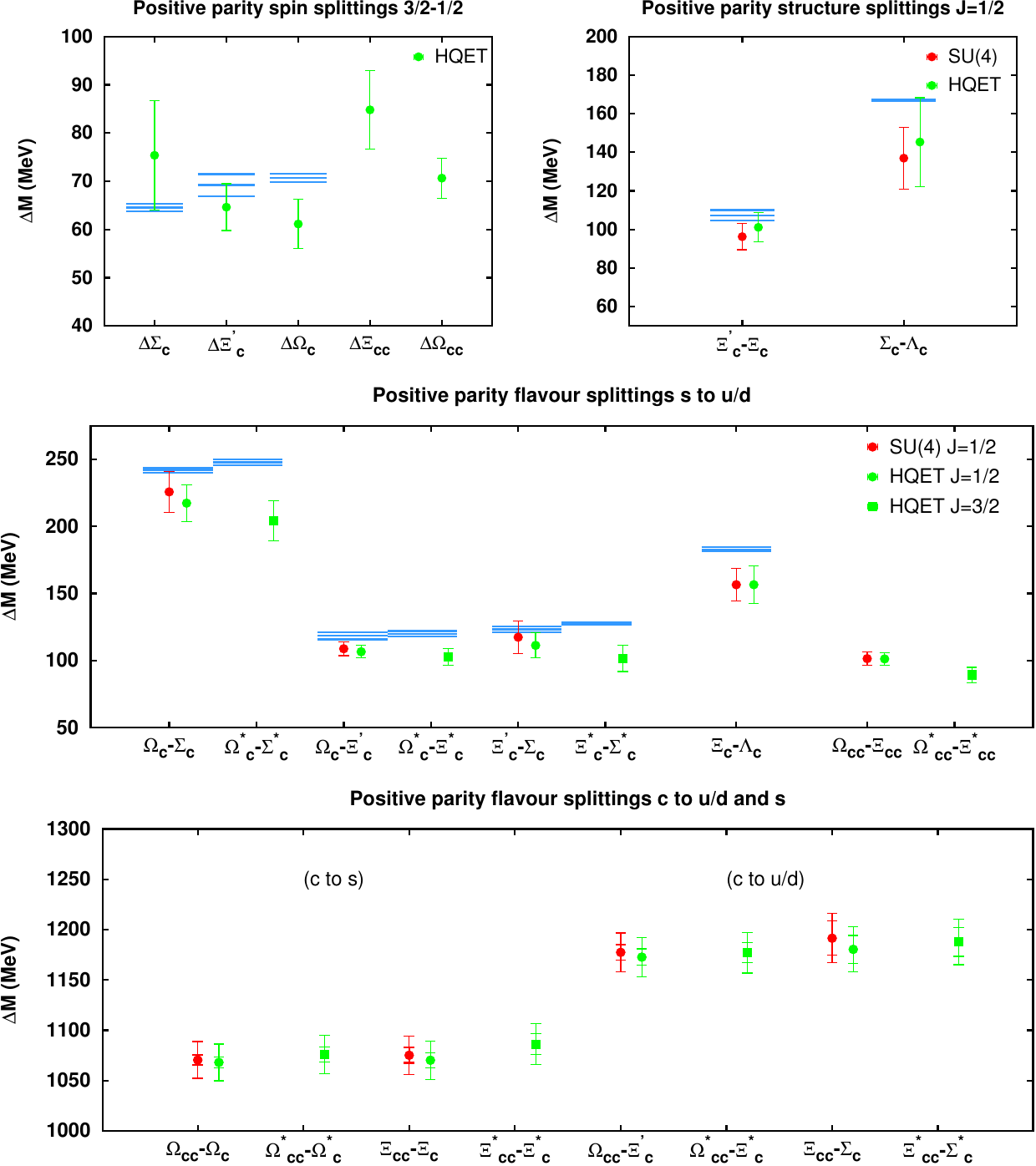}  
\caption{\label{lighttoc} The spin and flavor mass splittings for the
  positive parity charmed baryons on the $V=32^3\times 64$ ensembles.
  For the splittings where the charm quark is replaced by a strange or
  light quark~(bottom plot) two errors are shown indicating the
  uncertainty with and without the error on $\Delta X_{\rm
    multiplet}$; this shift cancels in the other splittings. The
  experimental values are indicated, where we take the $J^P$
  assignment suggested by the PDG in the cases where the quantum
  numbers have not been identified. The experimental errors are shown
  as bands. For those splittings for which a value is not quoted by
  the PDG, the error is calculated adding the uncertainties of
  the individual masses in quadrature. }
\end{figure}

Mass differences uncover specific aspects of the quark dynamics within
the charmed baryons. We focus on three types of positive parity ground
state mass differences, spin $(J=\frac 32)-(J=\frac 12)$,
``structure''~(between the sextet and anti-triplet multiplets) and
flavor~(within multiplets corresponding to changing $s\to u/d$ and
between multiplets for $c\to u/d$ and $c\to s$).  The splittings are
computed from the jackknifes of the individual masses and are
extrapolated to $\delta m_{\ell}^{\rm phys}$ using functional forms
derived by taking the differences of the expressions for the
masses. The parameterizations used for each channel are listed in
Table~\ref{extrap_splittings}. With the assumption that the singly
charmed multiplets all require shifts, $\Delta X_{\rm multiplet}$, of a
similar size to make contact with the physical point, only the $c\to
u/d$ and $c\to s$ flavor splittings require a correction. The final
results are displayed in Fig.~\ref{lighttoc}.  \bite
\item{ {\bf Spin splittings:} according to the predictions of
  HQET/pNRQCD, the $\frac 32-\frac12$ mass differences should be
  small, of ${\rm O}(\overline{\Lambda}^2/m_c)$, vanishing in the
  static limit. Figure~\ref{lighttoc} shows all spin splittings are
  indeed small, in the region of $60$--$85$~MeV. They are also similar
  in size, independent of the light quark content and the number of
  charm quarks. This can be understood as follows. For doubly charmed
  baryons, if the two charm quarks form a spatially small diquark
  ($r\ll\overline{\Lambda}^{-1}$), the spin difference should
  correspond to $3/4$ times the corresponding $D$ (or $D_s$) fine
  structure splitting, where a factor $3/2$ is from coupling a $s_d=1$
  or $s_d=0$ diquark, rather than a spin-$1/2$ antiquark, to the light
  quark. Another factor $1/2$ is due to the different color
  contraction within a baryon, relative to the meson color
  trace~\cite{Brambilla:2005yk}\footnote{ In pNRQCD this factor is due
    to the light quark interacting with two color charges, in HQET one
    would ascribe it to the mass of an effective $\overline Q = cc$
    antiquark.}.  Testing this HQET picture (or, equivalently, the
  short-distance pNRQCD picture), we have $\frac 34 (M_{D^*}-M_D) =
  95(8)$~MeV compared to $M_{\Xi_{cc}^*}-M_{\Xi_{cc}}=85(9)$~MeV and
  $\frac34 (M_{D_s^*}-M_{D_s}) = 92(2)$~MeV compared to
  $M_{\Omega_{cc}^*}-M_{\Omega_{cc}}=71(4)$~MeV. The $\Omega_{cc}$
  splitting agrees less well with this expectation, however, higher
  order corrections in HQET and (p)NRQCD of size
  $\overline{\Lambda}^3/m_c^2$ and $m_cv^6$, respectively, are not
  necessarily small at the charm quark mass. Brown et
  al.~\cite{Brown:2014ena}, making a similar comparison~(using the
  experimental meson splittings), find that agreement improves for
  bottom quarks. We remark that singly charmed baryons, where the two
  light quarks form a diquark, share the same spin- and color-factors
  of doubly charmed baryons ($\frac32\cdot\frac12=\frac34$), relative
  to charmed mesons. Therefore, also the $\Sigma_c$, $\Xi_c$ and
  $\Omega_c$ fine structure splittings are of similar sizes, see
  Fig.~\ref{lighttoc}.  Finally, for $r>\overline{\Lambda}^{-1}$,
  doubly charmed baryons become similar to charmonia. Again, the spin
  and color factors are the same but in this case one should compare
  to the charmonium fine structure,
  $\frac34(M_{J/\psi}-M_{\eta_c})=76(1)$~MeV. The situation is now
  reversed with the $\Omega_{cc}$ splitting being in closer
  agreement. Therefore, based on the fine structure, it is not
  possible to cleanly discriminate between the two cases (HQET or
  charmonium-like).  As discussed in Section~\ref{simdetails}, from
  the meson sector we expect discretization effects 
      in spin splittings of about $10$--$20$~MeV, while the final
  errors in the figure are around $5$--$10$~MeV. Nevertheless the
  sextet differences are reasonably consistent with experiment.}

\item{{\bf Flavor structure splittings}: these are differences
  between baryons with the same $J^P$ and quark content but different
  flavor structure. The $\Lambda_c$~($I=0$) and $\Sigma_c$~($I=1$)
  lie in the anti-triplet and sextet multiplets, respectively, and
  differ in isospin. Similarly,  the $\Xi^\prime_c$ and the $\Xi_c$ share the 
  same valence quark content but differ in terms of their SU(3) wavefunctions.
  Nature respects isospin symmetry reasonably well
  but not so SU(3); the physical $\Xi^\prime_c$ and $\Xi_c$ states  may be mixtures of
  the sextet and the anti-triplet states. While we have not studied this mixing we note
  that Brown et al.~\cite{Brown:2014ena} found that such effects are
  not significant.  In Fig.~\ref{lighttoc} one can see that the
  splittings are compatible with experiment.
}

\item{{\bf Flavor splittings}: in the middle panel of Fig.~\ref{lighttoc} 
  we show differences between masses of particles where one or two  strange 
  quarks are replaced by  light quarks. The mass differences for $ss \to \ell\ell$ 
  are roughly twice as large as those for $s\to \ell$. Agreement is found 
  with experiment for differences involving $J=\frac 12$ 
  singly charmed baryons, however, our results
  seem to be systematically~(slightly) lower for $J=\frac 32$.
  In the last panel of
  Fig.~\ref{lighttoc} we show differences between masses of doubly
  charmed baryons and their singly charmed counterparts, replacing
  either one charm quark by a strange quark or by an up/down
  quark. Clearly, $M_{\Omega_{cc}^*}-M_{\Omega_c^*} \approx
  M_{\Omega_{cc}}-M_{\Omega_c}$ since
  $M_{\Omega_{cc}^*}-M_{\Omega_{cc}}\approx
  M_{\Omega_c^*}-M_{\Omega_c}$ etc., as already observed
  above under ``Spin splittings". However, also $M_{\Omega_{cc}}-M_{\Omega_c} \approx
  M_{\Xi_{cc}}-M_{\Xi_c}$ and $M_{\Omega_{cc}}-M_{\Xi'_c} \approx
  M_{\Xi_{cc}}-M_{\Sigma_c}$, suggesting the dynamics of doubly
  charmed baryons to be closely linked to those of singly charmed
  baryons (and charmed mesons); in the first case the two heavy quarks
  seem to form a diquark (meaning $r\ll\overline{\Lambda}^{-1}$) while
  in the second case either the light quarks form a diquark,
  interacting with the remaining charm quark or one light quark
  interacts with a heavy-light core.  This would explain why the two
  systems resemble each other closely, up to constant shifts, due to
  differences of the respective (constituent) quark masses. Note that
  replacing the remaining charm quark by a light quark flavor will
  not result in constant energy shifts. For example, the fine
  structure of the light and heavy baryons are very
  different.   } \eite

\subsection{Comparison with previous results}
\label{compothers}

\begin{center}
\begin{figure}[h]
\includegraphics{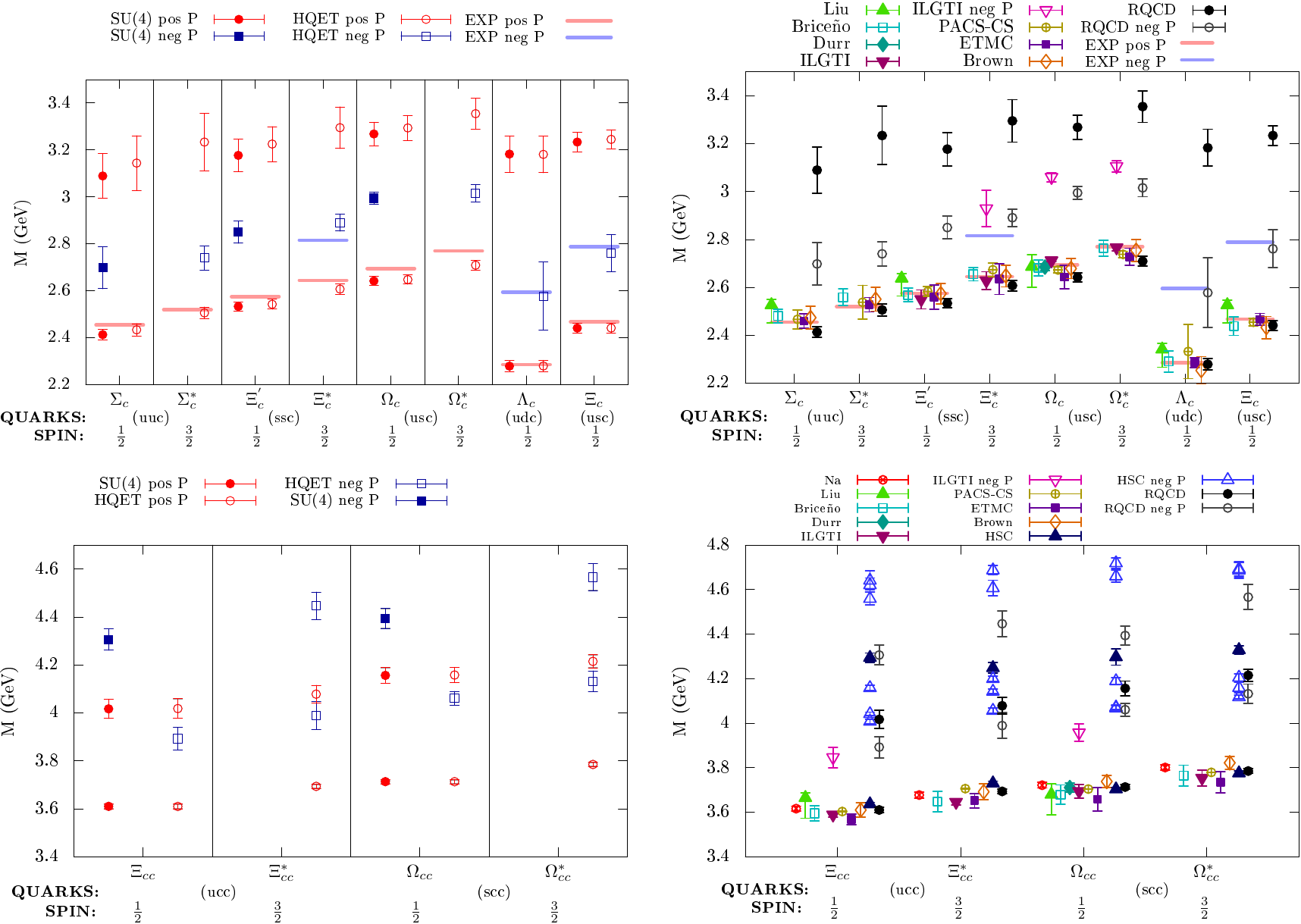}
\caption{\label{summary} \footnotesize{ Singly charmed (top) and
    doubly charmed (bottom) low lying baryon spectra. (Left) our
    results~(RQCD) from the $V = 32^3\times 64$ ensembles. (Right) a
    comparison with previous determinations from Na et
    al.~\cite{Na:2007pv}, Liu et al.~\cite{Liu:2009jc}, Brice\~no et
    al.~\cite{Briceno:2012wt}, D\"urr et al.~\cite{Durr:2012dw}, 
    ILGTI~\cite{Basak:2012py}, PACS-CS
    collaboration~\cite{Namekawa:2013vu},
    ETMC~\cite{Alexandrou:2014sha}, Brown et al.~\cite{Brown:2014ena}
    and HSC~\cite{Padmanath:2015jea}. Note that our results for the
    negative parity, spin-1/2 $\Omega_c$ and $\Xi_c^\prime$ are likely
    scattering states, see Section~\ref{varmethod}. The first 
    excitations may also contain such states.}}
\vspace{-1.2em}
    \end{figure}
\end{center}

Our final results for the individual masses are summarized
in~\refig{summary}, where for the negative parity channels only the
results for interpolators which gave a reliable signal are shown.
Overall, for the singly charmed baryons there is reasonable agreement
with experiment, as  also seen for the mesons with open and hidden charm,
which  is encouraging in terms of the size of
the remaining systematics. In particular, in some previous
calculations $M_{cqq}-\frac 12 M_{\eta_c}$ or $M_{ccq} - M_{\eta_c}$ 
splittings were computed, instead of the
individual baryon masses, in order to remove the dependence on the
charm quark mass and to reduce the size of discretization effects and
uncertainties arising from tuning the charm quark mass using the meson
sector. This has not been done here. However, states involving strange
quarks are systematically lower than experiment, in particular, the  $\Omega_c$. This
may be due to residual effects related to simulating along an
incorrect $\overline{m}={\rm constant}$ line. Furthermore, our finite
volume study suggests that the positive parity ground state singly
charmed baryons and excited state doubly charmed baryons will decrease
by a few tens of MeV in the infinite volume limit.

There have been a number of recent lattice studies of the charmed
baryon spectra, also displayed in Fig.~\ref{summary}, most of which
are restricted to positive parity ground states. Different systematics
apply in each case depending on,
\bite
\item the number of sea quarks included:
$N_f=2$~(D\"urr et al.~\cite{Durr:2012dw}), $N_f=2+1$~(Na et
al.~\cite{Na:2007pv}, Liu et al.~\cite{Liu:2009jc},
PACS-CS~\cite{Namekawa:2013vu}, Brown et al.~\cite{Brown:2014ena},
HSC~\cite{Padmanath:2015jea}) and $N_f=2+1+1$~(Brice\~no et
al.~\cite{Briceno:2012wt}, ILGTI~\cite{Basak:2012py},
ETMC~\cite{Alexandrou:2014sha}).
\item The charm quark action: the Fermilab
  action~\cite{ElKhadra:1996mp}~(Na et al., Liu et al.), a relativistic
  heavy quark action~\cite{Aoki:2001ra,Christ:2006us}~(Brice\~no et al., PACS-CS, Brown et al.), twisted
  mass~\cite{Frezzotti:2000nk}~(ETMC), anisotropic clover~\cite{Liu:2012ze}~(HSC), the Brillouin
  action~\cite{Durr:2012dw}~(D\"urr et al.) and the overlap action~\cite{Neuberger:1997fp}~(ILGTI).
\item The light valence and sea quark action: MILC Asqtad staggered
  sea~\cite{Bernard:2001av} and valence~(Na et al.), MILC Asqtad staggered 
  sea~\cite{Bernard:2001av} with domain
  wall valence~(Liu et al.), MILC HISQ sea~\cite{Bazavov:2010ru} 
  with tadpole-improved clover
  valence~(Brice\~no et al.), QCDSF non-perturbative 
  clover sea~\cite{Bietenholz:2010az} with
  Brillouin valence~(D\"urr et al.), MILC HISQ sea~\cite{Bazavov:2010ru}
   with overlap
  valence~(ILGTI), non-perturbative clover sea~\cite{Aoki:2009ix}
  and valence~(PACS-CS), twisted
  mass sea~\cite{Alexandrou:2014sha} and valence~(ETMC), RBC/UKQCD domain 
  wall sea~\cite{Aoki:2010dy} and
  valence~(Brown et al.) and anisotropic tadpole-improved
  clover sea~\cite{Padmanath:2015jea} and valence~(HSC).  
  \eite Systematics arise from finite lattice spacings,
  unphysical quark masses, finite volumes, excited state contamination
  and, among mixed action approaches, violations of unitarity. Notable
  are the Brice\~no et al.~\cite{Briceno:2012wt}, ETMC~\cite{Alexandrou:2014sha} 
  and Brown et al.~\cite{Brown:2014ena}  studies, involving
  both continuum and chiral limit extrapolations, the other works are
  predominately at a single lattice spacing.  Despite such varied
  approaches, there is general agreement for the spectra. In
  particular, the lattice results are approximately $80$~MeV above the
  SELEX measurement of
  $M_{\Xi_{cc}}=3518.7(1.7)$~\cite{Ocherashvili:2004hi}. The recent result
  of Borsanyi et al.,
  $M_{\Xi_{cc}^{++}}-M_{\Xi_{cc}^+}=2.16(11)(17)$~MeV~\cite{Borsanyi:2014jba}
  also contradicts the SELEX value of approximately $60$~MeV for this
  isospin splitting~\cite{Russ:2002bw}.

Of particular interest for our work are the determinations of ground
state negative parity particles by ILGTI~\cite{Basak:2012py} and the
ground state and excitations for both parities for the doubly charmed
spectrum by HSC~\cite{Padmanath:2015jea}. ILGTI use pion masses down
to 240~MeV and two lattice spacings, $a=0.06$~fm and $0.09$~fm, both
with a spatial extent $L=2.9$~fm while HSC employ a temporal lattice
spacing of $a_t\sim 0.035$~fm on anisotropic lattices with
$L=1.9\,{\rm fm}$ and a single pion mass, $M_\pi=391$~MeV. Considering
the different systematics involved in each study, Fig.~\ref{summary}
shows that the determinations for the ground state negative parity
channels are reasonably consistent, including those which we identify
as scattering states.  The picture is less satisfying comparing with
the HSC predictions for excited doubly charmed baryons.  Our positive
parity first excited states are significantly below those of
HSC\footnote{For improved visibility we have omitted the HSC results
  for positive parity higher excitations from Fig.~\ref{summary}.},
even for the $\Omega^{(*)}_{cc}$. This may be due to finite volume
effects and/or the lack of chiral extrapolation in the HSC
study. Certainly the discrepancy is smaller for the
$\Omega^{(*)}_{cc}$ compared to the $\Xi^{(*)}_{cc}$ and, as discussed
in Section~\ref{finitesize}, we expect finite volume effects to
increase the mass for smaller physical volumes. For the negative
parity states, HSC predict several levels~(as expected from
non-relativistic quark models with SU(6)$\times$O(3) symmetry, see
Ref.~\cite{Padmanath:2015jea} for a discussion) in the range
$4.0-4.8$~GeV. With only one interpolator~(but multiple smearings) we
are not able to resolve closely lying states and our excited states
are above the first three HSC levels. With our limited basis the first
excitation could also have an overlap with a state corresponding to
higher continuum spin. This would correspond to $J=\frac 72$ for
spin-1/2 interpolators and $J=\frac 52$ for spin-3/2. A more extensive
analysis is needed in order to make a closer comparison.

\section{\label{6}Conclusions}

In this work we have presented results for the ground states and first
excitations of singly and doubly charmed baryons. Through the use of
the variational method with a basis of three differently smeared
interpolators, reliable signals were obtained for both positive and
negative parities, where for the latter the first excitation was only
extracted for doubly charmed states. For spin-1/2 channels we
implemented both HQET and SU(4) inspired interpolators and found
consistent results for positive parities, while for negative parities
SU(4) interpolators had a much better overlap with the ground state
for sextet baryons~($\Sigma_c$, $\Xi_c^\prime$, $\Omega_c$) and HQET
interpolators for the anti-triplet~($\Lambda_c$, $\Xi_c$) and
triplet~($\Xi_{cc}$, $\Omega_{cc}$). Overall agreement was obtained
with experiment, where available, suggesting the remaining systematics
are not large. This is supported by the results for the lower lying
charmonium, $D$ and $D_s$ mesons, which also reproduced the gross
experimental spectra. Discretization effects can be significant in a
simulation at a single lattice spacing, for which $am_c\sim
0.4$.  Fine
structure is particularly sensitive to such effects, and in the meson
sector spin splittings were underestimated by $10$--$20$~MeV. However,
for the singly charmed baryons the spin splittings are consistent with
experiment within statistical uncertainties ranging from $5$ to
$11$~MeV.  Our study of finite volume effects suggests the
lightest members of the ground state singly charmed multiplets and
first excitations of the doubly charmed triplet, in the infinite
volume limit, could be around $1\sigma$ lower~(in terms of the final
quoted error).

The negative parity channels and first excitations of both parities
require careful study due to a number of decay channels with
thresholds close to our mass estimates. By looking for degeneracy of
the masses with the threshold values on each ensemble, over the range
of pion masses $M_\pi=259-461$~MeV, we identified the ground states in
the $\Omega_c$ and $\Xi^\prime_{c}$ negative parity channels as
scattering states. Interestingly, for the singly charmed anti-triplet
the HQET interpolators had a good overlap with the level consistent
with a bound state, while the SU(4) interpolators gave results
consistent with the corresponding thresholds. We also encountered
scattering states in the meson spectra, where, by including additional
interpolators in the variational basis, results consistent with experiment
for the $D_1$ and $D_{s1}$ were obtained. Cross-correlation
functions between the SU(4) and HQET interpolators would be needed to
repeat this analysis for the baryons.

One of the main aims of our study was to investigate the light flavor
structure of charmed baryons: employing ensembles with the average light quark mass fixed
enables expansions in the flavor symmetry breaking quark mass
difference, starting from a point where the lattice results are most
precise. The corresponding Gell-Mann--Okubo relations worked well for
almost all multiplets and the extrapolations were well constrained,
with the linear terms in $\delta m_{\ell}$ within each multiplet being
fixed by one coefficient. In some cases the quadratic terms were
statistically significant, however, the magnitudes of these contributions were still
small.  A notable exception was the singly charmed anti-triplet 
$(\Lambda_c, \Xi_c)$, which
did not fit the GMO pattern. This may indicate a more complex internal
structure. However, one can still perform a phenomenological fit using
an expansion in $\delta m_{\ell}$, where the linear coefficients are
not constrained. The errors on our final values were estimated
conservatively, including  both linear and quadratic
fits. An additional uncertainty is present in our results due to the
simulation trajectory missing the physical point in the mass
plane. This effect was corrected for in a consistent way and the
agreement found with experiment for extrapolated flavor singlet mass
combinations suggests the correction is small. However, we note that
our results for heavy baryons involving strange quarks are
consistently lower than experiment which may indicate residual
effects.

The dynamics of quarks within heavy baryons is interesting in terms of
possible similarities with the heavy-heavy or heavy-light mesons and
the applicability of various effective theories. In particular,
for doubly charmed baryons, there are two relevant limits, in terms of
a point-like heavy-heavy diquark~(the HQET or short distance pNRQCD
picture) or a more diffuse heavy-light diquark~(charmonium
picture). Flavor splittings where the charm quark is replaced by a
strange or up/down quark suggest the dynamics of doubly charmed
baryons are closely linked to those of singly charmed ones.

In the future we plan to perform a more extensive study on CLS
configurations~\cite{Bruno:2014jqa}, with a larger range of pion
masses, all with $LM_\pi\gtrsim 4$. This will enable a further
investigation of the magnitude of flavor symmetry breaking effects.
Initial studies show the physical trajectory has been reproduced
reasonably well~\cite{Soldner:2015oea}.

\subsection*{Acknowledgments}
We thank Issaku Kanamori, Nilmani Mathur and Johannes Najjar for
discussions.  The numerical calculations were performed on the
BlueGeneQ (FERMI) at CINECA as part of the PRACE project 2012071240
and the iDataCool cluster at Regensburg University.  The Chroma
software package \cite{Edwards:2004sx} was used extensively in the
analysis.  We thank our collaborators within QCDSF who generated the
$N_f=2+1$ ensembles analyzed here.  We also thank the International
Lattice DataGrid. This work was supported by the DFG (SFB/TRR 55)
and a Research Linkage Grant of the Alexander von Humboldt 
Foundation.
  
\appendix
\newpage

\section{\label{Appmesons} Meson effective masses}

\begin{figure}[ht!]
\includegraphics{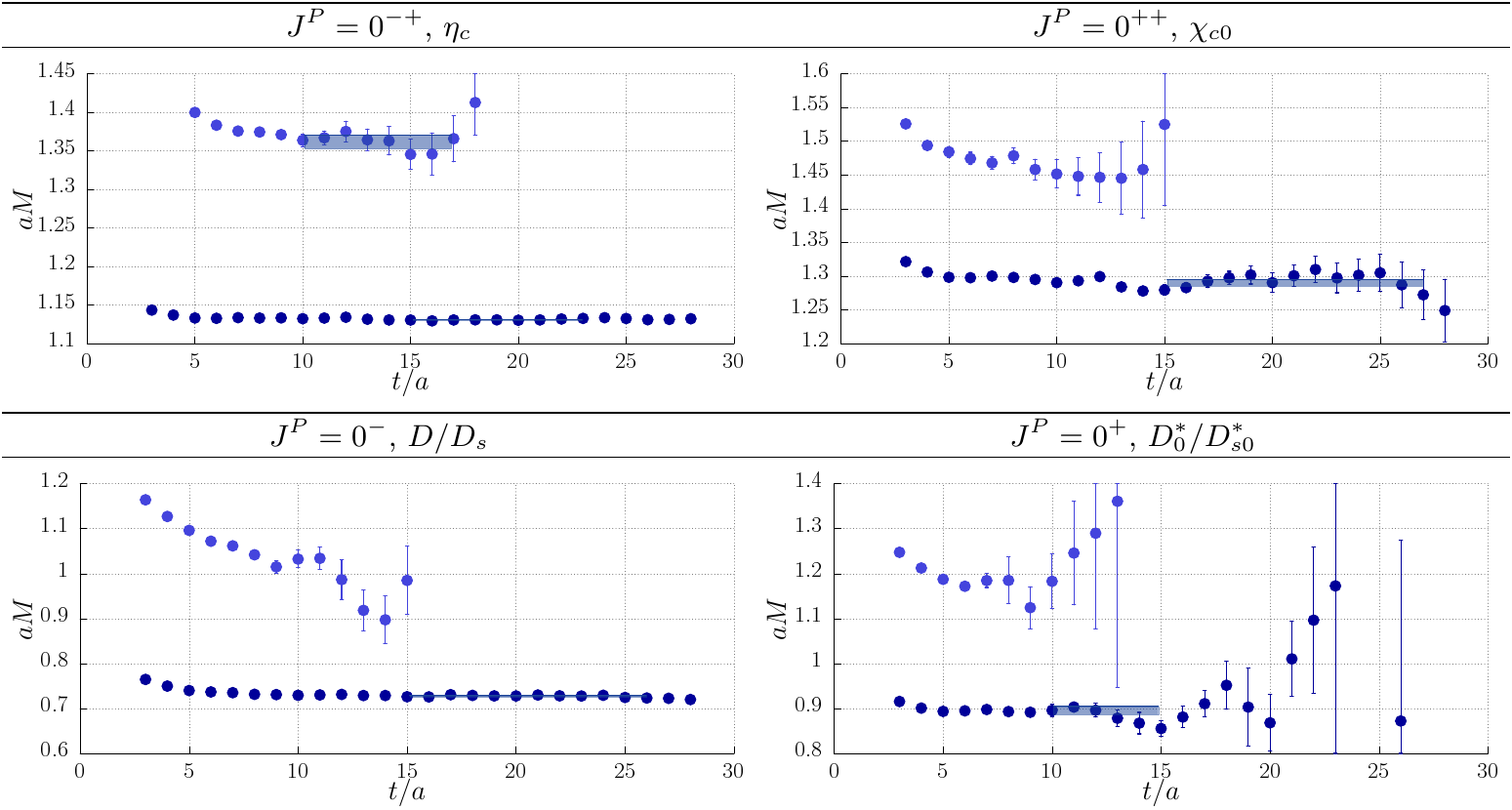}
\caption{\label{sample_mesons} Sample of effective masses of the
  charmonium and $D/D_s$ mesons for the symmetric ensemble with $V=32^3\times 64$. The filled
  regions indicate the fitting ranges chosen and the fit result
  including errors. }
\end{figure}

\begin{figure}[ht!]
\begin{center}
\includegraphics{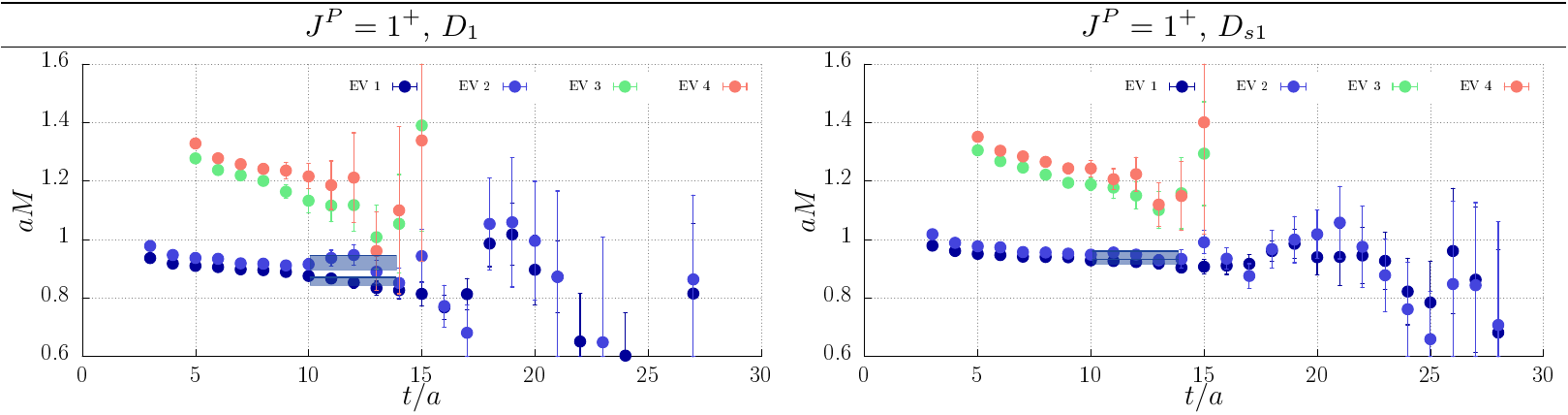}
\caption{\label{mesons_a1_b1} As in Fig.~\ref{sample_mesons} for the
 effective masses of the lowest four  eigenvalues~(EV) for the $1^+$
  channels on the $M_\pi=259$~MeV ensemble with $V=32^3\times 64$.
}
\end{center}
\end{figure}

Effective masses of the lowest two eigenvalues
 for a number of channels for charmonium and
$D/D_s$ are shown in Fig.~\ref{sample_mesons}
for the larger symmetric ensemble. The effective masses for
the lowest four eigenvalues for $D_1$ and $D_{s1}$ on the
$M_{\pi}=259$~MeV ensemble are displayed in Fig.~\ref{mesons_a1_b1}.

\vspace{-2mm}

\newpage

\section{\label{AppD} Finite volume effects}

The finite volume effects computed using the matched $V=24^3\times 48$
and $32^3\times 64$ ensembles are given in Tables~\ref{finitevols_pos}
and ~\ref{finitevols_neg} for positive and negative parity particles,
respectively.

\begin{table}[h]
\begin{tabular}{ccc|cc|cc|cc|cc}\hline\hline
Particle & $P$  && \multicolumn{4}{c}{First eigenvalue} & \multicolumn{4}{c}{Second eigenvalue}\\\hline
&         &       & \multicolumn{2}{|c}{ SU(4)} & \multicolumn{2}{|c}{ HQET}& \multicolumn{2}{|c}{ SU(4)} & \multicolumn{2}{|c}{ HQET}\\
&          &       & $\Delta M$~(MeV) &  $\frac{\Delta M}{M_{32}}$~(\%) & $\Delta M$~(MeV) &  $\frac{\Delta M}{M_{32}}$~(\%)& $\Delta M$~(MeV) &  $\frac{\Delta M}{M_{32}}$~(\%) & $\Delta M$~(MeV) &  $\frac{\Delta M}{M_{32}}$~(\%)\\\hline
$\Sigma_c$, $\Xi^\prime_c$, $\Omega_c$  &   +  & $\kappa_{\rm sym}$ & 33(10)  &  1.3 &  24(11)  &  0.9 &  -63(89)  &  -1.9 &  -59(81)  &  -1.8  \\\hline
$\Sigma_c$  &   +  & \multirow{3}{*}{$\kappa_{\rm asym}$} & 31(15)  &  1.2 &  15(16)  &  0.6 &  69(134)  &  2.1 &  -4(139)  &  -0.1\\
$\Xi^\prime_c$  &   +   & &  18(11)  &  0.7 &  16(11)  &  0.6 &  50(102)  &  1.5 &  -19(96)  &  -0.6\\
$\Omega_c$  &   +  & & 10(8)  &  0.4 &  8(9)  &  0.3 &  31(84)  &  1.0 &  -22(77)  &  -0.7\\\hline
$\Sigma^*_c$, $\Xi^*_c$, $\Omega^*_c$  &   + & $\kappa_{\rm sym}$ && &  39(12)  &  1.5 & &  &  -139(90)  &  -4.0 \\\hline
$\Sigma^*_c$  &   + &\multirow{3}{*}{$\kappa_{\rm asym}$}  && &  27(17)  &  1.0 &&&  129(164)  &  3.8 \\
$\Xi^*_c$  &   +  & & && 15(12)  &  0.6 && &  67(108)  &  2.0\\
$\Omega^*_c$  &   +  &&& &  6(10)  &  0.2 &&&  34(84)  &  1.0 \\\hline
$\Lambda_c$, $\Xi_c$  &  + &  $\kappa_{\rm sym}$ & 32(8)  &  1.3 &  33(8)  &  1.4 &  -229(98)  &  -7.0 &  -266(98)  &  -8.1 \\\hline
$\Lambda_c$  &  + & \multirow{2}{*}{$\kappa_{\rm asym}$} &  19(12)  &  0.8 &  20(13)  &  0.9 &  -139(187)  &  -4.4  &  -150(201)  &  -4.7\\
$\Xi_c$  &   +  & &  11(9)  &  0.5 &  12(9)  &  0.5 &  -63(121)  &  -1.9 & -72(126)  &  -2.2 \\\hline
$\Omega_{cc}$, $\Xi_{cc}$  &    + & $\kappa_{\rm sym}$ &  4(6)  &  0.1 &  3(6)  &  0.1 &  76(24)  &  1.9 &  79(24)  &  1.9\\\hline
$\Xi_{cc}$  &    +  & \multirow{2}{*}{$\kappa_{\rm asym}$} &  4(8)  &  0.1 &  6(7)  &  0.2 &  63(29)  &  1.5 &  56(29)  &  1.4\\
$\Omega_{cc}$  & +  &  & -1(6)  &  -0.0 &  0(5)  &  0.0 &  43(27)  &  1.0 &  39(27)  &  0.9\\\hline
$\Omega^*_{cc}$, $\Xi^*_{cc}$  &    +  & $\kappa_{\rm sym}$ & &&  9(7)  &  0.3 && &  87(26)  &  2.1 \\\hline
$\Xi^*_{cc}$  &   +  & \multirow{2}{*}{$\kappa_{\rm asym}$} & && 1(9)  & 0.0  && &  69(30)  &  1.7\\
$\Omega^*_{cc}$  &   +  & &&  &  -5(7)  &  -0.1 && &  46(27)  &  1.1 \\
\hline
\hline
\end{tabular}
\caption{\label{finitevols_pos} Mass differences, $\Delta M =
  M_{24}-M_{32}$, between the positive parity masses extracted from
  the $V=24^3\times 48$ and the $32^3\times 64$ ensembles,
  where the quark mass parameters~($\kappa$) are the same:
  the symmetric point $\kappa_{\rm sym}=0.1209$ and the asymmetric
  combination $\kappa_{\rm asym}=(0.12104,0.12062)$. The errors on the
  differences are computed by adding the individual uncertainties in quadrature. }
\end{table}

\begin{table}
\begin{tabular}{ccc|c|cc}\hline\hline
Particle & $P$  && Operator & $\Delta M$~(MeV) &  $\frac{\Delta M}{M_{32}}$~(\%) \\\hline
\multicolumn{6}{c}{First eigenvalue}\\\hline
$\Sigma_c$, $\Xi^\prime_c$, $\Omega_c$  &   -  & $\kappa_{\rm sym}$ & SU(4) &  -36(37)  & -1.2   \\\hline
$\Sigma_c$  &   -   & \multirow{3}{*}{$\kappa_{\rm asym}$} &\multirow{3}{*}{SU(4)} & -49(40)  &  -1.7   \\
$\Xi^\prime_c$  &   -   & & &  -37(31)  &  -1.3   \\
$\Omega_c$  &   -   &  & & -22(24)  &  -0.7  \\\hline
$\Lambda_c$, $\Xi_c$  &  -  &  $\kappa_{\rm sym}$ & HQET &   -4(40)   & -0.1   \\\hline
$\Lambda_c$  &  -  & \multirow{2}{*}{$\kappa_{\rm asym}$} & \multirow{2}{*}{HQET} &   122(48)  &  4.7 \\
$\Xi_c$  &   -   & &  &    63(36)  &  2.3 \\\hline
$\Omega_{cc}$, $\Xi_{cc}$  &    -  & $\kappa_{\rm sym}$ & HQET  &   9(23)  &  0.2 \\\hline
$\Xi_{cc}$  &    -   & \multirow{2}{*}{$\kappa_{\rm asym}$} & \multirow{2}{*}{HQET} &  -2(23)  &  -0.1 \\
$\Omega_{cc}$  & -   &  & &    -8(17)  &  -0.2 \\\hline
$\Omega^*_{cc}$, $\Xi^*_{cc}$  &    -  & $\kappa_{\rm sym}$ & HQET  & -19(23)  &  -0.5  \\\hline
$\Xi^*_{cc}$  &   -  & \multirow{2}{*}{$\kappa_{\rm asym}$} &\multirow{2}{*}{HQET} & -39(22)  &  -1.0 \\
$\Omega^*_{cc}$  &   -  & &  &  -31(16)  &  -0.8  \\\hline
\multicolumn{6}{c}{Second eigenvalue}\\\hline
$\Omega_{cc}$, $\Xi_{cc}$  &    -  & $\kappa_{\rm sym}$ & SU(4)  &  22(47)  &  0.5 \\\hline
$\Xi_{cc}$  &    -   & \multirow{2}{*}{$\kappa_{\rm asym}$} & \multirow{2}{*}{SU(4)} &  43(46)  &  1.0 \\
$\Omega_{cc}$  & -   &  & &    32(39)  &  0.7 \\\hline
$\Omega^*_{cc}$, $\Xi^*_{cc}$  &    -  & $\kappa_{\rm sym}$ & HQET  & -67(39)  &  -1.5  \\\hline
$\Xi^*_{cc}$  &   -  & \multirow{2}{*}{$\kappa_{\rm asym}$} &\multirow{2}{*}{HQET} & 79(52)  &  1.8 \\
$\Omega^*_{cc}$  &   -  & &  &  21(36)  &  0.5 \\
\hline
\hline
\end{tabular}
\caption{\label{finitevols_neg} As in Table~\ref{finitevols_pos} for
  the negative parity states.  }
\end{table}
\vspace{-1em}
\section{\label{AppB} Tensorial notation for SU(3) representations}
For convenience we review the tensor notation for SU(3) representations.
Consider a general SU(3) tensor $T^{a_1\dots a_n}_{b_1\dots b_m}$  
with $n$ contravariant indices and $m$ covariant indices~(i.e. rank $(n,m)$),
which transforms according to
\be
T^{a_1\dots a_n}_{b_1\dots b_m} \to  U^{a_1^\prime}_{a_1}\ldots U^{a_n^\prime}_{a_n} T^{a_1\dots a_n}_{b_1\dots b_m} (U^\dagger)^{b_1}_{b_1^\prime}\ldots(U^\dagger)^{b_n}_{b_n^\prime},
\ee
where $U$ is the SU(3) transformation.
We can construct lower rank tensors starting from $T^{a_1\dots a_n}_{b_1\dots
  b_m}$ by performing the following contractions,
\bea
\delta^{b_j}_{a_i}T^{a_1\dots a_n}_{b_1\dots b_m},&\quad {\rm for }  & 
\quad 1\leq i \leq n, 1 \leq j\leq m, \quad \mathrm{rank}\,\,(n-1,m-1), \nonumber \\
\epsilon_{a_ia_jb_{m+1}}T^{a_1\dots a_n}_{b_1\dots b_m}&\quad {\rm for }  & 
\quad 1 \leq i,j \leq n, \quad \mathrm{rank}\,\,(n-2,m+1),\\
\epsilon^{b_ib_ja_{n+1}}T^{a_1\dots a_n}_{b_1\dots b_m}&\quad {\rm for }  & 
\quad 1 \leq i,j \leq m, \quad \mathrm{rank}\,\,(n+1,m-2), \nonumber 
\eea
where $\delta^{b}_{a}$, $\epsilon_{abc}$ and $\epsilon^{abc}$ are
SU(3) invariant.  
Irreducible tensors cannot be expressed in terms of tensors of lower
rank, and hence, one obtains zero from any of the above
contractions. This means that the irreducible tensors are pairwise
symmetric in both the covariant and contravariant indices and
traceless on any index.  The lower dimensional irreducible tensors are
listed in \ret{irred}, with the dimension $D(m,n) = \tfrac 12
(n+1)(m+1)(n+m+2)$.  We are interested  in decomposing the product of two
irreducible representations. The tensor product is expressed in terms of the
irreducible tensors and $\delta^{b}_{a}$, $\epsilon_{abc}$ or $\epsilon^{abc}$.
For example, 
\begin{equation}
{\bf 3 \otimes \overline{3}= 1 \oplus
  8}, \quad q^a q_b = \frac 13 \delta^a_b S + O^a_b.
\end{equation}
Similarly, tensors of irreducible  representations can be expressed
in terms of the components of the fundamental representation,
\begin{equation}
S = q^cq_c \,\,\mathrm{and}\,\, \quad O^a_b = q^aq_b - \frac 13 (q^c q_c).
\end{equation}

\begin{table}
\begin{center}
\begin{tabular}{c|c|c}\hline\hline
Tensor       & rank & Representation \\\hline
$S$          & $(0,0)$ & {$\bf 1$} \\
$q^a$        & $(1,0)$ & {$\bf 3$} \\
$q_a$        & $(0,1)$ & {$\bf \overline 3$} \\
$O^a_b$      & $(1,1)$ & {$\bf 8$} \\
$S^{ab}$     & $(2,0)$ & {$\bf 6$} \\
$S_{ab}$     & $(0,2)$ & {$\bf \overline 6$} \\
$D^{abc}$    & $(3,0)$ & {$\bf 10$} \\
$D_{abc}$    & $(0,3)$ & {$\bf \overline{10}$} \\
$T_{ab}^{cd}$& $(2,2)$ & {$\bf 27$} \\\hline\hline
\end{tabular}
\caption{Tensors corresponding to the lower dimensional irreducible representations of SU(3).}
\label{irred}
\end{center}

\end{table}

\section{Gell-Mann--Okubo relations for charmed baryons}
\label{derive_gmo}
Below, we derive the expressions for the mass dependence of each
state within a multiplet up to first order in the flavor symmetry
violating parameter $\delta m_{\ell}$. In Eq.~(\ref{gmo_pert}) we have
matrix elements of tensor operators between tensor states
\begin{equation}
 \langle B_R | H | B_R\rangle, 
\end{equation}
where the baryons $B$ and the Hamiltonian $H$ are in the irreducible
representations $R$ and $T_H$, respectively. Of interest are $R={\bf
6}$, ${\bf 3},{\bf \bar{3}}$ and $T_H={\bf 1}, {\bf
  8}$. All the indices of the representations must be contracted,
i.e., in combination they form the trivial representation. Thus, the
number of terms each matrix element can depend on is equal to the
number of times the singlet representation appears in the
decomposition of the direct product ${\bf \overline{R}\otimes T_H \otimes
  R}$~\footnote{This is a generalization of the Wigner-Eckart
  theorem.}. For the combinations of interest, illustrated below, there is
only a single instance of $\identidad$ in the direct products.

In the following, for each representation we first consider ${\bf
  \overline{R}\otimes R}$ and then the direct product with $T_H$.
\begin{itemize}
\item{{\bf Sextet}}: corresponding to the singly charmed baryons with
  $J=\frac 12^\pm$ or $J=\frac 32^\pm$. We have
\begin{equation}
{\bf \overline{R}\otimes R} = {\bf 6} \otimes
  \overline{\bf 6} = {\bf 1} \oplus {\bf 8} \oplus {\bf 27}
\end{equation}
The ${\bf 6}$ representation is constructed from components of the
  fundamental representation, $S^{ab} = (q^aq^b + q^b q^a)$,
where $q^a=(u,d,s)$. Taking the example of the $J=\frac 12^+$ sextet,
we can assign the singly charmed baryons to the elements of $S^{ab}$:
\bea S^{11} = 2uu \sim 2 \Sigma_c^{++}, \hspace{6em}&\quad & S^{22} =
2dd \sim 2\Sigma_c^{0}, \\ \nonumber S^{12} = S^{21} = (ud +du) \sim
\sqrt{2} \Sigma_c^{+}, \hspace{0.5em}&& S^{23} = S^{32} = (ds + sd)
\sim \sqrt{2} \Xi_c^{'0},\\ S^{13} = S^{31} = (us+ su) \sim \sqrt2
\Xi_c^{'+}, && S^{33} = 2ss \sim 2 \Omega_c^{0}. \nonumber \eea
The tensor product of $S^{ab}$ with its conjugate representation, $S_{cd}$ 
can be decomposed in terms of a singlet~($S$), octet~($O^a_b$) and 27-plet~($T^{cd}_{ab}$).
\be
{\bf 6} \otimes \overline{\bf 6}: \quad S_{ab}S^{cd}=\tfrac{5}{72}\left\{\delta_a^c\delta_b^d+\delta_a^d\delta_b^c\right\}
S
+\tfrac16 \left\{\delta_a^c O_b^d + \delta_a^d O_b^c + 
\delta_b^c O_a^d + \delta_b^d O_a^c  \right\} + T^{cd}_{ab},\label{decomp_sext}
\ee
with
\bea
S &=& S_{ef} S^{fe} \nonumber \\
O_a^b &=& S_{ae}S^{eb} - \tfrac 13 \delta_a^b( S_{ef} S^{fe}) \\
T_{ab}^{cd} &=& S_{ab}S^{cd} - \tfrac16 \left\{ \delta_a^c (S_{be}S^{ed}) + 
\delta_a^d(S_{be}S^{ec}) + \delta_b^c(S_{ae}S^{ed}) + \delta_b^d(S_{ae}S^{ec} )\right\}
+ \tfrac{1}{24}\left\{\delta_a^c\delta_b^d+\delta_a^d\delta_b^c\right\}S, \nonumber 
\eea
where the summation over repeated indices is understood.

We can now identify the singlet representation that appears in the
product ${\bf \bar{6}\otimes T_H \otimes 6}$ and the expressions for
the matrix elements. For ${\bf T_H= 1}$ we simply obtain the singlet term
appearing in Eq.~(\ref{decomp_sext}). Thus, the lowest order term in the
expansion in $\delta m_{\ell}$~(cf. Eq.~(\ref{gmo_pert})) is given by
\begin{equation}
 \overline{ m}\langle B_{\bf \bar{6}} | H_0 | B_{\bf 6}\rangle =  
 M_0^{{\bf 6}} \left[ |\Sigma_c^{++}|^2 +  |\Sigma_c^{+}|^2 + |\Sigma_c^{0}|^2 + |\Xi_c^{'+}|^2 + |\Xi_c^{'0}|^2  + |\Omega_c^{0}|^2\right]\,.
\end{equation}
The coefficient $M^{{\bf 6}}_0$ is the mass of the sextet in the flavor symmetric limit.

For the next order term, $H_8\propto T_8\propto O^3_3$, where $T_8$ is
the second diagonal SU(3) generator and $O$ is the octet
representation~(${\bf T_H= 8 = \bar{8}}$). In the direct product ${\bf
  \bar{6}\otimes 8 \otimes 6 = 8 \otimes ({\bf 1} \oplus {\bf 8}
  \oplus {\bf 27})}$ the singlet appears in the ${\bf 8\otimes 8}$
term. More explicitly, one contracts $O_3^3$ from the Hamiltonian with
the $O_3^3 = S_{3c}S^{c3} - \tfrac 13 S_{cd}S^{dc}$ part of the tensor
product of the baryon representations in Eq.~(\ref{decomp_sext}):
\begin{equation}  
\delta m_{\ell} \langle B_{\bf \bar{6}} | H_8 | B_{\bf 6}\rangle = 
\delta m_{\ell} \frac{A_1}{2} \left[-\tfrac 43\left( |\Sigma_c^{++}|^2 +  |\Sigma_c^{+}|^2 + 
|\Sigma_c^{0}|^2 \right)+
\tfrac23 \left(|\Xi_c^{'+}|^2 + |\Xi_c^{'0}|^2 \right)  + 
\tfrac83|\Omega_c^{0}|^2 \right]. \end{equation}
Note that all particles in the sextet have the same coefficient,
$A_1$, for this ${\rm O}(\delta m_{\ell})$ term modulo factors that are
proportional to the light hypercharge.  Isolating the mass terms for each
particle up to first order in the Taylor expansion and taking the
isospin limit~(i.e. dropping the superscripts for the electric
charges), we arrive at \bea\label{sextet} M_{\Sigma_c} &=&
M_0^{{\bf 6}} - \tfrac 23 A_1\delta m_{\ell} + {\rm O} (\delta
m_{\ell}^2),\nonumber \\ M_{\Xi_c} &=& M_0^{{\bf 6}} + \tfrac 13
A_1\delta m_{\ell} + {\rm O} (\delta m_{\ell}^2), \\ M_{\Omega_c}
&=& M_0^{{\bf 6}} + \tfrac 43 A_1\delta m_{\ell} + {\rm O} (\delta m_{\ell}^2).
\nonumber \eea
The same expressions, but with different coefficients in each case, can
be used for the other sextets.
\item{{\bf Triplet}: corresponding to the doubly charmed baryons with
  $J=\frac 12^\pm$ or $J=\frac 32^\pm$.  We proceed in an analogous
  way to the above starting with ${\bf \overline{R}\otimes R} = {\bf 3} \otimes
  \overline{\bf 3} = {\bf 8} \oplus {\bf 1}$}. The assignment of the
  components of the fundamental~(${\bf 3}$) representation to the
  corresponding doubly charmed baryons is straightforward. For the example of 
  $J=\frac 12^+$, \be q^1 = u
  \sim \Xi_{cc}^{++}, \quad q^2 = d \sim \Xi_{cc}^{+}, \quad q^3 = s
  \sim \Omega_{cc}^{+}.  \ee The tensor product of $q^a$ with the
  conjugate representation $q_b$ takes the form \be {\bf 3} \otimes \overline{\bf 3}: \quad q^aq_b = \tfrac 13
  \delta^a_b S + O^a_b. \label{triplet_decomp}\ee 
For the lowest order matrix element, corresponding to ${\bf T_H= 1}$, we find
\begin{equation}
\overline{ m} \langle B_{\bf \bar{3}} | H_0 | B_{\bf 3}\rangle =  M_0^{{\bf 3}} \left[ |\Xi_{cc}^{++}|^2 +  |\Xi_{cc}^{+}|^2 + |\Omega_{cc}^{+}|^2\right]\,,
\end{equation}
corresponding to the singlet appearing in Eq.~(\ref{triplet_decomp}),
while at the next order with ${\bf T_H= 8}$
\begin{equation}
\delta m_{\ell} \langle B_{\bf \bar{3}} | H_8 | B_{\bf 3}\rangle = \delta m_{\ell} B_1 \left[- \tfrac 13( |\Xi_{cc}^{++}|^2 +  |\Xi_{cc}^{+}|^2) +\tfrac 23 |\Omega_{cc}^{+}|^2\right]\,,\label{firstorder_express}
\end{equation}
arising from contracting $O^3_3$ from the Hamiltonian with $O^3_3 = q^3q_3 - \frac 13 q^cq_c$ from Eq.~(\ref{triplet_decomp}).

Considering the individual states we arrive at the following
Gell-Mann--Okubo relations for the doubly charmed baryons,
\bea\label{triplet} M_{\Xi_{cc}} &=& M^{{\bf 3}}_{0} - \tfrac 13 B_1
\delta m_{\ell} + {\rm O}(\delta m_{\ell}^2), \nonumber \\ M_{\Omega_{cc}} &=&
M^{{\bf 3}}_0 + \tfrac 23 B_1 \delta m_{\ell} + {\rm O}(\delta m_{\ell}^2). \eea
As before, the same expressions apply to the other triplets
but with different coefficients.
\item{{\bf Anti-triplet}: corresponding to the singly charmed baryons
  with $J=\frac 12^\pm$. Here ${\bf \overline{R}\otimes R}$ is the
  same as for the triplet. We construct the ${\bf \bar{3}}$
  representation from the ${\bf 3}$, $q_c= \epsilon_{cab}q^a q^b$,
  leading to the components, \be q_1= (ds -sd) \sim \Xi_c^{0},\quad
  q_2= (us -su) \sim \Xi_c^{+}, \quad q_3= (ud - du)\sim
  \Lambda_c^{+}, \ee in terms of the corresponding singly charmed
  baryons. Using this and the tensor product, $q^aq_b$, given in
  Eqs.~(\ref{triplet_decomp}) we can derive the expressions, as above,
  for the Taylor expansion \bea\label{antitriplet} M_{\Lambda_{c}} &=&
  M_{0}^{{\bf \bar{3}}} - \tfrac 23 C_1 \delta m_{\ell} + {\rm
    O}(\delta m_{\ell}^2), \nonumber \\ M_{\Xi_{c}} &=& M_{0}^{{\bf
      \bar{3}}} + \tfrac 13 C_1 \delta m_{\ell} + {\rm O}(\delta
  m_{\ell}^2). \eea }

\end{itemize}

\newpage

\section{\label{AppC} Fitting and extrapolation details}

The fitting ranges chosen to extract the charmed baryon masses and the
charmonium, $D$ and $D_s$ spectra are given in
Tables~\ref{tablefitrange} and~\ref{tablefitmesons},
respectively. Also included are the fits employed in the extrapolation
to $\delta m_\ell^{\rm phys}$ that are used to compute the physical
point result, see Eqs.~(\ref{mave}) and (\ref{deltamave}).  Similarly,
for the baryon mass differences in Table~\ref{extrap_splittings}.
In Table~\ref{tablecoefficients}, we provide the
  coefficients extracted from linear and quadratic fits using the GMO
  formulae~(Eqs.~(\ref{ext_sextet}) to (\ref{ext_antitripletb})) and their
  less constrained counter-parts~(denoted LC1 and LC2) for the
  multiplets appearing in \refig{fan_plots}.

\begin{table}[ht!]

\begin{center}
\begin{tabular}{ccc|cc|cc}
\hline
\hline
& & & \multicolumn{2}{c}{First eigenvalue}& \multicolumn{2}{|c}{Second eigenvalue}\\
Particle & $P$ & Operator &   Fit-range    & Extrapolation &   Fit-range    & Extrapolation  \\ 
\hline
%%%%%%%%%%%%%%%%%%%%%% Positive parity %%%%%%%%%%%%%%%%%%%%%%%%%%
$\Sigma_c $ &$+$&\multirow{3}{*}{SU(4)} & $[10-24]_{\rm N_f=3 }$, $[10-18]_{\rm M_\pi=255\,{\rm MeV}}$  & \multirow{3}{*}{GMO2, LC1,LC2} & \multirow{3}{*}{$[9-13]$}   & \multirow{3}{*}{GMO1,GMO2}\\
$\Xi'_c$    &$+$& &\multirow{2}{*}{$[10-24]$}  &  & & \\
$\Omega_c$  &$+$& &&  & & \\\hline
$\Lambda_{c} $ &$+$&\multirow{2}{*}{SU(4)} & \multirow{2}{*}{$[11-22]$}  & \multirow{2}{*}{GMO1, GMO2} & \multirow{2}{*}{$[7-13]$}  & \multirow{2}{*}{GMO1, GMO2} \\
$\Xi_{c}$  &$+$& &&  & & \\\hline
$\Xi_{cc} $ &$+$&\multirow{2}{*}{SU(4)} & \multirow{2}{*}{$[11-24]$}  & \multirow{2}{*}{GMO2, LC1,LC2} & \multirow{2}{*}{$[10-20]$}  & \multirow{2}{*}{GMO1, GMO2} \\
$\Omega_{cc}$  &$+$& &&  & & \\\hline
$\Sigma_c $ &$+$&\multirow{3}{*}{HQET} & \multirow{3}{*}{$[10-19]$}  & \multirow{3}{*}{GMO1, GMO2} & \multirow{3}{*}{$[9-16]$}  & \multirow{3}{*}{GMO1, GMO2} \\
$\Xi'_c$    &$+$& &&   & & \\
$\Omega_c$  &$+$& &&  & & \\\hline
$\Lambda_{c} $ &$+$&\multirow{2}{*}{HQET} & \multirow{2}{*}{$[11-22]$}  & \multirow{2}{*}{GMO1, GMO2} & \multirow{2}{*}{$[7-13]$}  & \multirow{2}{*}{GMO1, GMO2} \\
$\Xi_{c}$  &$+$& &&  & & \\\hline
$\Xi_{cc} $ &$+$&\multirow{2}{*}{HQET} & \multirow{2}{*}{$[11-25]$}  & \multirow{2}{*}{GMO2, LC1, LC2} & \multirow{2}{*}{$[10-20]$}  & \multirow{2}{*}{GMO1, GMO2}\\
$\Omega_{cc}$  &$+$& &&  & & \\\hline
$\Sigma^*_c $ &$+$&\multirow{3}{*}{HQET} & \multirow{3}{*}{$[10-20]$}  & \multirow{3}{*}{GMO1, GMO2} &\multirow{3}{*}{$[9-13]$}  & \multirow{3}{*}{GMO1, GMO2} \\
$\Xi^*_c$    &$+$& &&   & & \\
$\Omega^*_c$  &$+$& &&  & & \\\hline
$\Xi^*_{cc} $ &$+$&\multirow{2}{*}{HQET} & \multirow{2}{*}{$[10-25]$}  & \multirow{2}{*}{GMO1, GMO2} & \multirow{2}{*}{$[10-20]$}  & \multirow{2}{*}{GMO1, GMO2}\\
$\Omega^*_{cc}$  &$+$& &&  & & \\\hline
\hline
%%%%%%%%%%%%%%%%%%%%%% Negative parity %%%%%%%%%%%%%%%%%%%%%%%%%%
$\Sigma_c $ &$-$&\multirow{3}{*}{SU(4)} & \multirow{3}{*}{$[10-15]$}  & \multirow{3}{*}{GMO1, GMO2} \\
$\Xi'_c$    &$-$& &  & &  &  \\
$\Omega_c$  &$-$& &&&  &  \\\hline
$\Xi_{cc} $ &$-$& \multirow{3}{*}{SU(4)}  &  & & \multirow{2}{*}{$[8-13]$}  & \multirow{2}{*}{GMO1, GMO2}\\
$\Omega_{cc}$  &$-$& && &  & \\\hline
$\Lambda_{c} $ &$-$&\multirow{2}{*}{HQET} & \multirow{2}{*}{$[11-15]$}  & \multirow{2}{*}{LC1, LC2} \\
$\Xi_{c}$  &$-$& && &  & \\\hline
$\Xi_{cc} $ &$-$&\multirow{2}{*}{HQET} & \multirow{2}{*}{$[10-16]$}  & \multirow{2}{*}{GMO2,LC1, LC2} & 
 & \\
$\Omega_{cc}$  &$-$& && &  & \\\hline
$\Sigma^*_c $ &$-$&\multirow{3}{*}{HQET} & \multirow{3}{*}{$[10-16]$}  & \multirow{3}{*}{GMO1, GMO2} \\
$\Xi^*_c$    &$-$& &&  &  & \\
$\Omega^*_c$  &$-$& && &  & \\\hline
$\Xi^*_{cc} $ &$-$&\multirow{2}{*}{HQET} & \multirow{2}{*}{$[9-15]$}  & \multirow{2}{*}{LC1, LC2} & \multirow{2}{*}{$[7-13]$}  & \multirow{2}{*}{LC1, LC2}\\
$\Omega^*_{cc}$  &$-$& && &  & \\\hline
\hline                             
\end{tabular}
\caption{\label{tablefitrange} Fitting ranges used to extract the
  charmed baryon masses and the fit functions employed for computing the
  final results at the physical point~(see Section~\ref{extrapfinal}) for the $32^3\times 64$
  ensembles. Note that for the negative parity states we only give 
  details for the operators for which  reliable signals 
  could be obtained.}
\end{center}
\vspace{-1em}
\end{table}

\newpage

\begin{table}[h!]
\begin{center}
\begin{tabular}{cc|cc|cc}
\hline
\hline
& & \multicolumn{2}{c}{First eigenvalue}& \multicolumn{2}{|c}{Second eigenvalue}\\
Particle & $J^{PC}$ & Fit-range    & Extrapolation &   Fit-range    & Extrapolation  \\ 
\hline
$\eta_c$ & $0^{-+}$     &15-23 &  FS1, FS2   & 10-17 & FS1, FS2\\
$J/\psi$ & $1^{--}$     &15-23 &  FS1, FS2   & 10-17 & FS1, FS2\\
$\chi_{c0}$ & $0^{++}$  &15-27 &  FS1, FS2  &  & \\
$\chi_{c1}$ & $1^{++}$  &15-27 &  FS1, FS2&  & \\
$h_c$       & $1^{+-}$  &13-27 &  FS1, FS2&  & \\\hline\hline
$D$   & \multirow{2}{*}{$0^-$} & \multirow{2}{*}{15-26} & \multirow{2}{*}{GMO2, LC1, LC2}  &  & \\
$D_s$ &                        &                        &   &  & \\\hline
$D^*$ & \multirow{2}{*}{$1^-$} & \multirow{2}{*}{15-26} & \multirow{2}{*}{GMO1, GMO2}  &  & \\
$(D^*_s)$&                       &                        &   &  & \\\hline
$\bar c l$& \multirow{2}{*}{$0^+$} & \multirow{2}{*}{10-15} & NC1, NC2  &  & \\
$\bar c s $ $ (D_{s0}^*)$&          &          & NC1, NC2  &  & \\\hline
$ (\bar c \gamma_i\gamma_5 l), (\bar c \epsilon_{ijk}\gamma_j\gamma_k l) $  
& \multirow{2}{*}{$1^+$}& \multirow{2}{*}{10-14} & NC1, NC2  & \multirow{2}{*}{10-14} & NC1, NC2\\
$(\bar c \gamma_i\gamma_5 s), (\bar c \epsilon_{ijk}\gamma_j\gamma_k s) $ & 
&                        & NC1, NC2  &  & NC1, NC2 \\
\hline                             
\hline
\end{tabular}
\caption{\label{tablefitmesons} As in Table~\ref{tablefitrange} for
  the lower lying meson channels. The extrapolations FS1 and FS2 refer
  to flavour singlet fits which include a constant term and a constant
  plus a quadratic~($\delta m_\ell^2$) term, respectively. Similarly, NC1
  and NC2 denote a linear fit to a single channel and a linear plus
  quadratic fit, respectively. The GMO and LC fits are defined in the
  Section~\ref{extrapmass}.}
\end{center}
\end{table}

\begin{table}[ht!]
\begin{center}
\begin{tabular}{c|c|c}
\hline
\hline
Difference  & Operator & Extrapolation \\ 
\hline
\multicolumn{3}{c}{Spin splittings}\\\hline
$\Sigma^*_c-\Sigma_c,\Xi^*_c-\Xi_c,\Omega^*_c-\Omega_c$&HQET & GMO1, GMO2 \\
$ \Xi^*_{cc} -   \Xi_{cc} ,\Omega^*_{cc} -\Omega_{cc}$ &    HQET & GMO2, LC1, LC2 \\\hline
\multicolumn{3}{c}{Structure splittings}\\\hline
% Positive parity structure splittings
$\Sigma_c-\Lambda_c,\Xi_c'-\Xi_c $ &  SU(4) & GMO1, GMO2 \\
$\Sigma_c-\Lambda_c,\Xi_c'-\Xi_c $ &  HQET & GMO1, GMO2 \\\hline
\multicolumn{3}{c}{Flavour splittings $s\to u/d$}\\\hline
% Positive parity flavour splittings s to u/d
$ \Xi'_{c} -   \Sigma_{c}, \Omega_{c} - \Xi'_{c},\Omega_{c}-\Sigma_{c}$ &
SU(4)& GMO1, GMO2 \\
$ \Xi'_{c} -   \Sigma_{c}, \Omega_{c} - \Xi'_{c},\Omega_{c}-\Sigma_{c}$ &
HQET& GMO1, GMO2 \\
$      \Xi_{c} -  \Lambda_{c}$  &     SU(4) & LC1, LC2 \\
$      \Xi_{c} -  \Lambda_{c}$  &     HQET & LC1, LC2 \\
$  \Omega_{cc} -     \Xi_{cc}$  &     SU(4) & LC1, LC2 \\
$  \Omega_{cc} -     \Xi_{cc}$  &     HQET & LC1, LC2 \\
$ \Xi^{*}_{c}-\Sigma^{*}_{c},\Omega^{*}_{c}-\Xi^{*}_{c},
\Omega^{*}_{c}-\Sigma^{*}_{c}$ &HQET & GMO1, GMO2 \\\hline
\multicolumn{3}{c}{Flavour splittings $c\to u/d$ and $c\to s$}\\\hline
% Positive parity flavour splittings c to u/d and s
% c to s
$\Omega_{cc}-\Omega_c,\Xi_{cc}-\Xi'_c,\Xi_{cc}  -  \Sigma_c,
\Omega_{cc}  -    \Xi'_c$&  SU(4) &GMO1, GMO2 \\ 
$\Omega_{cc}-\Omega_c,\Xi_{cc}-\Xi'_c,\Xi_{cc}  -  \Sigma_c,
\Omega_{cc}  -    \Xi'_c$&  HQET &GMO1, GMO2 \\ 
$  \Omega_{cc}^*-  \Omega_c^*, \Xi_{cc}^*-\Xi_c^*, 
 \Xi_{cc}^*-  \Sigma_c^*,\Omega_{cc}^*- \Xi_c^* $&HQET &GMO1, GMO2 \\
\hline
\hline
\end{tabular}
\caption{\label{extrap_splittings} The fit functions used for the
  extrapolation of the positive parity ground state differences to
  $\delta m_l^{phys}$. The splittings listed together were fitted
  simultaneously. Note that for $\Xi_c-\Lambda_c$ and
  $\Omega_{cc}-\Xi_{cc}$, there are no corresponding GMO expressions,
  since these are splittings between multiplets containing two states.}
\end{center}
\end{table}

\begin{table}[ht!] 
\begin{center}
   \renewcommand{\arraystretch}{1.4}
   \begin{tabular}{|c|cr|c|cr|c|cr|} 
   \multicolumn{9}{c}{Sextet SU(4) $J=\frac12^+$, $(\Sigma_c,\Xi'_{c},\Omega_c)$}\\
   \hline
   \hline
   \multirow{4}{*}{LC1} & $M^{\bf 6}_{0({\rm LC1}) }$ &$2.5530 (59)$ & 
   \multirow{7}{*}{LC2} & $M^{\bf 6}_{0({\rm LC2})}$ &$2.5502 (66)$  &
   \multirow{5}{*}{GMO2}& $M^{\bf 6}_{0({\rm GMO2})}$ &$2.5509 (57)$ \\
    &$A_{1({\rm LC1}) }$      &$-0.1428 (153)$&&$A_{1({\rm LC2}) }$    &$-0.0921 (601)$  &&$A_{1({\rm GMO2}) }$&$0.3006 (282)$ \\
    &$A^{'}_{1({\rm LC1}) }$  &$0.0232 (127)$ &&$A^{'}_{1({\rm LC2})}$ &$0.0720  (497)$  &&$A_{2({\rm GMO2}) }$&$-0.0688 (344)$\\
    &$A^{''}_{1({\rm LC1}) }$ &$0.1740 (115)$ &&$A^{''}_{1({\rm LC2}}$ &$0.2155  (422)$  &&$A_{3({\rm GMO2}) }$&$-0.0382 (181)$\\\cline{1-3}
   \multicolumn{3}{c|}{}                    &&$A_{2({\rm LC2})}$     &$-0.0744 (941)$  &&$A_{4({\rm GMO2}) }$&$-0.0372 (273)$\\\cline{7-9} 
   \multicolumn{3}{c|}{}                    &&$A_{3({\rm LC2})}$     &$-0.0715 (723)$  &\multicolumn{3}{c}{}\\
   \multicolumn{3}{c|}{}                    &&$A_{4({\rm LC2})}$     &$-0.0592 (576)$  &\multicolumn{3}{c}{}\\\cline{4-6}
   \multicolumn{9}{c}{}\\
   \multicolumn{6}{c}{Sextet SU(4) $J^P=\frac12^-$, $(\Sigma_c,\Xi'_{c},\Omega_c)$}&\multicolumn{3}{c}{}\\\hhline{======~~~}
   \multirow{2}{*}{GMO1} & $M^{\bf 6}_{0({\rm GMO1}) }$ &$2.8650 (160)$ & 
   \multirow{5}{*}{GMO2}& $M^{\bf 6}_{0({\rm GMO2})}$ & $2.9080 (239)$& \multicolumn{3}{c}{}\\
    &$A_{1({\rm GMO1}) }$  &$0.3383 (300)$&&$A_{1({\rm GMO2}) }$  &$0.0312 (1065)$ & \multicolumn{3}{c}{}\\\cline{1-3}
   \multicolumn{3}{c|}{}               &&$A_{2({\rm GMO2}) }$  &$-0.4212 (1391)$& \multicolumn{3}{c}{}\\
   \multicolumn{3}{c|}{}               &&$A_{3({\rm GMO2}) }$  &$-0.0875 (734)$ & \multicolumn{3}{c}{}\\
   \multicolumn{3}{c|}{}               &&$A_{4({\rm GMO2}) }$  &$0.2048 (1019)$ & \multicolumn{3}{c}{}\\\cline{4-6}
   \multicolumn{9}{c}{}\\
   \multicolumn{9}{c}{Triplet HQET $J^P=\frac12^+$, $(\Xi_{cc},\Omega_{cc})$}\\
   \hline
   \hline
   \multirow{3}{*}{LC1} & $M^{\bf 3}_{0({\rm LC1}) }$ &$3.6585 (44)$ & 
   \multirow{5}{*}{LC2} & $M^{\bf 3}_{0({\rm LC2})}$  &$3.6596 (48)$  &
   \multirow{4}{*}{GMO2}& $M^{\bf 3}_{0({\rm GMO2})}$ &$3.6559 (40)$ \\
    &$B_{1({\rm LC1}) }$      &$0.0723 (89)$  &&$B_{1({\rm LC2}) }$    &$0.0557(305)$    &&$B_{1({\rm GMO2}) }$&$-0.0471 (46)$  \\
    &$B^{''}_{1({\rm LC1}) }$ &$-0.0729 (99)$ &&$B^{''}_{1({\rm LC2})}$ &$-0.0955(369)$   &&$B_{2({\rm GMO2}) }$&$-0.0271 (164)$ \\\cline{1-3}
   \multicolumn{3}{c|}{}                      &&$B_{2({\rm LC2})}$     &$0.0242(420)$    &&$B_{3({\rm GMO2}) }$&$-0.0330 (199)$ \\\cline{7-9} 
   \multicolumn{3}{c|}{}                      &&$B_{3({\rm LC2})}$     &$0.0349 (551)$   &\multicolumn{3}{c}{}\\\cline{4-6}
   \multicolumn{9}{c}{}\\
   \multicolumn{9}{c}{Triplet HQET $J^P=\frac12^-$,  $(\Xi_{cc},\Omega_{cc})$}\\
   \hline
   \hline
   \multirow{3}{*}{LC1} & $M^{\bf 3}_{0({\rm LC1}) }$ &$4.0143 (191)$ & 
   \multirow{5}{*}{LC2} & $M^{\bf 3}_{0({\rm LC2})}$  &$4.0281 (216)$  &
   \multirow{4}{*}{GMO2}& $M^{\bf 3}_{0({\rm GMO2})}$ &$4.0158 (172)$ \\
    &$B_{1({\rm LC1}) }$      &$0.0559 (392)$ &&$B_{1({\rm LC2}) }$    &$-0.0817 (1274)$   &&$B_{1({\rm GMO2}) }$&$-0.0152 (227)$ \\
    &$B^{''}_{1({\rm LC1}) }$ &$-0.1580 (445)$&&$B^{''}_{1({\rm LC2})}$ &$-0.1611 (1568)$   &&$B_{2({\rm GMO2}) }$&$0.0435 (759)$  \\\cline{1-3}
   \multicolumn{3}{c|}{}                      &&$B_{2({\rm LC2})}$     &$0.1897 (1730)$    &&$B_{3({\rm GMO2}) }$&$-0.2641 (908)$ \\\cline{7-9} 
   \multicolumn{3}{c|}{}                      &&$B_{3({\rm LC2})}$     &$-0.0605 (2348)$   &\multicolumn{3}{c}{}\\\cline{4-6}
   \multicolumn{9}{c}{}\\
   \multicolumn{6}{c}{Antitriplet SU(4) $J^P=\frac12^+$,  $(\Lambda_c, \Xi_c)$} &\multicolumn{3}{c}{}\\\hhline{======~~~}
   \multirow{2}{*}{GMO1} & $M^{\overline{\bf 3}}_{0({\rm GMO1}) }$ &$2.4274 (40)$& 
   \multirow{4}{*}{GMO2}& $M^{\overline{\bf 3}}_{0({\rm GMO2})}$   &$2.4337 (56)$& \multicolumn{3}{c}{}\\
   &$C_{1({\rm GMO1}) }$   &$0.0756 (25)$&&$C_{1({\rm GMO2}) }$  &$0.0817 (90)$   & \multicolumn{3}{c}{}\\\cline{1-3}
   \multicolumn{3}{c|}{}               &&$C_{2({\rm GMO2}) }$  &$-0.0099 (537)$ & \multicolumn{3}{c}{}\\
   \multicolumn{3}{c|}{}               &&$C_{3({\rm GMO2}) }$  &$-0.0429 (205)$ & \multicolumn{3}{c}{}\\\cline{4-6}
   \multicolumn{9}{c}{}\\
   \multicolumn{6}{c}{Antitriplet HQET $J^P=\frac12^-$, $(\Lambda_c, \Xi_c)$} & \multicolumn{3}{c}{}\\\hhline{======~~~}
   \multirow{3}{*}{LC1} & $M^{\overline{\bf 3}}_{0({\rm LC1}) }$  & $2.7825 (332)$& 
   \multirow{5}{*}{LC2}& $M^{\overline{\bf 3}}_{0({\rm LC2})}$    & $2.8002 (356)$& \multicolumn{3}{c}{}\\
   &$C_{1({\rm LC1}) }$      &$-0.3500 (936)$  &&$C_{1({\rm LC2}) }$      &$-0.9128 (3261)$ & \multicolumn{3}{c}{}\\
   &$C^{'}_{1({\rm LC1}) }$  &$-0.0276 (783)$  &&$C^{'}_{1({\rm LC2}) }$  &$-0.3978 (2514)$ & \multicolumn{3}{c}{}\\\cline{1-3}
   \multicolumn{3}{c|}{}                         &&$C_{2({\rm LC2}) }$  &$1.0029 (5636)$  & \multicolumn{3}{c}{}\\
   \multicolumn{3}{c|}{}                         &&$C_{3({\rm LC2}) }$  &$0.5979 (3759)$  & \multicolumn{3}{c}{}\\\cline{4-6}
\end{tabular}
\caption{\label{tablecoefficients} The coefficients determined from
  linear and quadratic GMO and LC fits~(see
  Section~\ref{extrapmass}) to the ground state positive and negative
  parity multiplets displayed in Fig.~\ref{fan_plots}.}  
\renewcommand{\arraystretch}{1}
\end{center} 
\end{table}

\newpage

\bibliography{references}

%Merlin.mbs v4.21 2009-07-09.
\begin{thebibliography}{10}%
\makeatletter
\providecommand \@ifxundefined [1]{%
 \ifx #1\undefined \expandafter \@firstoftwo
 \else \expandafter \@secondoftwo
\fi
}%
\providecommand \@ifnum [1]{%
 \ifnum #1\expandafter \@firstoftwo
 \else \expandafter \@secondoftwo
\fi
}%
\providecommand \enquote [1]{``#1''}%
\providecommand \bibnamefont  [1]{#1}%
\providecommand \bibfnamefont [1]{#1}%
\providecommand \citenamefont [1]{#1}%
\providecommand\href[0]{\@sanitize\@href}%
\providecommand\@href[1]{\endgroup\@@startlink{#1}\endgroup\@@href}%
\providecommand\@@href[1]{#1\@@endlink}%
\providecommand \@sanitize [0]{\begingroup\catcode`\&12\catcode`\#12\relax}%
\@ifxundefined \pdfoutput {\@firstoftwo}{%
 \@ifnum{\z@=\pdfoutput}{\@firstoftwo}{\@secondoftwo}%
}{%
 \providecommand\@@startlink[1]{\leavevmode\special{html:<a href="#1">}}%
 \providecommand\@@endlink[0]{\special{html:</a>}}%
}{%
 \providecommand\@@startlink[1]{%
  \leavevmode
  \pdfstartlink
   attr{/Border[0 0 1 ]/H/I/C[0 1 1]}%
   user{/Subtype/Link/A<</Type/Action/S/URI/URI(#1)>>}%
  \relax
 }%
 \providecommand\@@endlink[0]{\pdfendlink}%
}%
\providecommand \url  [0]{\begingroup\@sanitize \@url }%
\providecommand \@url [1]{\endgroup\@href {#1}{\urlprefix}}%
\providecommand \urlprefix [0]{URL }%
\providecommand \Eprint[0]{\href }%
\@ifxundefined \urlstyle {%
  \providecommand \doi [1]{doi:\discretionary{}{}{}#1}%
}{%
  \providecommand \doi [0]{doi:\discretionary{}{}{}\begingroup
  \urlstyle{rm}\Url }%
}%
\providecommand \doibase [0]{http://dx.doi.org/}%
\providecommand \Doi[1]{\href{\doibase#1}}%
\providecommand \bibAnnote [3]{%
  \BibitemShut{#1}%
  \begin{quotation}\noindent
    \textsc{Key:}\ #2\\\textsc{Annotation:}\ #3%
  \end{quotation}%
}%
\providecommand \bibAnnoteFile [2]{%
  \IfFileExists{#2}{\bibAnnote {#1} {#2} {\input{#2}}}{}%
}%
\providecommand \typeout [0]{\immediate \write \m@ne }%
\providecommand \selectlanguage [0]{\@gobble}%
\providecommand \bibinfo [0]{\@secondoftwo}%
\providecommand \bibfield [0]{\@secondoftwo}%
\providecommand \translation [1]{[#1]}%
\providecommand \BibitemOpen[0]{}%
\providecommand \bibitemStop [0]{}%
\providecommand \bibitemNoStop [0]{.\EOS\space}%
\providecommand \EOS [0]{\spacefactor3000\relax}%
\providecommand \BibitemShut [1]{\csname bibitem#1\endcsname}%
%</preamble>
\bibitem{Isgur:1989vq}%
  \BibitemOpen
  \bibfield{author}{%
  \bibinfo {author} {\bibfnamefont{N.}~\bibnamefont{Isgur}}\ and\ \bibinfo
  {author} {\bibfnamefont{M.~B.}\ \bibnamefont{Wise}},\ }%
  \bibfield{journal}{%
  \Doi{10.1016/0370-2693(89)90566-2}{\bibinfo {journal} {Phys.~Lett.}}\ }%
  \textbf{\bibinfo {volume} {B232}},\ \bibinfo {pages} {113} (\bibinfo {year}
  {1989})%
  \bibAnnoteFile{NoStop}{Isgur:1989vq}%
%%CITATION = PHLTA,B232,113;%%
\bibitem{Eichten:1989zv}%
  \BibitemOpen
  \bibfield{author}{%
  \bibinfo {author} {\bibfnamefont{E.}~\bibnamefont{Eichten}}\ and\ \bibinfo
  {author} {\bibfnamefont{B.~R.}\ \bibnamefont{Hill}},\ }%
  \bibfield{journal}{%
  \Doi{10.1016/0370-2693(90)92049-O}{\bibinfo {journal} {Phys.~Lett.}}\ }%
  \textbf{\bibinfo {volume} {B234}},\ \bibinfo {pages} {511} (\bibinfo {year}
  {1990})%
  \bibAnnoteFile{NoStop}{Eichten:1989zv}%
%%CITATION = PHLTA,B234,511;%%
\bibitem{Caswell:1985ui}%
  \BibitemOpen
  \bibfield{author}{%
  \bibinfo {author} {\bibfnamefont{W.}~\bibnamefont{Caswell}}\ and\ \bibinfo
  {author} {\bibfnamefont{G.}~\bibnamefont{Lepage}},\ }%
  \bibfield{journal}{%
  \Doi{10.1016/0370-2693(86)91297-9}{\bibinfo {journal} {Phys. Lett.}}\ }%
  \textbf{\bibinfo {volume} {B167}},\ \bibinfo {pages} {437} (\bibinfo {year}
  {1986})%
  \bibAnnoteFile{NoStop}{Caswell:1985ui}%
%%CITATION = PHLTA,B167,437;%%
\bibitem{Bodwin:1994jh}%
  \BibitemOpen
  \bibfield{author}{%
  \bibinfo {author} {\bibfnamefont{G.~T.}\ \bibnamefont{Bodwin}}, \bibinfo
  {author} {\bibfnamefont{E.}~\bibnamefont{Braaten}},\ and\ \bibinfo {author}
  {\bibfnamefont{G.~P.}\ \bibnamefont{Lepage}},\ }%
  \bibfield{journal}{%
  \Doi{10.1103/PhysRevD.51.1125 10.1103/PhysRevD.55.5853,
  10.1103/PhysRevD.55.5853, 10.1103/PhysRevD.51.1125}{\bibinfo {journal} {Phys.
  Rev.}}\ }%
  \textbf{\bibinfo {volume} {D51}},\ \bibinfo {pages} {1125} (\bibinfo {year}
  {1995}),\ \Eprint{http://arxiv.org/abs/hep-ph/9407339}{arXiv:hep-ph/9407339
  [hep-ph]}%
  \bibAnnoteFile{NoStop}{Bodwin:1994jh}%
%%CITATION = HEP-PH/9407339;%%
\bibitem{Pineda:1997bj}%
  \BibitemOpen
  \bibfield{author}{%
  \bibinfo {author} {\bibfnamefont{A.}~\bibnamefont{Pineda}}\ and\ \bibinfo
  {author} {\bibfnamefont{J.}~\bibnamefont{Soto}},\ }%
  \bibfield{journal}{%
  \Doi{10.1016/S0920-5632(97)01102-X}{\bibinfo {journal} {Nucl. Phys. Proc.
  Suppl.}}\ }%
  \textbf{\bibinfo {volume} {64}},\ \bibinfo {pages} {428} (\bibinfo {year}
  {1998}),\ \Eprint{http://arxiv.org/abs/hep-ph/9707481}{arXiv:hep-ph/9707481
  [hep-ph]}%
  \bibAnnoteFile{NoStop}{Pineda:1997bj}%
%%CITATION = HEP-PH/9707481;%%
\bibitem{Brambilla:1999xf}%
  \BibitemOpen
  \bibfield{author}{%
  \bibinfo {author} {\bibfnamefont{N.}~\bibnamefont{Brambilla}}, \bibinfo
  {author} {\bibfnamefont{A.}~\bibnamefont{Pineda}}, \bibinfo {author}
  {\bibfnamefont{J.}~\bibnamefont{Soto}},\ and\ \bibinfo {author}
  {\bibfnamefont{A.}~\bibnamefont{Vairo}},\ }%
  \bibfield{journal}{%
  \Doi{10.1016/S0550-3213(99)00693-8}{\bibinfo {journal} {Nucl. Phys.}}\ }%
  \textbf{\bibinfo {volume} {B566}},\ \bibinfo {pages} {275} (\bibinfo {year}
  {2000}),\ \Eprint{http://arxiv.org/abs/hep-ph/9907240}{arXiv:hep-ph/9907240
  [hep-ph]}%
  \bibAnnoteFile{NoStop}{Brambilla:1999xf}%
%%CITATION = HEP-PH/9907240;%%
\bibitem{Agashe:2014kda}%
  \BibitemOpen
  \bibfield{author}{%
  \bibinfo {author} {\bibfnamefont{K.}~\bibnamefont{Olive}} \emph{et~al.}
  (\bibinfo {collaboration} {Particle Data Group}),\ }%
  \bibfield{journal}{%
  \Doi{10.1088/1674-1137/38/9/090001}{\bibinfo {journal} {Chin. Phys.}}\ }%
  \textbf{\bibinfo {volume} {C38}},\ \bibinfo {pages} {090001} (\bibinfo {year}
  {2014})%
  \bibAnnoteFile{NoStop}{Agashe:2014kda}%
%%CITATION = CHPHD,C38,090001;%%
\bibitem{Cazzoli:1975et}%
  \BibitemOpen
  \bibfield{author}{%
  \bibinfo {author} {\bibfnamefont{E.}~\bibnamefont{Cazzoli}} \emph{et~al.},\
  }%
  \bibfield{journal}{%
  \Doi{10.1103/PhysRevLett.34.1125}{\bibinfo {journal} {Phys.~Rev.~Lett.}}\ }%
  \textbf{\bibinfo {volume} {34}},\ \bibinfo {pages} {1125} (\bibinfo {year}
  {1975})%
  \bibAnnoteFile{NoStop}{Cazzoli:1975et}%
%%CITATION = PRLTA,34,1125;%%
\bibitem{Knapp:1976qw}%
  \BibitemOpen
  \bibfield{author}{%
  \bibinfo {author} {\bibfnamefont{B.}~\bibnamefont{Knapp}} \emph{et~al.},\ }%
  \bibfield{journal}{%
  \Doi{10.1103/PhysRevLett.37.882}{\bibinfo {journal} {Phys.~Rev.~Lett.}}\ }%
  \textbf{\bibinfo {volume} {37}},\ \bibinfo {pages} {882} (\bibinfo {year}
  {1976})%
  \bibAnnoteFile{NoStop}{Knapp:1976qw}%
%%CITATION = PRLTA,37,882;%%
\bibitem{Basile:1981wr}%
  \BibitemOpen
  \bibfield{author}{%
  \bibinfo {author} {\bibfnamefont{M.}~\bibnamefont{Basile}} \emph{et~al.},\ }%
  \bibfield{journal}{%
  \Doi{10.1007/BF02822406}{\bibinfo {journal} {Lett.~Nuovo Cim.}}\ }%
  \textbf{\bibinfo {volume} {31}},\ \bibinfo {pages} {97} (\bibinfo {year}
  {1981})%
  \bibAnnoteFile{NoStop}{Basile:1981wr}%
%%CITATION = NCLTA,31,97;%%
\bibitem{Mattson:2002vu}%
  \BibitemOpen
  \bibfield{author}{%
  \bibinfo {author} {\bibfnamefont{M.}~\bibnamefont{Mattson}} \emph{et~al.}
  (\bibinfo {collaboration} {SELEX}),\ }%
  \bibfield{journal}{%
  \Doi{10.1103/PhysRevLett.89.112001}{\bibinfo {journal} {Phys. Rev. Lett.}}\
  }%
  \textbf{\bibinfo {volume} {89}},\ \bibinfo {pages} {112001} (\bibinfo {year}
  {2002})%
  \bibAnnoteFile{NoStop}{Mattson:2002vu}%
%%CITATION = HEP-EX/0208014;%%
\bibitem{Ocherashvili:2004hi}%
  \BibitemOpen
  \bibfield{author}{%
  \bibinfo {author} {\bibfnamefont{A.}~\bibnamefont{Ocherashvili}}
  \emph{et~al.} (\bibinfo {collaboration} {SELEX}),\ }%
  \bibfield{journal}{%
  \Doi{10.1016/j.physletb.2005.09.043}{\bibinfo {journal} {Phys. Lett.}}\ }%
  \textbf{\bibinfo {volume} {B628}},\ \bibinfo {pages} {18} (\bibinfo {year}
  {2005}),\ \Eprint{http://arxiv.org/abs/hep-ex/0406033}{arXiv:hep-ex/0406033
  [hep-ex]}%
  \bibAnnoteFile{NoStop}{Ocherashvili:2004hi}%
%%CITATION = HEP-EX/0406033;%%
\bibitem{Russ:2002bw}%
  \BibitemOpen
  \bibfield{author}{%
  \bibinfo {author} {\bibfnamefont{J.}~\bibnamefont{Russ}} (\bibinfo
  {collaboration} {SELEX})}%
   (\bibinfo {year} {2002}),\
  \Eprint{http://arxiv.org/abs/hep-ex/0209075}{arXiv:hep-ex/0209075 [hep-ex]}%
  \bibAnnoteFile{NoStop}{Russ:2002bw}%
%%CITATION = HEP-EX/0209075;%%
\bibitem{Copley:1979wj}%
  \BibitemOpen
  \bibfield{author}{%
  \bibinfo {author} {\bibfnamefont{L.}~\bibnamefont{Copley}}, \bibinfo {author}
  {\bibfnamefont{N.}~\bibnamefont{Isgur}},\ and\ \bibinfo {author}
  {\bibfnamefont{G.}~\bibnamefont{Karl}},\ }%
  \bibfield{journal}{%
  \Doi{10.1103/PhysRevD.23.817.3, 10.1103/PhysRevD.20.768}{\bibinfo {journal}
  {Phys. Rev.}}\ }%
  \textbf{\bibinfo {volume} {D20}},\ \bibinfo {pages} {768} (\bibinfo {year}
  {1979})%
  \bibAnnoteFile{NoStop}{Copley:1979wj}%
%%CITATION = PHRVA,D20,768;%%
\bibitem{Capstick:1986bm}%
  \BibitemOpen
  \bibfield{author}{%
  \bibinfo {author} {\bibfnamefont{S.}~\bibnamefont{Capstick}}\ and\ \bibinfo
  {author} {\bibfnamefont{N.}~\bibnamefont{Isgur}},\ }%
  \bibfield{journal}{%
  \Doi{10.1103/PhysRevD.34.2809}{\bibinfo {journal} {Phys. Rev.}}\ }%
  \textbf{\bibinfo {volume} {D34}},\ \bibinfo {pages} {2809} (\bibinfo {year}
  {1986})%
  \bibAnnoteFile{NoStop}{Capstick:1986bm}%
%%CITATION = PHRVA,D34,2809;%%
\bibitem{Roncaglia:1995az}%
  \BibitemOpen
  \bibfield{author}{%
  \bibinfo {author} {\bibfnamefont{R.}~\bibnamefont{Roncaglia}}, \bibinfo
  {author} {\bibfnamefont{D.}~\bibnamefont{Lichtenberg}},\ and\ \bibinfo
  {author} {\bibfnamefont{E.}~\bibnamefont{Predazzi}},\ }%
  \bibfield{journal}{%
  \Doi{10.1103/PhysRevD.52.1722}{\bibinfo {journal} {Phys. Rev.}}\ }%
  \textbf{\bibinfo {volume} {D52}},\ \bibinfo {pages} {1722} (\bibinfo {year}
  {1995})%
  \bibAnnoteFile{NoStop}{Roncaglia:1995az}%
%%CITATION = HEP-PH/9502251;%%
\bibitem{SilvestreBrac:1996bg}%
  \BibitemOpen
  \bibfield{author}{%
  \bibinfo {author} {\bibfnamefont{B.}~\bibnamefont{Silvestre-Brac}},\ }%
  \bibfield{journal}{%
  \Doi{10.1007/s006010050028}{\bibinfo {journal} {Few Body Syst.}}\ }%
  \textbf{\bibinfo {volume} {20}},\ \bibinfo {pages} {1} (\bibinfo {year}
  {1996})%
  \bibAnnoteFile{NoStop}{SilvestreBrac:1996bg}%
%%CITATION = FBSYE,20,1;%%
\bibitem{Ebert:2002ig}%
  \BibitemOpen
  \bibfield{author}{%
  \bibinfo {author} {\bibfnamefont{D.}~\bibnamefont{Ebert}}, \bibinfo {author}
  {\bibfnamefont{R.}~\bibnamefont{Faustov}}, \bibinfo {author}
  {\bibfnamefont{V.}~\bibnamefont{Galkin}},\ and\ \bibinfo {author}
  {\bibfnamefont{A.}~\bibnamefont{Martynenko}},\ }%
  \bibfield{journal}{%
  \Doi{10.1103/PhysRevD.66.014008}{\bibinfo {journal} {Phys. Rev.}}\ }%
  \textbf{\bibinfo {volume} {D66}},\ \bibinfo {pages} {014008} (\bibinfo {year}
  {2002})%
  \bibAnnoteFile{NoStop}{Ebert:2002ig}%
%%CITATION = HEP-PH/0201217;%%
\bibitem{Ebert:2005xj}%
  \BibitemOpen
  \bibfield{author}{%
  \bibinfo {author} {\bibfnamefont{D.}~\bibnamefont{Ebert}}, \bibinfo {author}
  {\bibfnamefont{R.}~\bibnamefont{Faustov}},\ and\ \bibinfo {author}
  {\bibfnamefont{V.}~\bibnamefont{Galkin}},\ }%
  \bibfield{journal}{%
  \Doi{10.1103/PhysRevD.72.034026}{\bibinfo {journal} {Phys. Rev.}}\ }%
  \textbf{\bibinfo {volume} {D72}},\ \bibinfo {pages} {034026} (\bibinfo {year}
  {2005})%
  \bibAnnoteFile{NoStop}{Ebert:2005xj}%
%%CITATION = HEP-PH/0504112;%%
\bibitem{Roberts:2007ni}%
  \BibitemOpen
  \bibfield{author}{%
  \bibinfo {author} {\bibfnamefont{W.}~\bibnamefont{Roberts}}\ and\ \bibinfo
  {author} {\bibfnamefont{M.}~\bibnamefont{Pervin}},\ }%
  \bibfield{journal}{%
  \Doi{10.1142/S0217751X08041219}{\bibinfo {journal} {Int. J. Mod. Phys.}}\ }%
  \textbf{\bibinfo {volume} {A23}},\ \bibinfo {pages} {2817} (\bibinfo {year}
  {2008})%
  \bibAnnoteFile{NoStop}{Roberts:2007ni}%
%%CITATION = ARXIV:0711.2492;%%
\bibitem{Garcilazo:2007eh}%
  \BibitemOpen
  \bibfield{author}{%
  \bibinfo {author} {\bibfnamefont{H.}~\bibnamefont{Garcilazo}}, \bibinfo
  {author} {\bibfnamefont{J.}~\bibnamefont{Vijande}},\ and\ \bibinfo {author}
  {\bibfnamefont{A.}~\bibnamefont{Valcarce}},\ }%
  \bibfield{journal}{%
  \Doi{10.1088/0954-3899/34/5/014}{\bibinfo {journal} {J. Phys.}}\ }%
  \textbf{\bibinfo {volume} {G34}},\ \bibinfo {pages} {961} (\bibinfo {year}
  {2007})%
  \bibAnnoteFile{NoStop}{Garcilazo:2007eh}%
%%CITATION = HEP-PH/0703257;%%
\bibitem{Valcarce:2008dr}%
  \BibitemOpen
  \bibfield{author}{%
  \bibinfo {author} {\bibfnamefont{A.}~\bibnamefont{Valcarce}}, \bibinfo
  {author} {\bibfnamefont{H.}~\bibnamefont{Garcilazo}},\ and\ \bibinfo {author}
  {\bibfnamefont{J.}~\bibnamefont{Vijande}},\ }%
  \bibfield{journal}{%
  \Doi{10.1140/epja/i2008-10616-4}{\bibinfo {journal} {Eur. Phys. J.}}\ }%
  \textbf{\bibinfo {volume} {A37}},\ \bibinfo {pages} {217} (\bibinfo {year}
  {2008})%
  \bibAnnoteFile{NoStop}{Valcarce:2008dr}%
%%CITATION = ARXIV:0807.2973;%%
\bibitem{Jenkins:1993ta}%
  \BibitemOpen
  \bibfield{author}{%
  \bibinfo {author} {\bibfnamefont{E.~E.}\ \bibnamefont{Jenkins}},\ }%
  \bibfield{journal}{%
  \Doi{10.1016/0370-2693(93)91639-5}{\bibinfo {journal} {Phys. Lett.}}\ }%
  \textbf{\bibinfo {volume} {B315}},\ \bibinfo {pages} {447} (\bibinfo {year}
  {1993})%
  \bibAnnoteFile{NoStop}{Jenkins:1993ta}%
%%CITATION = HEP-PH/9307245;%%
\bibitem{Bagan:1991sc}%
  \BibitemOpen
  \bibfield{author}{%
  \bibinfo {author} {\bibfnamefont{E.}~\bibnamefont{Bagan}}, \bibinfo {author}
  {\bibfnamefont{M.}~\bibnamefont{Chabab}}, \bibinfo {author}
  {\bibfnamefont{H.~G.}\ \bibnamefont{Dosch}},\ and\ \bibinfo {author}
  {\bibfnamefont{S.}~\bibnamefont{Narison}},\ }%
  \bibfield{journal}{%
  \Doi{10.1016/0370-2693(92)90208-L}{\bibinfo {journal} {Phys. Lett.}}\ }%
  \textbf{\bibinfo {volume} {B278}},\ \bibinfo {pages} {367} (\bibinfo {year}
  {1992})%
  \bibAnnoteFile{NoStop}{Bagan:1991sc}%
%%CITATION = PHLTA,B278,367;%%
\bibitem{Bagan:1992tp}%
  \BibitemOpen
  \bibfield{author}{%
  \bibinfo {author} {\bibfnamefont{E.}~\bibnamefont{Bagan}}, \bibinfo {author}
  {\bibfnamefont{M.}~\bibnamefont{Chabab}}, \bibinfo {author}
  {\bibfnamefont{H.~G.}\ \bibnamefont{Dosch}},\ and\ \bibinfo {author}
  {\bibfnamefont{S.}~\bibnamefont{Narison}},\ }%
  \bibfield{journal}{%
  \Doi{10.1016/0370-2693(92)91896-H}{\bibinfo {journal} {Phys. Lett.}}\ }%
  \textbf{\bibinfo {volume} {B287}},\ \bibinfo {pages} {176} (\bibinfo {year}
  {1992})%
  \bibAnnoteFile{NoStop}{Bagan:1992tp}%
%%CITATION = PHLTA,B287,176;%%
\bibitem{Wang:2002ts}%
  \BibitemOpen
  \bibfield{author}{%
  \bibinfo {author} {\bibfnamefont{D.-W.}\ \bibnamefont{Wang}}, \bibinfo
  {author} {\bibfnamefont{M.-Q.}\ \bibnamefont{Huang}},\ and\ \bibinfo {author}
  {\bibfnamefont{C.-Z.}\ \bibnamefont{Li}},\ }%
  \bibfield{journal}{%
  \Doi{10.1103/PhysRevD.65.094036}{\bibinfo {journal} {Phys. Rev.}}\ }%
  \textbf{\bibinfo {volume} {D65}},\ \bibinfo {pages} {094036} (\bibinfo {year}
  {2002})%
  \bibAnnoteFile{NoStop}{Wang:2002ts}%
%%CITATION = HEP-PH/0201183;%%
\bibitem{Wang:2007sqa}%
  \BibitemOpen
  \bibfield{author}{%
  \bibinfo {author} {\bibfnamefont{Z.-G.}\ \bibnamefont{Wang}},\ }%
  \bibfield{journal}{%
  \Doi{10.1140/epjc/s10052-008-0521-x}{\bibinfo {journal} {Eur. Phys. J.}}\ }%
  \textbf{\bibinfo {volume} {C54}},\ \bibinfo {pages} {231} (\bibinfo {year}
  {2008})%
  \bibAnnoteFile{NoStop}{Wang:2007sqa}%
%%CITATION = ARXIV:0704.1106;%%
\bibitem{Zhang:2008rt}%
  \BibitemOpen
  \bibfield{author}{%
  \bibinfo {author} {\bibfnamefont{J.-R.}\ \bibnamefont{Zhang}}\ and\ \bibinfo
  {author} {\bibfnamefont{M.-Q.}\ \bibnamefont{Huang}},\ }%
  \bibfield{journal}{%
  \Doi{10.1103/PhysRevD.78.094007}{\bibinfo {journal} {Phys.~Rev.}}\ }%
  \textbf{\bibinfo {volume} {D78}},\ \bibinfo {pages} {094007} (\bibinfo {year}
  {2008}),\ \Eprint{http://arxiv.org/abs/0810.5396}{arXiv:0810.5396 [hep-ph]}%
  \bibAnnoteFile{NoStop}{Zhang:2008rt}%
%%CITATION = ARXIV:0810.5396;%%
\bibitem{Zhang:2008pm}%
  \BibitemOpen
  \bibfield{author}{%
  \bibinfo {author} {\bibfnamefont{J.-R.}\ \bibnamefont{Zhang}}\ and\ \bibinfo
  {author} {\bibfnamefont{M.-Q.}\ \bibnamefont{Huang}},\ }%
  \bibfield{journal}{%
  \Doi{10.1103/PhysRevD.78.094015}{\bibinfo {journal} {Phys.~Rev.}}\ }%
  \textbf{\bibinfo {volume} {D78}},\ \bibinfo {pages} {094015} (\bibinfo {year}
  {2008}),\ \Eprint{http://arxiv.org/abs/0811.3266}{arXiv:0811.3266 [hep-ph]}%
  \bibAnnoteFile{NoStop}{Zhang:2008pm}%
%%CITATION = ARXIV:0811.3266;%%
\bibitem{Bowler:1996ws}%
  \BibitemOpen
  \bibfield{author}{%
  \bibinfo {author} {\bibfnamefont{K.}~\bibnamefont{Bowler}} \emph{et~al.}
  (\bibinfo {collaboration} {UKQCD}),\ }%
  \bibfield{journal}{%
  \Doi{10.1103/PhysRevD.54.3619}{\bibinfo {journal} {Phys.~Rev.}}\ }%
  \textbf{\bibinfo {volume} {D54}},\ \bibinfo {pages} {3619} (\bibinfo {year}
  {1996}),\ \Eprint{http://arxiv.org/abs/hep-lat/9601022}{arXiv:hep-lat/9601022
  [hep-lat]}%
  \bibAnnoteFile{NoStop}{Bowler:1996ws}%
%%CITATION = HEP-LAT/9601022;%%
\bibitem{Flynn:2003vz}%
  \BibitemOpen
  \bibfield{author}{%
  \bibinfo {author} {\bibfnamefont{J.}~\bibnamefont{Flynn}}, \bibinfo {author}
  {\bibfnamefont{F.}~\bibnamefont{Mescia}},\ and\ \bibinfo {author}
  {\bibfnamefont{A.~S.~B.}\ \bibnamefont{Tariq}} (\bibinfo {collaboration}
  {UKQCD}),\ }%
  \bibfield{journal}{%
  \Doi{10.1088/1126-6708/2003/07/066}{\bibinfo {journal} {JHEP}}\ }%
  \textbf{\bibinfo {volume} {0307}},\ \bibinfo {pages} {066} (\bibinfo {year}
  {2003}),\ \Eprint{http://arxiv.org/abs/hep-lat/0307025}{arXiv:hep-lat/0307025
  [hep-lat]}%
  \bibAnnoteFile{NoStop}{Flynn:2003vz}%
%%CITATION = HEP-LAT/0307025;%%
\bibitem{Mathur:2002ce}%
  \BibitemOpen
  \bibfield{author}{%
  \bibinfo {author} {\bibfnamefont{N.}~\bibnamefont{Mathur}}, \bibinfo {author}
  {\bibfnamefont{R.}~\bibnamefont{Lewis}},\ and\ \bibinfo {author}
  {\bibfnamefont{R.}~\bibnamefont{Woloshyn}},\ }%
  \bibfield{journal}{%
  \Doi{10.1103/PhysRevD.66.014502}{\bibinfo {journal} {Phys. Rev.}}\ }%
  \textbf{\bibinfo {volume} {D66}},\ \bibinfo {pages} {014502} (\bibinfo {year}
  {2002}),\ \Eprint{http://arxiv.org/abs/hep-ph/0203253}{arXiv:hep-ph/0203253
  [hep-ph]}%
  \bibAnnoteFile{NoStop}{Mathur:2002ce}%
%%CITATION = HEP-PH/0203253;%%
\bibitem{Lewis:2001iz}%
  \BibitemOpen
  \bibfield{author}{%
  \bibinfo {author} {\bibfnamefont{R.}~\bibnamefont{Lewis}}, \bibinfo {author}
  {\bibfnamefont{N.}~\bibnamefont{Mathur}},\ and\ \bibinfo {author}
  {\bibfnamefont{R.}~\bibnamefont{Woloshyn}},\ }%
  \bibfield{journal}{%
  \Doi{10.1103/PhysRevD.64.094509}{\bibinfo {journal} {Phys. Rev.}}\ }%
  \textbf{\bibinfo {volume} {D64}},\ \bibinfo {pages} {094509} (\bibinfo {year}
  {2001}),\ \Eprint{http://arxiv.org/abs/hep-ph/0107037}{arXiv:hep-ph/0107037
  [hep-ph]}%
  \bibAnnoteFile{NoStop}{Lewis:2001iz}%
%%CITATION = HEP-PH/0107037;%%
\bibitem{Chiu:2005zc}%
  \BibitemOpen
  \bibfield{author}{%
  \bibinfo {author} {\bibfnamefont{T.-W.}\ \bibnamefont{Chiu}}\ and\ \bibinfo
  {author} {\bibfnamefont{T.-H.}\ \bibnamefont{Hsieh}},\ }%
  \bibfield{journal}{%
  \Doi{10.1016/j.nuclphysa.2005.03.090}{\bibinfo {journal} {Nucl. Phys.}}\ }%
  \textbf{\bibinfo {volume} {A755}},\ \bibinfo {pages} {471} (\bibinfo {year}
  {2005}),\ \Eprint{http://arxiv.org/abs/hep-lat/0501021}{arXiv:hep-lat/0501021
  [hep-lat]}%
  \bibAnnoteFile{NoStop}{Chiu:2005zc}%
%%CITATION = HEP-LAT/0501021;%%
\bibitem{Na:2007pv}%
  \BibitemOpen
  \bibfield{author}{%
  \bibinfo {author} {\bibfnamefont{H.}~\bibnamefont{Na}}\ and\ \bibinfo
  {author} {\bibfnamefont{S.~A.}\ \bibnamefont{Gottlieb}},\ }%
  \bibfield{journal}{%
  \bibinfo {journal} {PoS}\ }%
  \textbf{\bibinfo {volume} {LAT2007}},\ \bibinfo {pages} {124} (\bibinfo
  {year} {2007})%
  \bibAnnoteFile{NoStop}{Na:2007pv}%
%%CITATION = ARXIV:0710.1422;%%
\bibitem{Liu:2009jc}%
  \BibitemOpen
  \bibfield{author}{%
  \bibinfo {author} {\bibfnamefont{L.}~\bibnamefont{Liu}}, \bibinfo {author}
  {\bibfnamefont{H.-W.}\ \bibnamefont{Lin}}, \bibinfo {author}
  {\bibfnamefont{K.}~\bibnamefont{Orginos}},\ and\ \bibinfo {author}
  {\bibfnamefont{A.}~\bibnamefont{Walker-Loud}},\ }%
  \bibfield{journal}{%
  \Doi{10.1103/PhysRevD.81.094505}{\bibinfo {journal} {Phys. Rev.}}\ }%
  \textbf{\bibinfo {volume} {D81}},\ \bibinfo {pages} {094505} (\bibinfo {year}
  {2010})%
  \bibAnnoteFile{NoStop}{Liu:2009jc}%
%%CITATION = ARXIV:0909.3294;%%
\bibitem{Briceno:2012wt}%
  \BibitemOpen
  \bibfield{author}{%
  \bibinfo {author} {\bibfnamefont{R.~A.}\ \bibnamefont{Brice{\~n}o}}, \bibinfo
  {author} {\bibfnamefont{H.-W.}\ \bibnamefont{Lin}},\ and\ \bibinfo {author}
  {\bibfnamefont{D.~R.}\ \bibnamefont{Bolton}},\ }%
  \bibfield{journal}{%
  \Doi{10.1103/PhysRevD.86.094504}{\bibinfo {journal} {Phys. Rev.}}\ }%
  \textbf{\bibinfo {volume} {D86}},\ \bibinfo {pages} {094504} (\bibinfo {year}
  {2012}),\ \Eprint{http://arxiv.org/abs/1207.3536}{arXiv:1207.3536 [hep-lat]}%
  \bibAnnoteFile{NoStop}{Briceno:2012wt}%
%%CITATION = ARXIV:1207.3536;%%
\bibitem{Alexandrou:2012xk}%
  \BibitemOpen
  \bibfield{author}{%
  \bibinfo {author} {\bibfnamefont{C.}~\bibnamefont{Alexandrou}}, \bibinfo
  {author} {\bibfnamefont{J.}~\bibnamefont{Carbonell}}, \bibinfo {author}
  {\bibfnamefont{D.}~\bibnamefont{Christaras}}, \bibinfo {author}
  {\bibfnamefont{V.}~\bibnamefont{Drach}}, \bibinfo {author}
  {\bibfnamefont{M.}~\bibnamefont{Gravina}},\ and\ \bibinfo {author}
  {\bibfnamefont{M.}~\bibnamefont{Papinutto}} (\bibinfo {collaboration}
  {ETMC}),\ }%
  \bibfield{journal}{%
  \Doi{10.1103/PhysRevD.86.114501}{\bibinfo {journal} {Phys. Rev.}}\ }%
  \textbf{\bibinfo {volume} {D86}},\ \bibinfo {pages} {114501} (\bibinfo {year}
  {2012}),\ \Eprint{http://arxiv.org/abs/1205.6856}{arXiv:1205.6856 [hep-lat]}%
  \bibAnnoteFile{NoStop}{Alexandrou:2012xk}%
%%CITATION = ARXIV:1205.6856;%%
\bibitem{Basak:2012py}%
  \BibitemOpen
  \bibfield{author}{%
  \bibinfo {author} {\bibfnamefont{S.}~\bibnamefont{Basak}}, \bibinfo {author}
  {\bibfnamefont{S.}~\bibnamefont{Datta}}, \bibinfo {author}
  {\bibfnamefont{M.}~\bibnamefont{Padmanath}}, \bibinfo {author}
  {\bibfnamefont{P.}~\bibnamefont{Majumdar}},\ and\ \bibinfo {author}
  {\bibfnamefont{N.}~\bibnamefont{Mathur}} (\bibinfo {collaboration} {ILGTI}),\
  }%
  \bibfield{journal}{%
  \bibinfo {journal} {PoS}\ }%
  \textbf{\bibinfo {volume} {LATTICE2012}},\ \bibinfo {pages} {141} (\bibinfo
  {year} {2012}),\ \Eprint{http://arxiv.org/abs/1211.6277}{arXiv:1211.6277
  [hep-lat]}%
  \bibAnnoteFile{NoStop}{Basak:2012py}%
%%CITATION = ARXIV:1211.6277;%%
\bibitem{Durr:2012dw}%
  \BibitemOpen
  \bibfield{author}{%
  \bibinfo {author} {\bibfnamefont{S.}~\bibnamefont{D{\"u}rr}}, \bibinfo
  {author} {\bibfnamefont{G.}~\bibnamefont{Koutsou}},\ and\ \bibinfo {author}
  {\bibfnamefont{T.}~\bibnamefont{Lippert}},\ }%
  \bibfield{journal}{%
  \Doi{10.1103/PhysRevD.86.114514}{\bibinfo {journal} {Phys. Rev.}}\ }%
  \textbf{\bibinfo {volume} {D86}},\ \bibinfo {pages} {114514} (\bibinfo {year}
  {2012}),\ \Eprint{http://arxiv.org/abs/1208.6270}{arXiv:1208.6270 [hep-lat]}%
  \bibAnnoteFile{NoStop}{Durr:2012dw}%
%%CITATION = ARXIV:1208.6270;%%
\bibitem{Namekawa:2013vu}%
  \BibitemOpen
  \bibfield{author}{%
  \bibinfo {author} {\bibfnamefont{Y.}~\bibnamefont{Namekawa}} \emph{et~al.}
  (\bibinfo {collaboration} {PACS-CS}),\ }%
  \bibfield{journal}{%
  \Doi{10.1103/PhysRevD.87.094512}{\bibinfo {journal} {Phys. Rev.}}\ }%
  \textbf{\bibinfo {volume} {D87}},\ \bibinfo {pages} {094512} (\bibinfo {year}
  {2013}),\ \Eprint{http://arxiv.org/abs/1301.4743}{arXiv:1301.4743 [hep-lat]}%
  \bibAnnoteFile{NoStop}{Namekawa:2013vu}%
%%CITATION = ARXIV:1301.4743;%%
\bibitem{Alexandrou:2014sha}%
  \BibitemOpen
  \bibfield{author}{%
  \bibinfo {author} {\bibfnamefont{C.}~\bibnamefont{Alexandrou}}, \bibinfo
  {author} {\bibfnamefont{V.}~\bibnamefont{Drach}}, \bibinfo {author}
  {\bibfnamefont{K.}~\bibnamefont{Jansen}}, \bibinfo {author}
  {\bibfnamefont{C.}~\bibnamefont{Kallidonis}},\ and\ \bibinfo {author}
  {\bibfnamefont{G.}~\bibnamefont{Koutsou}} (\bibinfo {collaboration} {ETMC}),\
  }%
  \bibfield{journal}{%
  \Doi{10.1103/PhysRevD.90.074501}{\bibinfo {journal} {Phys. Rev.}}\ }%
  \textbf{\bibinfo {volume} {D90}},\ \bibinfo {pages} {074501} (\bibinfo {year}
  {2014}),\ \Eprint{http://arxiv.org/abs/1406.4310}{arXiv:1406.4310 [hep-lat]}%
  \bibAnnoteFile{NoStop}{Alexandrou:2014sha}%
%%CITATION = ARXIV:1406.4310;%%
\bibitem{Brown:2014ena}%
  \BibitemOpen
  \bibfield{author}{%
  \bibinfo {author} {\bibfnamefont{Z.~S.}\ \bibnamefont{Brown}}, \bibinfo
  {author} {\bibfnamefont{W.}~\bibnamefont{Detmold}}, \bibinfo {author}
  {\bibfnamefont{S.}~\bibnamefont{Meinel}},\ and\ \bibinfo {author}
  {\bibfnamefont{K.}~\bibnamefont{Orginos}},\ }%
  \bibfield{journal}{%
  \Doi{10.1103/PhysRevD.90.094507}{\bibinfo {journal} {Phys. Rev.}}\ }%
  \textbf{\bibinfo {volume} {D90}},\ \bibinfo {pages} {094507} (\bibinfo {year}
  {2014}),\ \Eprint{http://arxiv.org/abs/1409.0497}{arXiv:1409.0497 [hep-lat]}%
  \bibAnnoteFile{NoStop}{Brown:2014ena}%
%%CITATION = ARXIV:1409.0497;%%
\bibitem{Padmanath:2015jea}%
  \BibitemOpen
  \bibfield{author}{%
  \bibinfo {author} {\bibfnamefont{M.}~\bibnamefont{Padmanath}}, \bibinfo
  {author} {\bibfnamefont{R.~G.}\ \bibnamefont{Edwards}}, \bibinfo {author}
  {\bibfnamefont{N.}~\bibnamefont{Mathur}},\ and\ \bibinfo {author}
  {\bibfnamefont{M.}~\bibnamefont{Peardon}} (\bibinfo {collaboration} {HSC})}%
   (\bibinfo {year} {2015}),\
  \Eprint{http://arxiv.org/abs/1502.01845}{arXiv:1502.01845 [hep-lat]}%
  \bibAnnoteFile{NoStop}{Padmanath:2015jea}%
%%CITATION = ARXIV:1502.01845;%%
\bibitem{GellMann:1962xb}%
  \BibitemOpen
  \bibfield{author}{%
  \bibinfo {author} {\bibfnamefont{M.}~\bibnamefont{Gell-Mann}},\ }%
  \bibfield{journal}{%
  \Doi{10.1103/PhysRev.125.1067}{\bibinfo {journal} {Phys. Rev.}}\ }%
  \textbf{\bibinfo {volume} {125}},\ \bibinfo {pages} {1067} (\bibinfo {year}
  {1962})%
  \bibAnnoteFile{NoStop}{GellMann:1962xb}%
%%CITATION = PHRVA,125,1067;%%
\bibitem{Okubo:1961jc}%
  \BibitemOpen
  \bibfield{author}{%
  \bibinfo {author} {\bibfnamefont{S.}~\bibnamefont{Okubo}},\ }%
  \bibfield{journal}{%
  \Doi{10.1143/PTP.27.949}{\bibinfo {journal} {Prog. Theor. Phys.}}\ }%
  \textbf{\bibinfo {volume} {27}},\ \bibinfo {pages} {949} (\bibinfo {year}
  {1962})%
  \bibAnnoteFile{NoStop}{Okubo:1961jc}%
%%CITATION = PTPKA,27,949;%%
\bibitem{Durr:2014oba}%
  \BibitemOpen
  \bibfield{author}{%
  \bibinfo {author} {\bibfnamefont{S.}~\bibnamefont{D{\"u}rr}}}%
   (\bibinfo {year} {2014}),\
  \Eprint{http://arxiv.org/abs/1412.6434}{arXiv:1412.6434 [hep-lat]}%
  \bibAnnoteFile{NoStop}{Durr:2014oba}%
%%CITATION = ARXIV:1412.6434;%%
\bibitem{Bietenholz:2010jr}%
  \BibitemOpen
  \bibfield{author}{%
  \bibinfo {author} {\bibfnamefont{W.}~\bibnamefont{Bietenholz}} \emph{et~al.}
  (\bibinfo {collaboration} {QCDSF}),\ }%
  \bibfield{journal}{%
  \Doi{10.1016/j.physletb.2010.05.067}{\bibinfo {journal} {Phys. Lett.}}\ }%
  \textbf{\bibinfo {volume} {B690}},\ \bibinfo {pages} {436} (\bibinfo {year}
  {2010})%
  \bibAnnoteFile{NoStop}{Bietenholz:2010jr}%
\bibitem{Bietenholz:2011qq}%
  \BibitemOpen
  \bibfield{author}{%
  \bibinfo {author} {\bibfnamefont{W.}~\bibnamefont{Bietenholz}} \emph{et~al.}
  (\bibinfo {collaboration} {QCDSF}),\ }%
  \bibfield{journal}{%
  \Doi{10.1103/PhysRevD.84.054509}{\bibinfo {journal} {Phys. Rev.}}\ }%
  \textbf{\bibinfo {volume} {D84}},\ \bibinfo {pages} {054509} (\bibinfo {year}
  {2011})%
  \bibAnnoteFile{NoStop}{Bietenholz:2011qq}%
%%CITATION = ARXIV:1102.5300;%%
\bibitem{Bruno:2014jqa}%
  \BibitemOpen
  \bibfield{author}{%
  \bibinfo {author} {\bibfnamefont{M.}~\bibnamefont{Bruno}} \emph{et~al.},\ }%
  \bibfield{journal}{%
  \Doi{10.1007/JHEP02(2015)043}{\bibinfo {journal} {JHEP}}\ }%
  \textbf{\bibinfo {volume} {1502}},\ \bibinfo {pages} {043} (\bibinfo {year}
  {2015}),\ \Eprint{http://arxiv.org/abs/1411.3982}{arXiv:1411.3982 [hep-lat]}%
  \bibAnnoteFile{NoStop}{Bruno:2014jqa}%
%%CITATION = ARXIV:1411.3982;%%
\bibitem{Cundy:2009yy}%
  \BibitemOpen
  \bibfield{author}{%
  \bibinfo {author} {\bibfnamefont{N.}~\bibnamefont{Cundy}} \emph{et~al.},\ }%
  \bibfield{journal}{%
  \Doi{10.1103/PhysRevD.79.094507}{\bibinfo {journal} {Phys. Rev.}}\ }%
  \textbf{\bibinfo {volume} {D79}},\ \bibinfo {pages} {094507} (\bibinfo {year}
  {2009})%
  \bibAnnoteFile{NoStop}{Cundy:2009yy}%
\bibitem{Sommer:1993ce}%
  \BibitemOpen
  \bibfield{author}{%
  \bibinfo {author} {\bibfnamefont{R.}~\bibnamefont{Sommer}},\ }%
  \bibfield{journal}{%
  \Doi{10.1016/0550-3213(94)90473-1}{\bibinfo {journal} {Nucl.Phys.}}\ }%
  \textbf{\bibinfo {volume} {B411}},\ \bibinfo {pages} {839} (\bibinfo {year}
  {1994}),\ \Eprint{http://arxiv.org/abs/hep-lat/9310022}{arXiv:hep-lat/9310022
  [hep-lat]}%
  \bibAnnoteFile{NoStop}{Sommer:1993ce}%
%%CITATION = HEP-LAT/9310022;%%
\bibitem{Borsanyi:2012zs}%
  \BibitemOpen
  \bibfield{author}{%
  \bibinfo {author} {\bibfnamefont{S.}~\bibnamefont{Borsanyi}} \emph{et~al.},\
  }%
  \bibfield{journal}{%
  \Doi{10.1007/JHEP09(2012)010}{\bibinfo {journal} {JHEP}}\ }%
  \textbf{\bibinfo {volume} {1209}},\ \bibinfo {pages} {010} (\bibinfo {year}
  {2012}),\ \Eprint{http://arxiv.org/abs/1203.4469}{arXiv:1203.4469 [hep-lat]}%
  \bibAnnoteFile{NoStop}{Borsanyi:2012zs}%
%%CITATION = ARXIV:1203.4469;%%
\bibitem{Luscher:2010iy}%
  \BibitemOpen
  \bibfield{author}{%
  \bibinfo {author} {\bibfnamefont{M.}~\bibnamefont{L{\"u}scher}},\ }%
  \bibfield{journal}{%
  \Doi{10.1007/JHEP08(2010)071}{\bibinfo {journal} {JHEP}}\ }%
  \textbf{\bibinfo {volume} {1008}},\ \bibinfo {pages} {071} (\bibinfo {year}
  {2010}),\ \Eprint{http://arxiv.org/abs/1006.4518}{arXiv:1006.4518 [hep-lat]}%
  \bibAnnoteFile{NoStop}{Luscher:2010iy}%
%%CITATION = ARXIV:1006.4518;%%
\bibitem{Horsley:2013wqa}%
  \BibitemOpen
  \bibfield{author}{%
  \bibinfo {author} {\bibfnamefont{R.}~\bibnamefont{Horsley}}, \bibinfo
  {author} {\bibfnamefont{J.}~\bibnamefont{Najjar}}, \bibinfo {author}
  {\bibfnamefont{Y.}~\bibnamefont{Nakamura}}, \bibinfo {author}
  {\bibfnamefont{H.}~\bibnamefont{Perlt}}, \bibinfo {author}
  {\bibfnamefont{D.}~\bibnamefont{Pleiter}}, \bibinfo {author}
  {\bibfnamefont{P.~E.~L.}\ \bibnamefont{Rakow}}, \bibinfo {author}
  {\bibfnamefont{G.}~\bibnamefont{Schierholz}}, \bibinfo {author}
  {\bibfnamefont{A.}~\bibnamefont{Schiller}}, \bibinfo {author}
  {\bibfnamefont{H.}~\bibnamefont{St{\"u}ben}},\ and\ \bibinfo {author}
  {\bibfnamefont{J.~M.}\ \bibnamefont{Zanotti}} (\bibinfo {collaboration}
  {QCDSF}),\ }%
  \bibfield{journal}{%
  \bibinfo {journal} {PoS}\ }%
  \textbf{\bibinfo {volume} {LATTICE2013}},\ \bibinfo {pages} {249} (\bibinfo
  {year} {2013}),\ \Eprint{http://arxiv.org/abs/1311.5010}{arXiv:1311.5010
  [hep-lat]}%
  \bibAnnoteFile{NoStop}{Horsley:2013wqa}%
%%CITATION = ARXIV:1311.5010;%%
\bibitem{Bali:1998pi}%
  \BibitemOpen
  \bibfield{author}{%
  \bibinfo {author} {\bibfnamefont{G.~S.}\ \bibnamefont{Bali}}\ and\ \bibinfo
  {author} {\bibfnamefont{P.}~\bibnamefont{Boyle}},\ }%
  \bibfield{journal}{%
  \Doi{10.1103/PhysRevD.59.114504}{\bibinfo {journal} {Phys.~Rev.}}\ }%
  \textbf{\bibinfo {volume} {D59}},\ \bibinfo {pages} {114504} (\bibinfo {year}
  {1999}),\ \Eprint{http://arxiv.org/abs/hep-lat/9809180}{arXiv:hep-lat/9809180
  [hep-lat]}%
  \bibAnnoteFile{NoStop}{Bali:1998pi}%
%%CITATION = HEP-LAT/9809180;%%
\bibitem{Nowak:1992um}%
  \BibitemOpen
  \bibfield{author}{%
  \bibinfo {author} {\bibfnamefont{M.~A.}\ \bibnamefont{Nowak}}, \bibinfo
  {author} {\bibfnamefont{M.}~\bibnamefont{Rho}},\ and\ \bibinfo {author}
  {\bibfnamefont{I.}~\bibnamefont{Zahed}},\ }%
  \bibfield{journal}{%
  \Doi{10.1103/PhysRevD.48.4370}{\bibinfo {journal} {Phys.~Rev.}}\ }%
  \textbf{\bibinfo {volume} {D48}},\ \bibinfo {pages} {4370} (\bibinfo {year}
  {1993}),\ \Eprint{http://arxiv.org/abs/hep-ph/9209272}{arXiv:hep-ph/9209272
  [hep-ph]}%
  \bibAnnoteFile{NoStop}{Nowak:1992um}%
%%CITATION = HEP-PH/9209272;%%
\bibitem{Bardeen:1993ae}%
  \BibitemOpen
  \bibfield{author}{%
  \bibinfo {author} {\bibfnamefont{W.~A.}\ \bibnamefont{Bardeen}}\ and\
  \bibinfo {author} {\bibfnamefont{C.~T.}\ \bibnamefont{Hill}},\ }%
  \bibfield{journal}{%
  \Doi{10.1103/PhysRevD.49.409}{\bibinfo {journal} {Phys.~Rev.}}\ }%
  \textbf{\bibinfo {volume} {D49}},\ \bibinfo {pages} {409} (\bibinfo {year}
  {1994}),\ \Eprint{http://arxiv.org/abs/hep-ph/9304265}{arXiv:hep-ph/9304265
  [hep-ph]}%
  \bibAnnoteFile{NoStop}{Bardeen:1993ae}%
%%CITATION = HEP-PH/9304265;%%
\bibitem{Ebert:1994tv}%
  \BibitemOpen
  \bibfield{author}{%
  \bibinfo {author} {\bibfnamefont{D.}~\bibnamefont{Ebert}}, \bibinfo {author}
  {\bibfnamefont{T.}~\bibnamefont{Feldmann}}, \bibinfo {author}
  {\bibfnamefont{R.}~\bibnamefont{Friedrich}},\ and\ \bibinfo {author}
  {\bibfnamefont{H.}~\bibnamefont{Reinhardt}},\ }%
  \bibfield{journal}{%
  \Doi{10.1016/0550-3213(94)00456-O}{\bibinfo {journal} {Nucl.~Phys.}}\ }%
  \textbf{\bibinfo {volume} {B434}},\ \bibinfo {pages} {619} (\bibinfo {year}
  {1995}),\ \Eprint{http://arxiv.org/abs/hep-ph/9406220}{arXiv:hep-ph/9406220
  [hep-ph]}%
  \bibAnnoteFile{NoStop}{Ebert:1994tv}%
%%CITATION = HEP-PH/9406220;%%
\bibitem{Mohler:2012na}%
  \BibitemOpen
  \bibfield{author}{%
  \bibinfo {author} {\bibfnamefont{D.}~\bibnamefont{Mohler}}, \bibinfo {author}
  {\bibfnamefont{S.}~\bibnamefont{Prelovsek}},\ and\ \bibinfo {author}
  {\bibfnamefont{R.}~\bibnamefont{Woloshyn}},\ }%
  \bibfield{journal}{%
  \Doi{10.1103/PhysRevD.87.034501}{\bibinfo {journal} {Phys.~Rev.}}\ }%
  \textbf{\bibinfo {volume} {D87}},\ \bibinfo {pages} {034501} (\bibinfo {year}
  {2013}),\ \Eprint{http://arxiv.org/abs/1208.4059}{arXiv:1208.4059 [hep-lat]}%
  \bibAnnoteFile{NoStop}{Mohler:2012na}%
%%CITATION = ARXIV:1208.4059;%%
\bibitem{Lang:2014yfa}%
  \BibitemOpen
  \bibfield{author}{%
  \bibinfo {author} {\bibfnamefont{C.}~\bibnamefont{Lang}}, \bibinfo {author}
  {\bibfnamefont{L.}~\bibnamefont{Leskovec}}, \bibinfo {author}
  {\bibfnamefont{D.}~\bibnamefont{Mohler}}, \bibinfo {author}
  {\bibfnamefont{S.}~\bibnamefont{Prelovsek}},\ and\ \bibinfo {author}
  {\bibfnamefont{R.}~\bibnamefont{Woloshyn}},\ }%
  \bibfield{journal}{%
  \Doi{10.1103/PhysRevD.90.034510}{\bibinfo {journal} {Phys.~Rev.}}\ }%
  \textbf{\bibinfo {volume} {D90}},\ \bibinfo {pages} {034510} (\bibinfo {year}
  {2014}),\ \Eprint{http://arxiv.org/abs/1403.8103}{arXiv:1403.8103 [hep-lat]}%
  \bibAnnoteFile{NoStop}{Lang:2014yfa}%
%%CITATION = ARXIV:1403.8103;%%
\bibitem{Crede:2013kia}%
  \BibitemOpen
  \bibfield{author}{%
  \bibinfo {author} {\bibfnamefont{V.}~\bibnamefont{Crede}}\ and\ \bibinfo
  {author} {\bibfnamefont{W.}~\bibnamefont{Roberts}},\ }%
  \bibfield{journal}{%
  \Doi{10.1088/0034-4885/76/7/076301}{\bibinfo {journal} {Rept. Prog. Phys.}}\
  }%
  \textbf{\bibinfo {volume} {76}},\ \bibinfo {pages} {076301} (\bibinfo {year}
  {2013}),\ \Eprint{http://arxiv.org/abs/1302.7299}{arXiv:1302.7299 [nucl-ex]}%
  \bibAnnoteFile{NoStop}{Crede:2013kia}%
%%CITATION = ARXIV:1302.7299;%%
\bibitem{Brambilla:2005yk}%
  \BibitemOpen
  \bibfield{author}{%
  \bibinfo {author} {\bibfnamefont{N.}~\bibnamefont{Brambilla}}, \bibinfo
  {author} {\bibfnamefont{A.}~\bibnamefont{Vairo}},\ and\ \bibinfo {author}
  {\bibfnamefont{T.}~\bibnamefont{Rosch}},\ }%
  \bibfield{journal}{%
  \Doi{10.1103/PhysRevD.72.034021}{\bibinfo {journal} {Phys. Rev.}}\ }%
  \textbf{\bibinfo {volume} {D72}},\ \bibinfo {pages} {034021} (\bibinfo {year}
  {2005}),\ \Eprint{http://arxiv.org/abs/hep-ph/0506065}{arXiv:hep-ph/0506065
  [hep-ph]}%
  \bibAnnoteFile{NoStop}{Brambilla:2005yk}%
%%CITATION = HEP-PH/0506065;%%
\bibitem{Michael:1985ne}%
  \BibitemOpen
  \bibfield{author}{%
  \bibinfo {author} {\bibfnamefont{C.}~\bibnamefont{Michael}},\ }%
  \bibfield{journal}{%
  \Doi{10.1016/0550-3213(85)90297-4}{\bibinfo {journal} {Nucl. Phys.}}\ }%
  \textbf{\bibinfo {volume} {B259}},\ \bibinfo {pages} {58} (\bibinfo {year}
  {1985})%
  \bibAnnoteFile{NoStop}{Michael:1985ne}%
\bibitem{Luscher:1990ck}%
  \BibitemOpen
  \bibfield{author}{%
  \bibinfo {author} {\bibfnamefont{M.}~\bibnamefont{L{\"u}scher}}\ and\
  \bibinfo {author} {\bibfnamefont{U.}~\bibnamefont{Wolff}},\ }%
  \bibfield{journal}{%
  \Doi{10.1016/0550-3213(90)90540-T}{\bibinfo {journal} {Nucl. Phys.}}\ }%
  \textbf{\bibinfo {volume} {B339}},\ \bibinfo {pages} {222} (\bibinfo {year}
  {1990})%
  \bibAnnoteFile{NoStop}{Luscher:1990ck}%
\bibitem{Blossier:2009kd}%
  \BibitemOpen
  \bibfield{author}{%
  \bibinfo {author} {\bibfnamefont{B.}~\bibnamefont{Blossier}}, \bibinfo
  {author} {\bibfnamefont{M.}~\bibnamefont{Della~Morte}}, \bibinfo {author}
  {\bibfnamefont{G.}~\bibnamefont{von Hippel}}, \bibinfo {author}
  {\bibfnamefont{T.}~\bibnamefont{Mendes}},\ and\ \bibinfo {author}
  {\bibfnamefont{R.}~\bibnamefont{Sommer}},\ }%
  \bibfield{journal}{%
  \Doi{10.1088/1126-6708/2009/04/094}{\bibinfo {journal} {JHEP}}\ }%
  \textbf{\bibinfo {volume} {0904}},\ \bibinfo {pages} {094} (\bibinfo {year}
  {2009}),\ \Eprint{http://arxiv.org/abs/0902.1265}{arXiv:0902.1265 [hep-lat]}%
  \bibAnnoteFile{NoStop}{Blossier:2009kd}%
%%CITATION = ARXIV:0902.1265;%%
\bibitem{Gusken:1989ad}%
  \BibitemOpen
  \bibfield{author}{%
  \bibinfo {author} {\bibfnamefont{S.}~\bibnamefont{G{\"u}sken}}, \bibinfo
  {author} {\bibfnamefont{U.}~\bibnamefont{L{\"o}w}}, \bibinfo {author}
  {\bibfnamefont{K.}~\bibnamefont{M{\"u}tter}}, \bibinfo {author}
  {\bibfnamefont{R.}~\bibnamefont{Sommer}}, \bibinfo {author}
  {\bibfnamefont{A.}~\bibnamefont{Patel}},\ and\ \bibinfo {author}
  {\bibfnamefont{K.}~\bibnamefont{Schilling}},\ }%
  \bibfield{journal}{%
  \Doi{10.1016/S0370-2693(89)80034-6}{\bibinfo {journal} {Phys. Lett.}}\ }%
  \textbf{\bibinfo {volume} {B227}},\ \bibinfo {pages} {266} (\bibinfo {year}
  {1989})%
  \bibAnnoteFile{NoStop}{Gusken:1989ad}%
\bibitem{Gusken:1989qx}%
  \BibitemOpen
  \bibfield{author}{%
  \bibinfo {author} {\bibfnamefont{S.}~\bibnamefont{G{\"u}sken}},\ }%
  \bibfield{journal}{%
  \Doi{10.1016/0920-5632(90)90273-W}{\bibinfo {journal} {Nucl. Phys. Proc.
  Suppl.}}\ }%
  \textbf{\bibinfo {volume} {17}},\ \bibinfo {pages} {361} (\bibinfo {year}
  {1990})%
  \bibAnnoteFile{NoStop}{Gusken:1989qx}%
\bibitem{Falcioni:1984ei}%
  \BibitemOpen
  \bibfield{author}{%
  \bibinfo {author} {\bibfnamefont{M.}~\bibnamefont{Falcioni}}, \bibinfo
  {author} {\bibfnamefont{M.}~\bibnamefont{Paciello}}, \bibinfo {author}
  {\bibfnamefont{G.}~\bibnamefont{Parisi}},\ and\ \bibinfo {author}
  {\bibfnamefont{B.}~\bibnamefont{Taglienti}},\ }%
  \bibfield{journal}{%
  \Doi{10.1016/0550-3213(85)90280-9}{\bibinfo {journal} {Nucl. Phys.}}\ }%
  \textbf{\bibinfo {volume} {B251}},\ \bibinfo {pages} {624} (\bibinfo {year}
  {1985})%
  \bibAnnoteFile{NoStop}{Falcioni:1984ei}%
\bibitem{Albanese:1987ds}%
  \BibitemOpen
  \bibfield{author}{%
  \bibinfo {author} {\bibfnamefont{M.}~\bibnamefont{Albanese}} \emph{et~al.}
  (\bibinfo {collaboration} {APE}),\ }%
  \bibfield{journal}{%
  \Doi{10.1016/0370-2693(87)91160-9}{\bibinfo {journal} {Phys. Lett.}}\ }%
  \textbf{\bibinfo {volume} {B192}},\ \bibinfo {pages} {163} (\bibinfo {year}
  {1987})%
  \bibAnnoteFile{NoStop}{Albanese:1987ds}%
\bibitem{Wolff:2003sm}%
  \BibitemOpen
  \bibfield{author}{%
  \bibinfo {author} {\bibfnamefont{U.}~\bibnamefont{Wolff}} (\bibinfo
  {collaboration} {ALPHA}),\ }%
  \bibfield{journal}{%
  \Doi{10.1016/S0010-4655(03)00467-3, 10.1016/j.cpc.2006.12.001}{\bibinfo
  {journal} {Comput. Phys. Commun.}}\ }%
  \textbf{\bibinfo {volume} {156}},\ \bibinfo {pages} {143} (\bibinfo {year}
  {2004}),\ \Eprint{http://arxiv.org/abs/hep-lat/0306017}{arXiv:hep-lat/0306017
  [hep-lat]}%
  \bibAnnoteFile{NoStop}{Wolff:2003sm}%
%%CITATION = HEP-LAT/0306017;%%
\bibitem{Schaefer:2010hu}%
  \BibitemOpen
  \bibfield{author}{%
  \bibinfo {author} {\bibfnamefont{S.}~\bibnamefont{Sch{\"a}fer}}, \bibinfo
  {author} {\bibfnamefont{R.}~\bibnamefont{Sommer}},\ and\ \bibinfo {author}
  {\bibfnamefont{F.}~\bibnamefont{Virotta}} (\bibinfo {collaboration}
  {ALPHA}),\ }%
  \bibfield{journal}{%
  \Doi{10.1016/j.nuclphysb.2010.11.020}{\bibinfo {journal} {Nucl. Phys.}}\ }%
  \textbf{\bibinfo {volume} {B845}},\ \bibinfo {pages} {93} (\bibinfo {year}
  {2011}),\ \Eprint{http://arxiv.org/abs/1009.5228}{arXiv:1009.5228 [hep-lat]}%
  \bibAnnoteFile{NoStop}{Schaefer:2010hu}%
%%CITATION = ARXIV:1009.5228;%%
\bibitem{ElKhadra:1996mp}%
  \BibitemOpen
  \bibfield{author}{%
  \bibinfo {author} {\bibfnamefont{A.~X.}\ \bibnamefont{El-Khadra}}, \bibinfo
  {author} {\bibfnamefont{A.~S.}\ \bibnamefont{Kronfeld}},\ and\ \bibinfo
  {author} {\bibfnamefont{P.~B.}\ \bibnamefont{Mackenzie}},\ }%
  \bibfield{journal}{%
  \Doi{10.1103/PhysRevD.55.3933}{\bibinfo {journal} {Phys. Rev.}}\ }%
  \textbf{\bibinfo {volume} {D55}},\ \bibinfo {pages} {3933} (\bibinfo {year}
  {1997}),\ \Eprint{http://arxiv.org/abs/hep-lat/9604004}{arXiv:hep-lat/9604004
  [hep-lat]}%
  \bibAnnoteFile{NoStop}{ElKhadra:1996mp}%
%%CITATION = HEP-LAT/9604004;%%
\bibitem{Aoki:2001ra}%
  \BibitemOpen
  \bibfield{author}{%
  \bibinfo {author} {\bibfnamefont{S.}~\bibnamefont{Aoki}}, \bibinfo {author}
  {\bibfnamefont{Y.}~\bibnamefont{Kuramashi}},\ and\ \bibinfo {author}
  {\bibfnamefont{S.-i.}\ \bibnamefont{Tominaga}},\ }%
  \bibfield{journal}{%
  \Doi{10.1143/PTP.109.383}{\bibinfo {journal} {Prog. Theor. Phys.}}\ }%
  \textbf{\bibinfo {volume} {109}},\ \bibinfo {pages} {383} (\bibinfo {year}
  {2003}),\ \Eprint{http://arxiv.org/abs/hep-lat/0107009}{arXiv:hep-lat/0107009
  [hep-lat]}%
  \bibAnnoteFile{NoStop}{Aoki:2001ra}%
%%CITATION = HEP-LAT/0107009;%%
\bibitem{Christ:2006us}%
  \BibitemOpen
  \bibfield{author}{%
  \bibinfo {author} {\bibfnamefont{N.~H.}\ \bibnamefont{Christ}}, \bibinfo
  {author} {\bibfnamefont{M.}~\bibnamefont{Li}},\ and\ \bibinfo {author}
  {\bibfnamefont{H.-W.}\ \bibnamefont{Lin}},\ }%
  \bibfield{journal}{%
  \Doi{10.1103/PhysRevD.76.074505}{\bibinfo {journal} {Phys. Rev.}}\ }%
  \textbf{\bibinfo {volume} {D76}},\ \bibinfo {pages} {074505} (\bibinfo {year}
  {2007}),\ \Eprint{http://arxiv.org/abs/hep-lat/0608006}{arXiv:hep-lat/0608006
  [hep-lat]}%
  \bibAnnoteFile{NoStop}{Christ:2006us}%
%%CITATION = HEP-LAT/0608006;%%
\bibitem{Frezzotti:2000nk}%
  \BibitemOpen
  \bibfield{author}{%
  \bibinfo {author} {\bibfnamefont{R.}~\bibnamefont{Frezzotti}}, \bibinfo
  {author} {\bibfnamefont{P.~A.}\ \bibnamefont{Grassi}}, \bibinfo {author}
  {\bibfnamefont{S.}~\bibnamefont{Sint}},\ and\ \bibinfo {author}
  {\bibfnamefont{P.}~\bibnamefont{Weisz}} (\bibinfo {collaboration} {ALPHA}),\
  }%
  \bibfield{journal}{%
  \bibinfo {journal} {JHEP}\ }%
  \textbf{\bibinfo {volume} {0108}},\ \bibinfo {pages} {058} (\bibinfo {year}
  {2001}),\ \Eprint{http://arxiv.org/abs/hep-lat/0101001}{arXiv:hep-lat/0101001
  [hep-lat]}%
  \bibAnnoteFile{NoStop}{Frezzotti:2000nk}%
%%CITATION = HEP-LAT/0101001;%%
\bibitem{Liu:2012ze}%
  \BibitemOpen
  \bibfield{author}{%
  \bibinfo {author} {\bibfnamefont{L.}~\bibnamefont{Liu}}, \bibinfo {author}
  {\bibfnamefont{G.}~\bibnamefont{Moir}}, \bibinfo {author}
  {\bibfnamefont{M.}~\bibnamefont{Peardon}}, \bibinfo {author}
  {\bibfnamefont{S.~M.}\ \bibnamefont{Ryan}}, \bibinfo {author}
  {\bibfnamefont{C.~E.}\ \bibnamefont{Thomas}}, \bibinfo {author}
  {\bibfnamefont{P.}~\bibnamefont{Vilaseca}}, \bibinfo {author}
  {\bibfnamefont{J.~J.}\ \bibnamefont{Dudek}}, \bibinfo {author}
  {\bibfnamefont{R.~G.}\ \bibnamefont{Edwards}}, \bibinfo {author}
  {\bibfnamefont{B.}~\bibnamefont{Joo}},\ and\ \bibinfo {author}
  {\bibfnamefont{D.~G.}\ \bibnamefont{Richards}} (\bibinfo {collaboration}
  {HSC}),\ }%
  \bibfield{journal}{%
  \Doi{10.1007/JHEP07(2012)126}{\bibinfo {journal} {JHEP}}\ }%
  \textbf{\bibinfo {volume} {1207}},\ \bibinfo {pages} {126} (\bibinfo {year}
  {2012}),\ \Eprint{http://arxiv.org/abs/1204.5425}{arXiv:1204.5425 [hep-ph]}%
  \bibAnnoteFile{NoStop}{Liu:2012ze}%
%%CITATION = ARXIV:1204.5425;%%
\bibitem{Neuberger:1997fp}%
  \BibitemOpen
  \bibfield{author}{%
  \bibinfo {author} {\bibfnamefont{H.}~\bibnamefont{Neuberger}},\ }%
  \bibfield{journal}{%
  \Doi{10.1016/S0370-2693(97)01368-3}{\bibinfo {journal} {Phys. Lett.}}\ }%
  \textbf{\bibinfo {volume} {B417}},\ \bibinfo {pages} {141} (\bibinfo {year}
  {1998}),\ \Eprint{http://arxiv.org/abs/hep-lat/9707022}{arXiv:hep-lat/9707022
  [hep-lat]}%
  \bibAnnoteFile{NoStop}{Neuberger:1997fp}%
%%CITATION = HEP-LAT/9707022;%%
\bibitem{Bernard:2001av}%
  \BibitemOpen
  \bibfield{author}{%
  \bibinfo {author} {\bibfnamefont{C.}~\bibnamefont{Bernard}}, \bibinfo
  {author} {\bibfnamefont{T.}~\bibnamefont{Burch}}, \bibinfo {author}
  {\bibfnamefont{T.~A.}\ \bibnamefont{DeGrand}}, \bibinfo {author}
  {\bibfnamefont{S.}~\bibnamefont{Datta}}, \bibinfo {author}
  {\bibfnamefont{C.}~\bibnamefont{DeTar}}, \bibinfo {author}
  {\bibfnamefont{S.}~\bibnamefont{Gottlieb}}, \bibinfo {author}
  {\bibfnamefont{U.~M.}\ \bibnamefont{Heller}}, \bibinfo {author}
  {\bibfnamefont{K.}~\bibnamefont{Orginos}}, \bibinfo {author}
  {\bibfnamefont{R.}~\bibnamefont{Sugar}},\ and\ \bibinfo {author}
  {\bibfnamefont{D.}~\bibnamefont{Toussaint}},\ }%
  \bibfield{journal}{%
  \Doi{10.1103/PhysRevD.64.054506}{\bibinfo {journal} {Phys.~Rev.}}\ }%
  \textbf{\bibinfo {volume} {D64}},\ \bibinfo {pages} {054506} (\bibinfo {year}
  {2001}),\ \Eprint{http://arxiv.org/abs/hep-lat/0104002}{arXiv:hep-lat/0104002
  [hep-lat]}%
  \bibAnnoteFile{NoStop}{Bernard:2001av}%
%%CITATION = HEP-LAT/0104002;%%
\bibitem{Bazavov:2010ru}%
  \BibitemOpen
  \bibfield{author}{%
  \bibinfo {author} {\bibfnamefont{A.}~\bibnamefont{Bazavov}} \emph{et~al.}
  (\bibinfo {collaboration} {MILC}),\ }%
  \bibfield{journal}{%
  \Doi{10.1103/PhysRevD.82.074501}{\bibinfo {journal} {Phys.~Rev.}}\ }%
  \textbf{\bibinfo {volume} {D82}},\ \bibinfo {pages} {074501} (\bibinfo {year}
  {2010}),\ \Eprint{http://arxiv.org/abs/1004.0342}{arXiv:1004.0342 [hep-lat]}%
  \bibAnnoteFile{NoStop}{Bazavov:2010ru}%
%%CITATION = ARXIV:1004.0342;%%
\bibitem{Bietenholz:2010az}%
  \BibitemOpen
  \bibfield{author}{%
  \bibinfo {author} {\bibfnamefont{W.}~\bibnamefont{Bietenholz}}, \bibinfo
  {author} {\bibfnamefont{M.}~\bibnamefont{G{\"o}ckeler}}, \bibinfo {author}
  {\bibfnamefont{R.}~\bibnamefont{Horsley}}, \bibinfo {author}
  {\bibfnamefont{Y.}~\bibnamefont{Nakamura}}, \bibinfo {author}
  {\bibfnamefont{D.}~\bibnamefont{Pleiter}}, \bibinfo {author}
  {\bibfnamefont{P.~E.~L.}\ \bibnamefont{Rakow}}, \bibinfo {author}
  {\bibfnamefont{G.}~\bibnamefont{Schierholz}},\ and\ \bibinfo {author}
  {\bibfnamefont{J.~M.}\ \bibnamefont{Zanotti}},\ }%
  \bibfield{journal}{%
  \Doi{10.1016/j.physletb.2010.03.063}{\bibinfo {journal} {Phys.~Lett.}}\ }%
  \textbf{\bibinfo {volume} {B687}},\ \bibinfo {pages} {410} (\bibinfo {year}
  {2010}),\ \Eprint{http://arxiv.org/abs/1002.1696}{arXiv:1002.1696 [hep-lat]}%
  \bibAnnoteFile{NoStop}{Bietenholz:2010az}%
%%CITATION = ARXIV:1002.1696;%%
\bibitem{Aoki:2009ix}%
  \BibitemOpen
  \bibfield{author}{%
  \bibinfo {author} {\bibfnamefont{S.}~\bibnamefont{Aoki}} \emph{et~al.}
  (\bibinfo {collaboration} {PACS-CS}),\ }%
  \bibfield{journal}{%
  \Doi{10.1103/PhysRevD.81.074503}{\bibinfo {journal} {Phys.~Rev.}}\ }%
  \textbf{\bibinfo {volume} {D81}},\ \bibinfo {pages} {074503} (\bibinfo {year}
  {2010}),\ \Eprint{http://arxiv.org/abs/0911.2561}{arXiv:0911.2561 [hep-lat]}%
  \bibAnnoteFile{NoStop}{Aoki:2009ix}%
%%CITATION = ARXIV:0911.2561;%%
\bibitem{Aoki:2010dy}%
  \BibitemOpen
  \bibfield{author}{%
  \bibinfo {author} {\bibfnamefont{Y.}~\bibnamefont{Aoki}} \emph{et~al.}
  (\bibinfo {collaboration} {RBC, UKQCD}),\ }%
  \bibfield{journal}{%
  \Doi{10.1103/PhysRevD.83.074508}{\bibinfo {journal} {Phys.~Rev.}}\ }%
  \textbf{\bibinfo {volume} {D83}},\ \bibinfo {pages} {074508} (\bibinfo {year}
  {2011}),\ \Eprint{http://arxiv.org/abs/1011.0892}{arXiv:1011.0892 [hep-lat]}%
  \bibAnnoteFile{NoStop}{Aoki:2010dy}%
%%CITATION = ARXIV:1011.0892;%%
\bibitem{Borsanyi:2014jba}%
  \BibitemOpen
  \bibfield{author}{%
  \bibinfo {author} {\bibfnamefont{S.}~\bibnamefont{Borsanyi}} \emph{et~al.}}%
   (\bibinfo {year} {2014}),\
  \Eprint{http://arxiv.org/abs/1406.4088}{arXiv:1406.4088 [hep-lat]}%
  \bibAnnoteFile{NoStop}{Borsanyi:2014jba}%
%%CITATION = ARXIV:1406.4088;%%
\bibitem{Soldner:2015oea}%
  \BibitemOpen
  \bibfield{author}{%
  \bibinfo {author} {\bibfnamefont{W.}~\bibnamefont{S{\"o}ldner}} (\bibinfo
  {collaboration} {RQCD})}%
   (\bibinfo {year} {2015}),\
  \Eprint{http://arxiv.org/abs/1502.05481}{arXiv:1502.05481 [hep-lat]}%
  \bibAnnoteFile{NoStop}{Soldner:2015oea}%
%%CITATION = ARXIV:1502.05481;%%
\bibitem{Edwards:2004sx}%
  \BibitemOpen
  \bibfield{author}{%
  \bibinfo {author} {\bibfnamefont{R.~G.}\ \bibnamefont{Edwards}}\ and\
  \bibinfo {author} {\bibfnamefont{B.}~\bibnamefont{Jo{\`o}}} (\bibinfo
  {collaboration} {SciDAC , LHPC , UKQCD}),\ }%
  \bibfield{journal}{%
  \Doi{10.1016/j.nuclphysbps.2004.11.254}{\bibinfo {journal} {Nucl. Phys. Proc.
  Suppl.}}\ }%
  \textbf{\bibinfo {volume} {140}},\ \bibinfo {pages} {832} (\bibinfo {year}
  {2005})%
  \bibAnnoteFile{NoStop}{Edwards:2004sx}%
\end{thebibliography}%

\end{document}